\documentclass[usenatbib]{mn2e}
\usepackage{natbib}
\usepackage{apjfonts}
\usepackage{amsmath}

\usepackage{epsfig}
\usepackage{graphicx}
\usepackage{amssymb}
\usepackage{aas_macros}
\usepackage{color}
\newcommand{\conv}{\mathbin{*}}

\newcommand{\kms}{\,{\rm km \, s^{-1}}}
\newcommand{\pc}{\,{\rm pc}}
\newcommand{\kpc}{\,{\rm kpc}}

\newcommand{\AU}{\,{\rm AU}}
\newcommand{\oversim}[2]{\protect{\mbox{\lower0.5ex\vbox{%
   \baselineskip=0pt\lineskip=0.2ex
   \ialign{$\mathsurround=0pt #1\hfil##\hfil$\crcr#2\crcr\sim\crcr}}}}} 
\newcommand{\bb}[1]{\ifmmode \mbox{\boldmath $ #1$} \else  \mbox{\boldmath $#1$} \fi}
\catcode`\"=\active\let"=\"  

\def\3{{\ss} }

\def\c12{{1\over 2}}   
\def\erf{{\rm erf}}   
   
\def\d{{\rm d}}   
   
\def\plusplus{\raise 0.3ex\hbox{${\scriptstyle ++}$}{}}

\def\and{{{\rm M}31}}

\bibliography{biblio}

\begin{document}   
\title[Stochastic tidal heating]{Stochastic tidal heating by random interactions with extended substructures}

\author[Jorge Pe\~{n}arrubia]{Jorge Pe\~{n}arrubia$^{1}$\thanks{jorpega@roe.ac.uk}\\
$^1$Institute for Astronomy, University of Edinburgh, Royal Observatory, Blackford Hill, Edinburgh EH9 3HJ, UK\\
}
\maketitle  

\begin{abstract}
  Gravitating systems surrounded by a dynamic sea of substructures experience fluctuations of the local tidal field which inject kinetic energy into the internal motions.
  This paper uses stochastic calculus techniques to describe `tidal heating' as a random walk of orbital velocities that leads to diffusion in a 4-dimensional energy--angular momentum space.
In spherical, static potentials we derive analytical solutions for the Green's propagators directly from the number density and velocity distribution of substructures with known mass \& size functions without arbitrary cuts in forces or impact parameters.
 Furthermore, a Monte-Carlo method is presented, which samples velocity 'kicks' from a probability function and can be used to model orbital scattering in fully generic potentials.
  For illustration, we follow the evolution of planetary orbits in a clumpy environment.
   We show that stochastic heating of (mass-less) discs in a Keplerian potential leads to the formation, and subsequent ``evaporation'' of Oort-like clouds, and derive analytical expressions for the escape rate and the fraction of comets on retrograde orbits as a function of time.
Extrapolation of the subhalo mass function of Milky Way-like haloes down to the WIMP free-streaming length suggests that objects in the outer Solar system experience repeated interactions with dark microhaloes on dynamical time-scales.
\end{abstract}

\section{Introduction}\label{sec:intro}
Understanding the evolution of gravitating systems embedded in a clumpy medium is both theoretically and numerically challenging. First, because gravitational forces cannot be shielded, which thwarts any clear-cut definition of a physical `boundary' between the system and the surrounding background.
Second, because dynamical equilibrium can never be attained in regions where the dynamical time is comparable to the time-scale on which the external field fluctuates.
The analytical hurdles involved in the description of non-equilibrium systems subject to long-range forces (see Padmanabhan 1990; Lynden-Bell 1999 for reviews) often calls for the heuristic assumption of ``isolation'', whereby the contribution of distant objects to the local acceleration is ignored. 
However, this approximation breaks down on long time-scales, as the cumulative effect of repeated interactions with background objects dominates over the secular evolution of the system.

The first attempt to construct a statistical theory for the response of a single, free-moving, mass-less particle to a fluctuating external force is due to Chandrasekhar. He proposed three different approaches to tackle this problem.
In his first classical paper, Chandrasekhar (1941a) divides the force acting on a tracer particle into two components: a force that changes very slowly and can be expressed as the gradient of a smooth potential, plus a random contribution of ``chance stellar encounters'' of short duration . The computation of the random part relies on two assumptions (i) an infinite homogeneous medium, and (ii) independent encounters, such that the average effect of the clumpy background will be the sum of the effects of separate two-body encounters. Alas, Chandrasekhar finds that these simplifications lead to a total energy variation $\sum \Delta E^2$, where $\Delta E$ is the energy exchanged during a single encounter, that {\it diverges} when the integral over the minimum two-body separation (the so-called ``impact parameter'') is taken to distances comparable to the intra-particle separation, $D$, as well as when it tends to infinity. 
Both issues are typically resolved by truncating the range of impact parameters at small and large distances $b_{\rm min}$ and $b_{\rm max}$, respectively, which results in an energy variation that is proportional to an ill-defined Coulomb logarithm $\ln (b_{\rm max}/b_{\rm min})$, whose computation is a long-standing matter of debate (e.g. Just \& Pe\~narrubia 2005 and references therein).

Due to these shortcomings, Chandrasekhar (1941b) proposes to abandon the two-body approximation altogether, and argues in favour of stochastic methods that describe the {\it combined force} exerted by the background onto a test star, $\mathbfit F=\sum_{i=1}^N\mathbfit{f}_i$, which depends on the instantaneous relative position of $N\gg 1$ particles and is therefore subject to fluctuations. 
In this framework, the average acceleration experienced by a test star during a time interval $t>0$ leads to a net variation of the velocity $\langle \Delta \mathbfit{v}\rangle =\langle \mathbfit{F}\,T(F)\rangle$, and a variance $\langle | \Delta \mathbfit{v}|^2\rangle =t\,\langle F^2\,T(F)\rangle$, where brackets denote averages over the probability function $p(\mathbfit{F})$ to experience a force in the interval $\mathbfit{F},\mathbfit{F}+\d\mathbfit{F}$, and $T(F)$ is the mean-life of a fluctuation. Chandrasekhar (1941b) derives $p(\mathbfit{F})$ using a method originally devised by Holtsmark (1919) to study the motion of charged particles in a plasma, and suggests that $T(F)$ may correspond to Smoluchowski's (1916) time-scale associated with a stochastic system, yet also warning that ``Smoluchowski's ideas cannot be applied without further deep generalizations of them''. To this aim, Chandrasekhar \& von Neumann (1942; 1943) define the mean life of fluctuations as $T(F)=|\mathbfit{F}|/\sqrt{\langle |\d \mathbfit{F}/\d t|^2\rangle_F}$, where brackets denote an average over the bivariate distribution $W(\mathbfit{F},\d \mathbfit{F}/\d t)$, which determines the simultaneous probability to experience a force $\mathbfit{F}$ and an associated rate of change $\d \mathbfit{F}/\d t$.
Kandrup (1980) shows that Smoluchowski's and Chandrasekhar \& von Neumann's derivations of $T(F)$ yield consistent results modulo numerical constants of order unity. We return to this issue in \S\ref{sec:duration}.

A third approach was proposed by Chandrasekhar (1944), in which the response of tracer particles to fluctuating forces is treated as a Brownian motion in velocity space, where the first and second moments of the velocity increments correspond to the drift and diffusion coefficients, respectively. In an isotropic medium the drift coefficient vanishes by symmetry, $\langle \Delta {\mathbfit v}\rangle=0$, while the diffusion coefficient is computed as $\langle |\Delta {\mathbfit v}|^2\rangle=2t\int_0^\infty\d \tau \,\langle {\mathbfit F}_0 \cdot {\mathbfit F}_\tau\rangle$. Here brackets denote averages over the autocorrelation function $W(\mathbfit{F}_0,\mathbfit{F}_t)$, which determines the probability to feeling a force $\mathbfit{F}_0$ at an initial time $t_0=0$, and a force $\mathbfit{F}_t$ at a later time $t$. Chandrasekhar (1944) finds that when background objects are assumed to move on straight lines during a short time interval ($\tau$) the autocorrelation function is such that $\langle {\mathbfit F}_0 \cdot {\mathbfit F}_\tau\rangle\sim \tau^{-1}$, which leads to a diffusion coefficient that diverges logarithmically. Lee (1968) shows that the divergence arises from the assumption of an infinite medium, and argues in favour of truncating the range of impact parameters at large and small distances, which re-introduces the Coulomb logarithm found by Chandrasekhar (1941a) in the theory.

An appealing aspect of the stochastic analysis of Chandrasekhar (1941b) is the cancelation of the net force contribution of particles separated by large distances, which eliminates the need of an arbitrary cut-off at weak forces. Yet, this approach still requires a truncation at short separations, or strong forces, due to the divergent force generated by point-mass particles at $r\ll D$ (Chandrasekhar 1941b; Cohen et al. 1950; Kandrup 1980). Recently, Pe\~narrubia (2018) shows that this last hindrance can be removed from the theory by considering a background of {\em extended} substructures with individual forces $\mathbfit{f}$ that (i) are not centrally divergent, and (ii) approach a Keplerian limit $f\sim 1/r^2$ at distances $r\gg c$, where the $c$ is the substructure size  

This paper extends the work of Chandrasekhar to tracer particles moving in a smooth potential $\Phi_s$ subject to random fluctuations of an external tidal field, $\d \mathbfit {F}/\d \mathbfit{r}$.
The random component is generated by the combined gravitational field of a large number of extended substructures orbiting within a host potential $\Phi_g$ in dynamical equilibrium. Section~2 summarizes the results of Pe\~narrubia (2018), who computes the spectrum of tidal fluctuations generated by an homogeneous distribution of extended objects using Holtsmark (1919) statistical technique. 
Section 3 applies autocorrelation (Chandrasekhar 1944) and stochastic (Chandrasekhar 1941b) methods to derive the drift and diffusion coefficients ($\langle \Delta {\mathbfit v}\rangle$ and $\langle |\Delta {\mathbfit v}|^2\rangle$, respectively) from the number density and velocity distribution of substructures with fixed mass and size without arbitrary cuts in forces or impact parameters.
Both derivations are shown to provide consistent results modulus a numerical factor of order unity that arises from the different assumptions on which the two methods rest. However, while the derivation of the autocorrelation function assumes that background objects move on straight-line trajectories, the stochastic approach requires no a priori information on the motion of substructures in the host potential. The method proposed by Chandrasekhar (1941a), which treats individual encounters separately, is not explored here, as it requires a non-trivial treatment of the 3-body problem (e.g. Heggie \& Rasio 1996).
In Chandrasekhar's theories, the computation of diffusion coefficients relies on the impulsive approximation, which assumes that the location of a tracer particle does not vary appreciably during an encounter. In Section 3.3 we apply {\it adiabatic corrections} computed by Weinberg (1994a,b,c), which allow for the motion of tracer particles during the duration of a tidal fluctuation. 

Section~4 uses the probability theory presented by Pe\~narrubia (2015) to describe the non-equilibrium state of gravitating systems subject to random tidal interactions as a diffusion process in the integral-of-motion space. Analytical expressions for Green's functions are given for spherical, static potentials, where the integrals correspond to the energy ($E$), and the three components of the angular momentum ($\mathbfit{L}$). The derivation of the Green's functions is described in detail in Appendix B, and accounts for the fact that gravitationally-bound particles can only diffuse in a confined region of the integral-of-motion space. To this end, we set an {\it absorbing} boundary at $E=0$, such that particles with $E\ge 0$ escape from a gravitating system, and treat the angular momentum volume as a cubic box with {\it reflecting} surfaces placed at the angular momentum of circular orbits with a fixed energy, $L_c(E)$.

Section~5 tests our analytical framework by running $N$-body experiments where (i) the tidal tensor is computed directly from the relative positions of a population of extended substructures in dynamical equilibrium within a host potential, and (ii) $N$-body particles experience velocity ``kicks'' which are sampled from a probability function using Monte-Carlo techniques.

As an application of the theoretical methods, Section~6 follows the evolution of idealized mass-less discs at $t_0=0$ in a Keplerian potential. The time-dependent distribution of tracer particles in the 4-dimensional integral-of-motion space, $N(E,\mathbfit{L},t)$, is calculated as a convolution of the Green's functions with the initial distribution $N(E_0,\mathbfit{L}_0,t_0)$. Tidal evaporation rates, and the rate of production of retrograde orbits are derived analytically from the flux of particles crossing the boundaries $E=0$ and $L_z=0$, respectively. The results are compared against Monte-Carlo $N$-body models.

Following up on the results of Pe\~narrubia (2018), Section~7 discusses the effects of dark matter microhaloes on weakly-bound objects in the outskirts of the Solar system. Our analysis relies on bold extrapolations of the subhalo properties found in Milky Way-like haloes down to free-streaming mass-scales, and must be therefore taken with caution. At face value, the results suggest that objects in the Oort cloud may experience a very large number of interactions with microhaloes during a single orbital period, which opens up the interesting possibility to use Trans-Neptunian Objects (TNO's) to probe the (local) subhalo mass function on sub-solar mass scales.

\section{Stochastic fluctuations of the tidal field}\label{sec:forces}
This Section briefly summarizes the analytical framework of Pe\~narrubia (2018), hereafter Paper I, who introduce a statistical technique for deriving the spectrum of random fluctuations of the tidal field generated by a large population of extended substructures. 

\subsection{Probability theory}\label{sec:prob}
As a starting point, consider a tracer particle at a distance ${\mathbfit R}$ from the centre of a small, system with a self-gravitating potential $\Phi_s$ located at a galactocentric radius ${\mathbfit r}_s$ from a larger host galaxy. For simplicity, we shall work in the collision-less, mean-field limit, where the granularity in the system may be ignored. Hence, in the reference frame of the host galaxy
the gravitational acceleration experienced by a tracer particle can be written as
\begin{align}\label{eq:eqmot1}
\frac{\d^2 ({\mathbfit R}+{\mathbfit r}_s)}{\d t^2}=-\nabla\Phi_s({\mathbfit R})-\nabla\Phi_g({\mathbfit R}+{\mathbfit r}_s) + \sum_{i=1}^{N} {\mathbfit f}_i({\mathbfit R}+{\mathbfit r}_s),
\end{align}
where $\Phi_g$ is the mean-field gravitational potential of host galaxy, and
\begin{align}\label{eq:F}
{\mathbfit F}\equiv \sum_{i=1}^{N}{\mathbfit f}_i,
\end{align}
is the specific force induced by a set of {\it extended} substructures in a state of dynamical equilibrium within the host galaxy.

Similarly, the equations of motion that define the trajectory of the smaller system about the parent galaxy can be written as
\begin{align}\label{eq:eqmot2}
\frac{\d^2 {\mathbfit r}_s}{\d t^2}=-\nabla\Phi_g({\mathbfit r}_s) + \sum_{i=1}^{N} {\mathbfit f}_i({\mathbfit r}_s).
\end{align}

Taylor-expanding~(\ref{eq:eqmot1}) at first order and subtracting~(\ref{eq:eqmot2}) yields the well-known {\it tidal approximation}
\begin{align}\label{eq:eqmots}
\frac{{\d^2 \mathbfit R}}{\d t^2}=-\nabla\Phi_s({\mathbfit R}) + T_g\cdot {\mathbfit R} + \sum_{i=1}^{N} t_i\cdot {\mathbfit R} + \mathcal{O}(R/r_s),
\end{align}
where $T_g$ and ${t}_i$ are 3$\times$3 tidal tensors evaluated at the centre of the self-gravitating potential $\Phi_s$. The smooth component has a form
\begin{align}\label{eq:tt_s}
  {T}_{g}^{jk}\equiv -\frac{\partial^2 \Phi_g}{\partial x_j\partial x_k}, 
\end{align}
while the stochastic tidal tensor  
\begin{align}\label{eq:tt}
 {T}^{jk}\equiv  \sum_{i=1}^{N}{t}_i^{jk}= \sum_{i=1}^{N} \frac{\partial f^k_{i}}{\partial x^j}=\frac{\partial }{\partial x^j}\sum_{i=1}^{N} f^k_{i}=\frac{\partial F^k}{\partial x^j},
\end{align}
arises from the gradient in the combined tidal force induced by a set of $N-${substructures} distributed across the host galaxy.

Paper I shows that the stochastic term of the tidal force can be approximately written as
\begin{align}\label{eq:Flambda}
    {\mathbfit F}_t\equiv\sum_{i=1}^N t_i\cdot {\mathbfit R} \approx R\sum_{i=1}^N {\bb \lambda}_i,
\end{align}
where $\lambda ={\rm Trace}( {t}_e)$ is the sum of eigenvalues of the effective tidal tensor $t_e=t+  \nabla f_c$, and ${\mathbfit f}_c={\bb \Omega}\times ({\bb \Omega}\times {\mathbfit R})$ is the centrifugal force component in a frame that co-rotates with the angular velocity of individual substructures, ${\bb \Omega}$ (see also Renaud et al. 2011). Note that the vectors ${\bb \lambda}_i$ have random directions if substructures are isotropically distributed around the test particle. As discussed in Paper I, Equation~(\ref{eq:Flambda}) neglects the Euler and Coriolis terms appearing in the non-inertial rest frame, which is a reasonable approximation in an impulsive regime, where one can assume that the angular frequency ${\bb \Omega}$ remains constant during the encounter duration. We will return to this issue in Section~\ref{sec:heating}.

In analogy with Equation~(\ref{eq:F}), it is useful to define the tidal vector 
\begin{align}\label{eq:Lambda}
  {\bb \Lambda}\equiv \sum_{i=1}^N {\bb \lambda}_i,
\end{align}
such that the combined tidal force~(\ref{eq:Flambda}) becomes ${\mathbfit F}_t=\bb \Lambda R$.
A large population of extended substructures homogeneously distributed within a volume $V'=4\pi d^3/3$ around a test particle generates a stochastic tidal field that is fully specified by the probability density $p(\bb \Lambda)$, which defines the probability of experiencing a tidal vector in the interval ${\bb \Lambda}, {\bb \Lambda}+\d{\bb \Lambda}$. 
Following up the method of Holtsmark (1919), which was originally devised to study the motion of charged particles in a plasma, Paper I derives the spectrum of tidal fluctuations as
\begin{align}\label{eq:plam}
  p({\bb \Lambda})&=\frac{1}{V'}\int \d^3 r_1 \times ...\times \frac{1}{V'}\int \d^3 r_N\delta\big({\bb \Lambda}-\sum_i{\bb \lambda}_i\big)\\\nonumber
  & \approx \frac{1}{(2\pi)^3}\int \d^3k \exp\big[-i{\mathbfit k}\cdot{\bb \Lambda}-\phi({\mathbfit k})\big]~~~~ {\rm for}~~~~  N\gg 1,
\end{align}
where $\delta$ is the Dirac's delta function and
\begin{align}\label{eq:phi}
  \phi({\mathbfit k})\equiv n\int_{V'} \d^3 r\big(1-e^{i {\mathbfit k}\cdot {\bb \lambda}({\mathbfit r}) }\big),
\end{align}
with $n\equiv N/V'$ denoting the number density of substructures. The above derivation can be generalized to inhomogeneous substructure distributions as (Chandrasekhar 1941b; Kandrup 1980; Chavanis 2009)
\begin{align}\label{eq:phi_inhomog}
  \phi_{\rm in}({\mathbfit k})= \int\d^3 r\big(1-e^{i{\mathbfit k}\cdot\bb \lambda({\mathbfit r})}\big) n({\mathbfit r}),
\end{align}
where $n({\mathbfit r})$ is the number density profile, and ${\mathbfit r}$ is centred at the location of the test particle. In the {\it local approximation} the number density can be assumed to be roughly constant, $n({\mathbfit R}+{\mathbfit r}_s)\approx n({\mathbfit r}_s)=n$, thus recovering~(\ref{eq:phi}). 
Paper I shows that this approximation holds in regions where the number density profile varies on scales that are much larger than the averaged separation between subhaloes, i.e. $|\nabla n/n|^{-1}\gg D$, where the distance $D$ can be measured from the probability of finding the closest substructure within the volume $V'$ 
\begin{align}\label{eq:pr_closest}
  p({\mathbfit r})\d^3 r \sim \exp\big(-\frac{4}{3}\pi r^3 n\big )4\pi r^2 n \d r,
    \end{align}
which peaks at $D\equiv (2\pi n)^{-1/3}$. 

In general, the probability density $p({\bb \Lambda})$ can be rarely expressed in an analytical form. An interesting exception with broad applications in cosmology corresponds to a large population of Hernquist (1990) spheres with a density profile
$$\rho(r)= \frac{M}{2\pi c^3}\frac{1}{(r/c)(1+r/c)^3},$$
where $M$ and $c$ are the substructure mass and scale length, respectively. Individual substructures induce a specific tidal force
\begin{align}\label{eq:fh}
{\mathbfit f}=-\frac{GM}{(r+c)^2} \hat{\mathbfit r},
\end{align}
which approaches the Keplerian (`particle') limit as $c\to 0$. The sum of eigenvalues associated with the force~(\ref{eq:fh}) can be straightforwardly calculated using Renaud et al. (2011) formalism
\begin{align}\label{eq:lambdah}
\bb \lambda=\frac{2GM}{(r+c)^3}\hat{\mathbfit u},
\end{align}
where $\hat{\mathbfit u}$ is a unit vector pointing in a random direction. The maximum value of Equation~(\ref{eq:lambdah}) is $\lambda_0\equiv 2GM/c^3$ at $r=0$.

Given that an ensemble of self-gravitating, overlapping objects is not dynamically stable, in what follows we limit our analysis to populations of Hernquist (1990) spheres with sizes much smaller than their typical separation ($c\ll D$). In this regime, which Paper I characterizes as 'rarefied' in analogy with the kinetic theory of gases, the combined tidal vector ${\bb \Lambda}$ fluctuates stochastically with a probability density~(\ref{eq:plam}) that can be expressed analytically as
\begin{align}\label{eq:plam_pm}
  p({\bb \Lambda})\simeq \frac{C}{\pi^2q^3}\frac{1}{(1+\xi^2)^2}\big[1-\big(q/\lambda_0\big)^{1/3}\xi^{1/3}\big]^2~~~~~~{\rm for}~~\Lambda< \lambda_0,
\end{align}
where $\xi\equiv \Lambda/q$ is a dimension-less quantity, $q\equiv \frac{2\pi^2}{3}GMn=(\pi/3)(GM/D^3)$, and $C$ is a normalization constant that guarantees $\int \d^3\Lambda p(\bb \Lambda)=1$. The distribution~(\ref{eq:plam_pm}) obeys the isotropic condition $p(\bb \Lambda)=p(\Lambda)$, which implies that the tidal vectors ${\bb \Lambda}$ point in random directions. Note that in the {\it weak-force limit} ($\Lambda\ll q)$ the probability function approaches asymptotically a constant value, which implies that the effect of having an increasing number of particles at large distances exactly balances with the declining force, such that $p(\bb \Lambda)$ becomes flat at small accelerations. The {\it strong-force limit} ($\Lambda\gg q)$ is dominated by the contribution of the nearest object (see Paper I), thus the probability density is truncated at the maximum force derivative exerted by an individual substructure~(\ref{eq:lambdah}), such that $p(\bb \Lambda)=0$ for $\Lambda>\lambda_0=2GM/c^3$.

As we shall see below, the second moment of $p({\bb \Lambda})$ is particularly relevant for understanding the dynamical response of a tracer particle subject to a stochastic tidal field. After some algebra, Paper~I finds that
\begin{align}\label{eq:lamvar2}
  \langle \Lambda^2\rangle &= \int_0^{\lambda_0}\d^3\Lambda\,p({\bb \Lambda})\Lambda^2\\ \nonumber
  &=\frac{4C}{\pi}q^2\int_0^{\lambda_0/q}\d\xi\frac{\xi^4}{(1+\xi^2)^2}\big[1-\big(q/\lambda_0\big)^{1/3}\xi^{1/3}\big]^2\\ \nonumber 
  &\simeq \frac{8\pi}{15}\frac{(GM)^2n}{c^3},
\end{align}
which diverges in the particle (point-masss) limit $c\to 0$. In practice, the divergence of $\langle \Lambda^2\rangle$ means that as the time progresses the maximum tidal force experienced by a tracer particle can grow up to arbitrarily-large values.

\section{Tidal heating}\label{sec:heating}
A test particle surrounded by a large population of moving substructures experiences random fluctuations of the local tidal force due to the rapid change of the (relative) position of nearby objects.
Over a sufficiently long interval of time the cumulative effect of multiple encounters contributes to the randomization of the peculiar velocity.
If the velocity impulses are small, $|\Delta{\mathbfit V}|<<|{\mathbfit V}|$, the fluctuations can be thought to occur independently, and the average effect of a series of fluctuations will be the sum of the expectation values of the effect of a single fluctuation. With these simplifications in place, the formalism of Brownian motion can be used to describe the response of self-gravitating objects to stochastic variations of a gravitational field (e.g. Chandrasekhar 1943), which reduces the problem of tidal heating to the computation of diffusion coefficients.

The averaged velocity increments acquired by tracer particles over short intervals of time, $\langle \Delta {\mathbfit V}\rangle$ and $\langle |\Delta {\mathbfit V}|^2\rangle$ -- here brackets denote averages over multiple interactions-- play a key role in the Brownian motion theory.
Below we explore two independent methods to calculate these quantities. The first method (\S\ref{sec:line}) follows the usual derivation of velocity increments, where one assumes that (i) the speed of fluctuations is much faster than the orbital velocity of the tracer particle in the potential $\Phi_s$, and (ii) external substructures move on linear trajectories within the host potential $\Phi_g$.  The second approach is presented in \S\ref{sec:duration} and follows up on the arguments of Chandrasekhar (1941b) and Kandrup (1980) to derive the speed of fluctuations directly from the phase-space distribution of substructures, without making assumptions on the orbital motion of these objects. Section~\ref{sec:adiab} extends this formalism to encounters in a non-impulsive regime by applying Weinberg (1994a,b,c) adiabatic corrections to the coefficients obtained in~\S\ref{sec:duration}.

\subsection{Impulse/Straight-line approximation}\label{sec:line}
The classical derivation of $\langle \Delta {\mathbfit V}\rangle$ and $\langle |\Delta {\mathbfit V}|^2\rangle$ relies on two key assumptions: (i) substructures move on straight lines, and (ii) the typical duration of tidal fluctuations is much shorter than the orbital period of the tracer particle about the potential $\Phi_s$. Under these conditions it is relatively straightforward to extend the analysis of Lee (1968) to velocity increments arising from stochastic fluctuations of an external tidal field.

Consider first a test particle orbiting in a potential $\Phi_s(\mathbfit{R})$ with an orbital frequency, $w\equiv V/R$, where $R$ and $V$ are the moduli of the position and velocity vectors, respectively. During a short time interval, $t \ll w^{-1}$, the particle position does not change appreciably. In contrast, from Equation~(\ref{eq:Flambda}) the net contribution of the stochastic tidal forces leads to a velocity variation 
\begin{align}\label{eq:delv}
   \Delta{\mathbfit V}=\int_0^t \d s\,{\mathbfit F}_t(s)\approx R\int_0^t\d s\, {\bb \Lambda}_s.
\end{align}
where ${\bb \Lambda}_s=\bb \Lambda(s)$ is the tidal vector acting on the tracer particle at the time $s$. Equation~(\ref{eq:delv}) is typically known as the {\it impulse approximation}. 
For a distribution of substructures uniformly distributed around the tracer particle the distribution of tidal fluctuations is isotropic, $p(\bb \Lambda)=p(\Lambda)$. Hence, by symmetry, the average velocity increment is
\begin{align}\label{eq:delva}
  \langle \Delta{\mathbfit V}\rangle &= R\int_0^t\d s\, \langle {\bb \Lambda}_s\rangle \\ \nonumber
  &=R\int_0^t\d s\, \int\d^3 \Lambda_s p(\Lambda_s){\bb \Lambda}_s=0.
\end{align}

The computation of the squared velocity increment is considerably more involved. From Equation~(\ref{eq:delv})
\begin{align}\label{eq:delv2}
   |\Delta{\mathbfit V}|^2=R^2\int_0^t\d s \int_0^t \d s'{\bb \Lambda}_s\cdot {\bb \Lambda}_{s'},
\end{align}
where ${\bb \Lambda}_s$ and ${\bb \Lambda}_{s'}$ are tidal vectors acting on a tracer particle at the times $s$ and $s'$, respectively.
Recall that our working assumptions are that (i) test particles suffer a large number of individual encounters with neighbour substructures during a time interval $t$, and (ii) velocity increments induced by subsequent encounters are statistically uncorrelated. Under those conditions, the product of tidal vectors only depends on the length of the time interval $\tau=s'-s$, such that ${\bb \Lambda}_s\cdot {\bb \Lambda}_{s+\tau}={\bb \Lambda}_0\cdot {\bb \Lambda}_\tau$. If the function ${\bb \Lambda}_0\cdot {\bb \Lambda}_\tau$ decreases more rapidly than $\tau^{-1}$ for large $\tau$, as it happens in most cases of astrophysical interest, then one can take the limit $t\to\infty$. Hence, the averaged squared velocity increment~(\ref{eq:delv2}) can be written as a single integral (see Appendix A of Lee 1968 for a detailed derivation)
\begin{align}\label{eq:delv2_2}
  \langle |\Delta{\mathbfit V}|^2\rangle\simeq R^2 2 t\int_0^\infty \d \tau\, \langle {\bb \Lambda}_0\cdot {\bb \Lambda}_\tau\rangle.
\end{align}
To compute $\langle {\bb \Lambda}_0\cdot {\bb \Lambda}_\tau\rangle$ we must take into account that the tidal vectors ${\bb \Lambda}_0$ and ${\bb \Lambda}_\tau$ are not statistically independent. Indeed, these quantities are related through the trajectories of individual substructures in the host potential. Following Chandrasekhar (1944), let us define the autocorrelation function $W({\bb \Lambda}_0,{\bb \Lambda}_\tau)$, which gives the probability that a tracer particle experiences a tidal vector ${\bb \Lambda}_0$ at $t=0$, and  ${\bb \Lambda}_\tau$ at a later time $t=\tau$. In the straight-line approximation, a substructure with an initial position vector ${\mathbfit r}$ and a relative velocity ${\mathbfit v}$ will be located at ${\mathbfit r}'={\mathbfit r}+{\mathbfit v}\tau$ at $t=\tau$. 
Hence, for a population of substructures with a distribution function $n({\mathbfit r},{\mathbfit v})=n f({\mathbfit v})$, where $n\int_{r<d}\d^3 r\int ^3 v \,f({\mathbfit v})=N$, the autocorrelation function can be written as (see Appendix A)
\begin{align}\label{eq:W}
  W({\bb \Lambda}_0,{\bb \Lambda}_\tau)=\int\frac{\d^3 k_0}{(2\pi)^3}\int\frac{\d^3 k_\tau}{(2\pi)^3} \exp\bigg[-i({\mathbfit k}_0\cdot\bb\Lambda_0+{\mathbfit k}_\tau\cdot\bb\Lambda_\tau) - \phi({\mathbfit k}_0,{\mathbfit k}_\tau)\bigg],
\end{align}
where
\begin{align}\label{eq:phiw}
  \phi({\mathbfit k}_0,{\mathbfit k}_\tau)= n\int \d^3v f({\mathbfit v})\int_{V'}\d^3 r \bigg[1-\exp\bigg(i {\mathbfit k}_0\cdot {\bb \lambda}({\mathbfit r}) +  i{\mathbfit k}_\tau\cdot {\bb \lambda}({\mathbfit r}+{\mathbfit v}\tau)\bigg) \bigg].
\end{align}

The explicit evaluation of $W({\bb \Lambda}_0,{\bb \Lambda}_\tau)$ is difficult, but the second moment is readily found from its Fourier transform $\tilde W({\mathbfit k}_0,{\mathbfit k}_\tau)$ (see Appendix A) as
\begin{align}\label{eq:l0lt}
  \langle {\bb \Lambda}_0\cdot {\bb \Lambda}_\tau\rangle&=-\bigg(\frac{\partial^2 \tilde W}{\partial {\mathbfit k}_0\cdot \partial {\mathbfit k}_\tau}\bigg)_{|{\mathbfit k}_0|\to 0, |{\mathbfit k}_t|\to 0}\\ \nonumber
  &=-\bigg(\frac{\partial \phi}{\partial {\mathbfit k}_0}\cdot \frac{\partial \phi}{\partial {\mathbfit k}_\tau} + \frac{\partial^2 \phi}{\partial {\mathbfit k}_0\cdot \partial {\mathbfit k}_\tau}\bigg)_{|{\mathbfit k}_0|\to 0, |{\mathbfit k}_t|\to 0}\\ \nonumber
    &= n\int \d^3 v f({\mathbfit v})\int_{V'} \d^3 r \,\bb\lambda({\mathbfit r})\cdot\bb\lambda({\mathbfit r} +{\mathbfit v}\tau),
\end{align}
with $\partial \phi/\partial {\mathbfit k}_0=\partial \phi/\partial {\mathbfit k}_\tau=0$ by symmetry. For most cases of astrophysical interest, $\lambda(r)$ decreases more rapidly than $r^{-3/2}$ at large radii, and one can let $V'\to \infty$ for simplicity.

For a random population of Hernquist (1990) spheres, the averaged squared velocity increment~(\ref{eq:delv2_2}) can be expressed analytically under a proper choice of coordinates.
Let us adopt a coordinate frame in which the tangential velocity of the substructure is parallel to the position vector ${\mathbfit r}$ at $t=0$. During a short time interval, $\tau\ll w^{-1}$, the norm of the orbital plane, $\hat{\bb \Omega}$, remains approximately constant, whereas the substructure moves to a new location ${\mathbfit r}'={\mathbfit r}+{\mathbfit v}\tau=(r+v\tau)\hat{\mathbfit r}$. Using the framework of \S\ref{sec:prob}, it straightforward to show that in this configuration the direction of the centrifugal acceleration remains invariant $\hat{\mathbfit f}_c'=\hat{\bb \Omega}'\times[\hat{\bb \Omega}'\times \hat{\mathbfit r}']=\hat{\bb \Omega}\times[\hat{\bb \Omega}\times \hat{\mathbfit r}]$, and one can set $\hat {\bb \lambda}\cdot \hat {\bb \lambda}'\approx 1$ in~(\ref{eq:l0lt}). Combination of~(\ref{eq:l0lt}),~(\ref{eq:delv2_2}) and~(\ref{eq:lambdah}) then yields
\begin{align}\label{eq:delv2_3}
  \langle |\Delta{\mathbfit V}|^2\rangle&= t\,R^2 2 n   \int_0^\infty \d \tau \int \d^3 v f({\mathbfit v})\int \d^3 r \frac{2 GM}{(r+c)^3}\frac{2 GM}{(r+ v\tau+c)^3} \\ \nonumber
  &= t\,R^2 8 (GM)^2 n \int \d^3 v f({\mathbfit v})\int  \frac{\d^3 r}{(r+c)^3}\int_0^\infty  \frac{\d \tau}{(r+v\tau+c)^3} \\ \nonumber
  &= t\,R^2 16\pi (GM)^2n  \int \d^3 v \frac{f({\mathbfit v})}{v}\int_0^\infty  \d r\frac{ r^2}{(r+c)^5}\\\nonumber
  &= t\,R^2 \frac{4\pi}{3} \frac{(GM)^2}{c^2}n \int \d^3 v \frac{f({\mathbfit v})}{v},
  \end{align}
which also diverges in the particle limit $c\to 0$. The quantity $\langle 1/v\rangle=\int\d^3 v f({\mathbfit v})/v$ is the reciprocal of a characteristic velocity whose meaning is discussed in \S\ref{sec:duration}.

\begin{figure}
\begin{center}
\includegraphics[width=84mm]{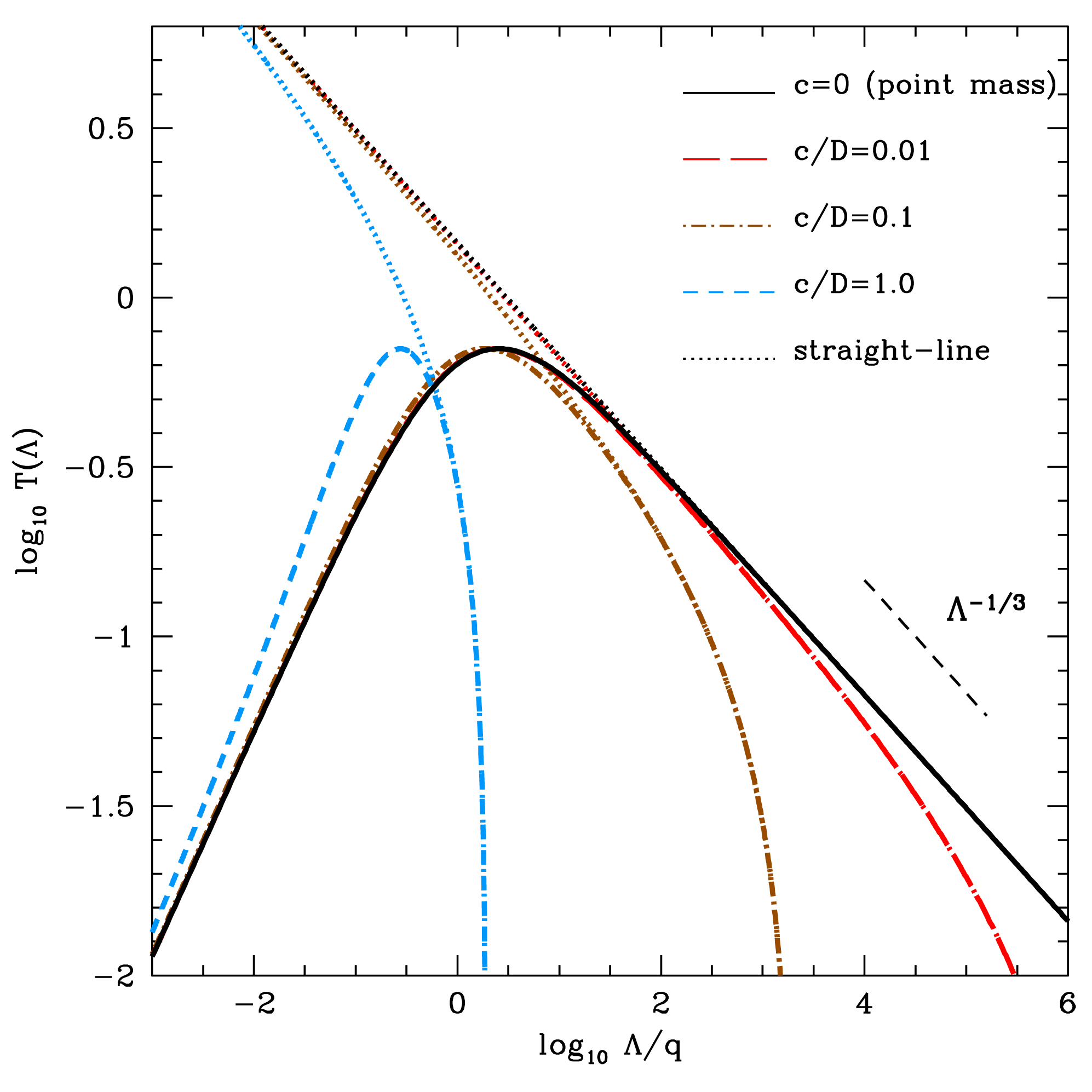}
\end{center}
\caption{Duration of tidal fluctuations~(\ref{eq:TL}) induced by a random population of extended substructures with a mass $G=M=1$ and scale length $c$, moving with velocities $\langle v^2\rangle=1$ and separated by an average distance $D=(2\pi n)^{-1/3}=1$.  The modulus of the tidal vector $\Lambda$ is given in units of $q=2\pi^2 G M n/3\simeq GM/D^3$. Note that the straight-line approximation, Equation~(\ref{eq:TL_sl}) (dotted lines), is accurate at large forces ($\Lambda\gg q$), but seriously over-estimates the duration of weak fluctuations ($\Lambda\ll q$). These curves peak at a characteristic time-scale $T(\Lambda)=T_{\rm ch}$ given by Equation~(\ref{eq:Tch}).}
\label{fig:TL}
\end{figure}

\subsection{Stochastic approach}\label{sec:duration}
The autocorrelation method described in \S\ref{sec:line} rests upon the assumption substructures move on straight-line trajectories. This condition is not required in a stochastic approach, where fluctuations of the combined tidal field arise from the presence at some point of time of substructures in the vicinity of the test particle.

Suppose that at $t=0$ there are $N$ particles distributed within a volume $V'=4\pi r^3/3$ around the test particle. The number of neighbours inside $V'$ will change either because one of the substructures exists this volume, or because another substructure enters from outside $V'$. Smoluchowski (1916) shows that the probability $P_N(t)$ that at some later time there are still $N$ substructures inside $V'$ may be written as $P_N(t)\d t =e^{-t/T} \d t/T$. Here, the time interval $T$ corresponds to the mean life of a state in which the number of substructures within $V'$ remains constant. 
For a uniform distribution of substructures with a number density $n=N/V'$ and a mean squared (relative) velocity $\langle v^2\rangle$, Smoluchowski (1916) finds (see also Chandrasekhar 1941b)
\begin{align}\label{eq:TR}
  T(r)={\sqrt\frac{2\pi}{3 \langle v^2\rangle} }\frac{r}{\frac{4\pi }{3}r^3n+1}.
\end{align}
Equation~(\ref{eq:TR}) has simple asymptotic behaviours. At small distances, $4\pi r^3 n/3 = (2/3) (r/D)^3\ll 1$, it recovers the time-scale derived from the straight-line approximation, $T\sim r/\sqrt{\langle v^2\rangle}$, which simply corresponds to the time that a substructure moving at a constant speed $\sqrt{\langle v^2\rangle}$ would take to cross the radius $r$. 
At large radii, however, the probability that a substructure leaves/enters $V'$ becomes proportional to the number of substructures within this volume. 
Accordingly, in the limit $4\pi r^3 n/3 \gg 1$ Smoluchowski's timescale becomes inversely proportional to $N=n V'$, such that $T\sim r/\sqrt{\langle v^2\rangle}N^{-1}\sim r^{-2}$, which vanishes in the limit $r\to \infty$.

Following Chandrasekhar (1941b) and Paper I, let us assume that the tidal field acting on a tracer particle is entirely dominated by the nearest substructure. Hence, from Equation~(\ref{eq:lambdah}) one can identify $r+c=(2GM/\Lambda)^{1/3}$. The duration of the tidal fluctuation~(\ref{eq:TR}) then becomes
\begin{align}\label{eq:TL}
  T(\Lambda)
  &={\sqrt\frac{2\pi}{3 \langle v^2\rangle}}\frac{(2GM/\Lambda)^{1/3}-c}{\frac{4\pi}{3}[(2GM/\Lambda)^{1/3}-c]^3n+1}\\ \nonumber
   &=T_0\frac{[1-(\Lambda/\lambda_0)^{1/3}](\Lambda/q)^{2/3}}{(\frac{4}{\pi})[1-(\Lambda/\lambda_0)^{1/3}]^3 +\Lambda/q},
 \end{align}
where the time-scale 
$T_0\equiv D\big(\frac{6}{\pi}\big)^{1/3}\sqrt\frac{2\pi}{3 \langle v^2\rangle}\approx 1.8 \frac{D}{\sqrt{\langle v^2\rangle}}$ approximately corresponds to the average time that it takes a substructure to travel twice the distance $D$.

In the straight-line limit, $4\pi nr^3/3\ll 1$, Equation~(\ref{eq:TL}) has a simple form
\begin{align}\label{eq:TL_sl}
    T_{sl}(\Lambda)= {\sqrt\frac{2\pi}{3 \langle v^2\rangle}}r= T_0\frac{1-(\Lambda/\lambda_0)^{1/3}}{(\Lambda/q)^{1/3}},
\end{align}
which diverges as $T_{sl}\sim \Lambda^{-1/3}$ in the limit $\Lambda\to0$.

For point-mass particles, $c\to 0\, (\lambda_0\to \infty)$, Equation~(\ref{eq:TL}) reduces to
\begin{align}\label{eq:TL_pm}
  T_{c=0}(\Lambda)=T_0\frac{(\Lambda/q)^{2/3}}{\frac{4}{\pi} +\Lambda/q},
\end{align}
Thus, in the strong-force regime $\Lambda\gg q$ the duration of the fluctuations scales as a power-law $T_{c=0}\sim \Lambda^{-1/3}$, which matches the straight-line asymptotic behaviour of Equation~(\ref{eq:TL_sl}). In contrast, for weak forces, $\Lambda\ll q$, Equation~(\ref{eq:TL_pm}) scales as $T_{c=0}\sim \Lambda^{2/3}$, thus vanishing in the limit $\Lambda\to 0$, which strongly deviates from the {\it divergent} duration one would expect for substructures moving on straight-line trajectories. This result was first noticed by Chandrasekhar \& von Neumann (1942), and is discussed in detail by Kandrup (1980), who warns that adopting the straight-line approximation leads to a serious overestimation of the contribution of distant encounters to the velocity increments derived in \S\ref{sec:line}.

To illustrate these results, Fig.~\ref{fig:TL} shows the duration of fluctuations induced by substructures with a mass $G=M=1$ and an average separation $D=1$, moving with a relative velocity dispersion $\langle v^2\rangle =1$. As expected, comparison between the duration expected in the straight-line approximation (dotted lines) and Smoluchowski's time-scale $T(\Lambda)$ shows good agreement at large forces ($\Lambda\gg q$), but strongly disagree in the weak-force regime $\Lambda\ll q$, where the straight-line assumption leads to a divergent $T(\Lambda)$. For extended substructures, the duration of tidal fluctuations vanishes in the limits $\Lambda\to 0$ and $\Lambda\to \lambda_0$. In the particle-limit $\lambda_0\to \infty$, it approaches a power-law behaviour $T\sim \Lambda^{-1/3}$ at $\Lambda\gg q$.

The maximum of $T(\Lambda)$ defines a characteristic time span, $T_{ch}$, associated with the longest, and therefore most likely, tidal fluctuation, $\Lambda_{ch}$. After some algebra, one can show that the solution to $\frac{\d T}{\d \Lambda}(\Lambda_{ch})=0$, with $T(\Lambda)$ given by~(\ref{eq:TL}), is
\begin{align}\label{eq:Lch}
 \Lambda_{ch}=\frac{48q }{\pi (6^{1/3} + 2c/D)^3}.
\end{align}
Note that the characteristic amplitude~(\ref{eq:Lch}) shifts towards smaller values as the ratio $c/D$ increases. Inserting~(\ref{eq:Lch}) into~(\ref{eq:TL}) returns the characteristic duration of tidal fluctuations
\begin{align}\label{eq:Tch}
 T_{ch}=T(\Lambda_{ch})=\frac{\pi^{1/3}}{3}T_0\approx  0.88\frac{D}{\sqrt{\langle v^2\rangle}},
\end{align}
which does not depend on substructure mass or size.

\begin{figure*}
\begin{center}
\includegraphics[width=162mm]{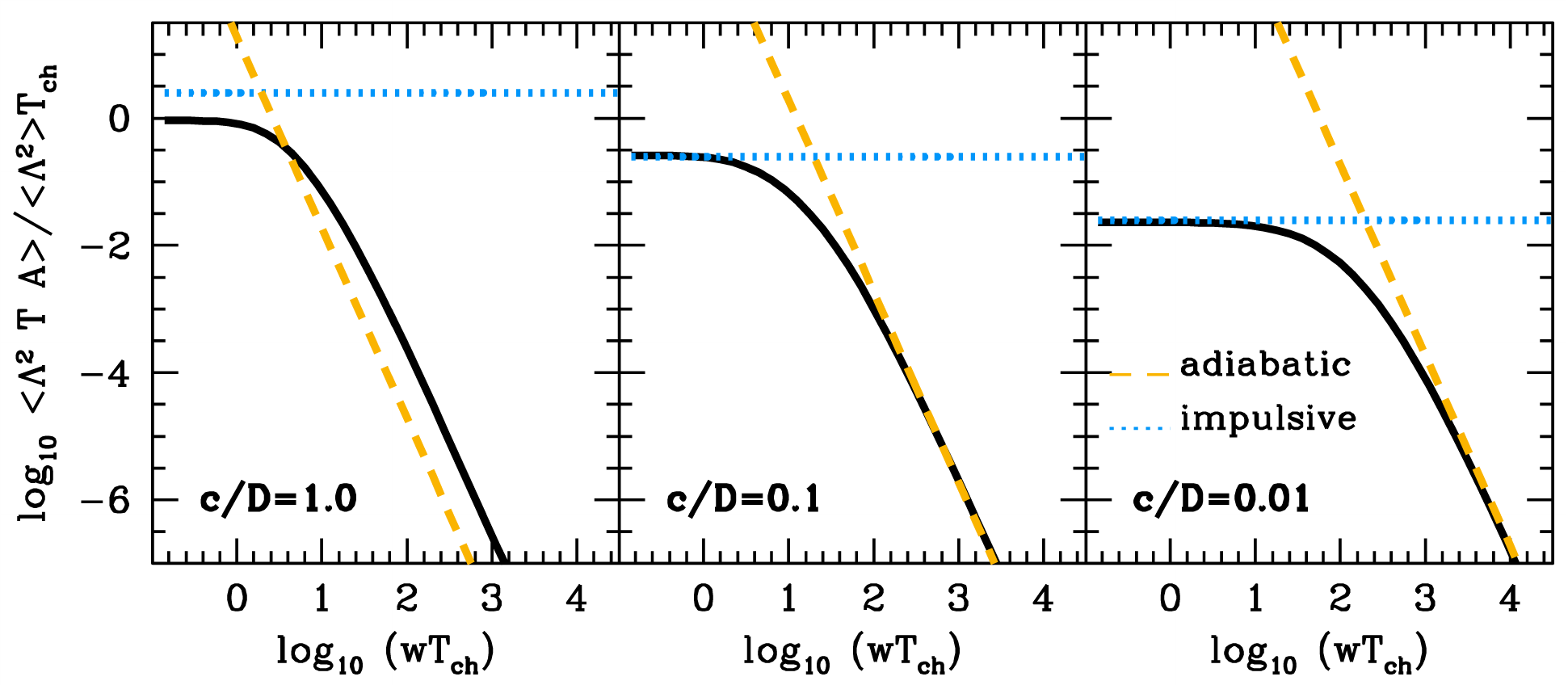}
\end{center}
\caption{Numerically-evaluated $\langle \Lambda^2 T A\rangle$ (black solid lines, Equation~\ref{eq:L2TA}) as a function of the dimensionless frequency $w T_{ch}$, where $T_{ch}$ is the characteristic duration of tidal fluctuations~(\ref{eq:Tch}), and $w=V/R$ is the orbital frequency of a test particle in the potential $\Phi_s$. The y-axis is given in units of $\langle \Lambda^2\rangle T_{ch}$, with $\langle \Lambda^2\rangle$ given by Equation~(\ref{eq:lamvar2}). For this plot we use $G=M=D=\langle v^2\rangle =1$. 
  Blue-dotted and orange-dashed lines show the impulsive and adiabatic approximations derived from Equations~(\ref{eq:L2T}) and~(\ref{eq:L2Ad}), respectively. Note that the impulsive regime extends over an increasingly larger range of orbital frequencies as the size-to-distance ratio $c/D$ decreases.}
\label{fig:L2TA}
\end{figure*}

Following Chandrasekhar (1941b), let us model impulsive velocity increments induced by a fluctuating tidal field as random-walk process in velocity space. For a substructure population distributed homogeneously around the tracer particle, the distribution of velocity impulses is isotropic and has a Gaussian form (Chandrasekhar 1943; Kandrup 1980)
\begin{align}\label{eq:Psi}
 \Psi({\mathbfit V}, \Delta {\mathbfit V},t)=\frac{1}{(\frac{2\pi}{3}\langle |\Delta{\mathbfit V}|^2\rangle)^{3/2}}\exp\big[-\frac{(\Delta{\mathbfit V}-\langle \Delta{\mathbfit V}\rangle)^2}{\frac{2}{3} \langle |\Delta{\mathbfit V}|^2\rangle}\big];
\end{align}
where $\Psi({\mathbfit V}, \Delta {\mathbfit V},t)$ denotes the probability that a test particle with a velocity ${\mathbfit V}$ will experience a velocity impulse $\Delta{\mathbfit V}$ within a time interval $t$.
From Equation~(\ref{eq:delva}), the average velocity increment vanishes by symmetry, $\langle \Delta{\mathbfit V}\rangle=0$, whereas the variance given by Equation~(\ref{eq:delv2_2}) can be re-written as\footnote{See Equation~(62) of Chandrasekhar (1941b).}
\begin{align}\label{eq:delv2_T}
  \langle |\Delta{\mathbfit V}|^2\rangle=t \,R^2 \langle \Lambda^2 T\rangle=
  t\,R^2 \int \d^3\Lambda p(\bb \Lambda)\, \Lambda^2 T(\Lambda).
\end{align}
Note that Equation~(\ref{eq:delv2_T}) obeys the ergodic property, as the contribution of a tidal fluctuation with a magnitude $\Lambda$ to the averaged velocity impulse is weighted by the {\it duration} of the fluctuation, $T(\Lambda)$.

In order to find an analytical solution to Equation~(\ref{eq:delv2_T}) it is convenient to approximate $T(\Lambda)\simeq T_{sl}(\Lambda)$. 
As shown in Fig.~\ref{fig:TL}, this approximation is accurate at large forces, $\Lambda\gg q$, which dominate the behaviour of the variance $\langle \Lambda^2\rangle$ (see Paper I).
In addition, it is useful to introduce the dimensionless quantity $\chi=q/\lambda_0=(\pi/6)(c/D)^3$, which approaches asymptotically $\chi\to 0$ as $c\to 0$. Combination of~(\ref{eq:plam_pm}) and~(\ref{eq:TL_sl}) then yields
\begin{align}\label{eq:L2T_aux}
  \langle \Lambda^2 T\rangle&\approx\langle \Lambda^2 T_{sl}\rangle \\ \nonumber
  &=T_0\frac{4 q^2 C}{\pi}\int_0^{1/\chi}\d\xi\,\frac{\xi^4}{(1+\xi^2)^2}\frac{[1-(\xi\chi)^{1/3}]^3}{\xi^{1/3}}\\ \nonumber
  &=T_0\frac{4 q^2 C}{\pi}\big[\frac{3}{20\chi^{2/3}}-\frac{4\pi}{3\sqrt{3}}-\frac{5\pi\chi^{2/3}}{\sqrt{3}}+\frac{11\pi \chi}{6}+\mathcal{O}(\chi^{4/3})\big].
\end{align}
In the limit $\chi\ll 1$ one has $C\to 1$, hence at leading order one can write~(\ref{eq:L2T_aux}) as
\begin{align}\label{eq:L2T}
  \langle \Lambda^2 T\rangle\approx \langle \Lambda^2 T_{sl}\rangle &\simeq T_0\frac{4 q^2 }{\pi}\frac{3}{20\chi^{2/3}} \\ \nonumber
  &=\frac{4\pi}{5}\bigg(\frac{GM}{c}\bigg)^2 n {\sqrt\frac{2\pi}{3 \langle v^2\rangle}}.
\end{align}
Thus, the variance of the velocity increments~(\ref{eq:delv2_T})` becomes
\begin{align}\label{eq:delv2_Tsl}
  \langle |\Delta{\mathbfit V}|^2\rangle\simeq
  t\,R^2 \frac{4\pi}{5}\bigg(\frac{GM}{c}\bigg)^2 n{\sqrt\frac{2\pi}{3 \langle v^2\rangle}}.
\end{align}
Expression~(\ref{eq:delv2_Tsl}) is remarkably similar to the result derived from the autocorrelation function, Equation~(\ref{eq:delv2_3}), which suggests that the characteristic speed in Smoluchowski's equation, $\sqrt{\langle v^2\rangle}$, corresponds to the inverse of the averaged reciprocal velocity $\langle 1/v\rangle^{-1}$, modulo a numerical factor of order unity that arises from the different assumptions on which the two methods rest\footnote{To be precise, equating~(\ref{eq:delv2_3}) to~(\ref{eq:delv2_Tsl}) yields $\langle v^2\rangle^{-1/2}=5\langle 1/v\rangle/\sqrt{6\pi}\simeq 1.15 \langle 1/v\rangle$.}.

To gain further physical insight onto the physical meaning of these quantities, let us for example consider a population of extended substructures with an isotropic Maxwellian velocity distribution displaced by a velocity ${\mathbfit V}$
\begin{align}\label{eq:vgauss}
f({\mathbfit v})=\frac{1}{(2\pi \sigma^2)^{3/2}}\exp\big[-\frac{({\mathbfit v+V})^2}{2\sigma^2}\big],
\end{align}
where ${\mathbfit V}$ is the velocity of a tracer particle measured from the centre of the self-gravitating potential $\Phi_s$. The average of the reciprocal velocity is
\begin{align}\label{eq:1_v}
  \langle 1/v \rangle  
  &=\frac{2\pi}{(2\pi \sigma^2)^{3/2}}\int_0^\infty\d v \,v\int_{-1}^{+1}\d x \,\exp\big[-\frac{v^2}{2\sigma^2}-\frac{V^2}{2\sigma^2}-\frac{Vvx}{\sigma^2}\big] \\ \nonumber
   &=\frac{1}{\sqrt{2\pi} \sigma}\frac{1}{V}\int_0^\infty\d v \,\exp[-\frac{(V+v)^2}{2\sigma^2}\big[\exp\big(\frac{2v V}{\sigma^2}\big)-1\big]\\ \nonumber
 &=\frac{1}{V}\erf\big[\frac{V}{\sqrt{2}\sigma}\big],
\end{align}
here $\erf(x)$ is the error function, which scales as $\erf(x)\simeq 2 x/\sqrt{\pi}$ for $x\ll 1$, and converges towards $\erf(x)\simeq 1$ at $x\gg 1$.
On the other hand, the averaged squared velocity is
\begin{align}\label{eq:v2av}
  \langle v^2 \rangle &=\frac{2\pi}{(2\pi \sigma^2)^{3/2}}\int_0^\infty\d v \,v^4\int_{-1}^{+1}\d x \,\exp\big[-\frac{v^2}{2\sigma^2}-\frac{V^2}{2\sigma^2}-\frac{Vvx}{\sigma^2}\big]\\ \nonumber
   &=\frac{1}{\sqrt{2\pi} \sigma^3}\frac{1}{V}\int_0^\infty\d v \,v^3\exp[-\frac{(V+v)^2}{2\sigma^2}\big[\exp\big(\frac{2v V}{\sigma^2}\big)-1\big]\\ \nonumber
 &=V^2 + 3\sigma^2.
\end{align}
It is straightforward to show that $\langle 1/v \rangle^{-1}$ and $\sqrt{\langle v^2\rangle }$ have practically the same asymptotic limits. Indeed, if the tracer particle moves slowly with respect to the substructure population, $V\ll \sigma$, the inverse of the averaged reciprocal velocity becomes $\langle 1/v\rangle^{-1}\simeq \sqrt{\pi}\sigma$, whereas Smoluchowski's characteristic velocity scales as $\sqrt{\langle v^2\rangle}\simeq \sqrt{3}\sigma$. In contrast, if the tracer particle moves with a large velocity, $V\gg \sigma$, both theories asymptotically approach $\sqrt{\langle v^2\rangle} = \langle 1/v\rangle^{-1}=V$.

\subsection{Adiabatic corrections}\label{sec:adiab}
The results obtained Sections~\ref{sec:line} and~\ref{sec:duration} assume that tidal interactions occur in an impulsive regime, wherein the location of a tracer particle does not vary appreciably during the duration of a flyby encounter/tidal fluctuation. This approximation is generally valid in the outskirts of self-gravitating systems, where the orbital period can be much longer than the time-scale on which the tidal field varies. However, it is bound to fail in the inner regions of the self-gravitating potential $\Phi_s$, where particles complete their orbits on short time-scales and thus  react adiabatically to external perturbations. In this Section we introduce {\it adiabatic corrections} that allow for the motion of stars during tidal field fluctuations.  In particular, we use the corrections found by Weinberg (1994a,b,c) (see also Gnedin \& Ostriker 1999)
\begin{align}\label{eq:A}
  A(T)=\frac{1}{[1 + (wT)^2]^{3/2}},
\end{align}    
where $A$ corresponds to the ratio between the actual energy change and the value found under the impulse approximation, $w=V/R$ is the orbital frequency of the tracer particle in the potential $\Phi_s$, and $T$ is the duration of tidal fluctuations.
As expected, the correction becomes increasingly insignificant ($A\approx 1$) in the outskirts of the system, where orbital frequencies are short, $w T\ll 1$. In contrast, the correction is strong in the inner regions of the potential, where orbital frequencies become much higher, $w T\gg 1$, and Equation~(\ref{eq:A}) approaches a power-law asymptotic form $A\sim (wT)^{-3}$.

In order to account for the adiabatic response of tracer particles to relatively long tidal fluctuations, we `correct' the impulsive variation of the kinetic energy~(\ref{eq:delv2_T}) as
\begin{align}\label{eq:delv2_TA}
  \langle |\Delta{\mathbfit V}|^2\rangle=t\,R^2\langle \Lambda^2 T A\rangle=
  t\,R^2 \int \d^3\Lambda p(\bb \Lambda)\, \Lambda^2 T(\Lambda) A(\Lambda),
\end{align}
where $A(\Lambda)=A[T(\Lambda)]$. For $c\ll D$, combination of~(\ref{eq:plam_pm}) and~(\ref{eq:TL}) yields
\begin{align}\label{eq:L2TA}
  \langle \Lambda^2 T A\rangle=\frac{4 q^2 C}{\pi}T_0\int_0^{1/\chi}\d\xi\frac{\xi^{4}}{(1+\xi^2)^2}\frac{(1-\chi^{1/3}\xi^{1/3})^3\xi^{2/3}}{\frac{4}{\pi}(1-\chi^{1/3}\xi^{1/3})^3+\xi}\\\nonumber
  \times \bigg\{1+(w T_0)^2\frac{(1-\chi^{1/3}\xi^{1/3})^2\xi^{4/3}}{[\frac{4}{\pi}(1-\chi^{1/3}\xi^{1/3})^3+\xi]^2}\bigg\}^{-3/2},
\end{align}
which must be solved numerically.

The strong dependence of the adiabatic correction~(\ref{eq:A}) with the orbital frequency of the tracer particle introduces two relevant regimes for the asymptotic behaviour of~(\ref{eq:L2TA}):
\begin{itemize}
\item {\bf Impulsive regime} arises when the characteristic duration of tidal fluctuations is much shorter than the orbital period of the tracer particle in the self-gravitating potential $(T_{ch}\ll w^{-1})$. In this case $A\approx 1$, Equation~(\ref{eq:L2TA}) approaches asymptotically~(\ref{eq:L2T}), and $\langle \Lambda^2 T A\rangle_{\rm imp}\simeq \langle \Lambda^2 T_{sl}\rangle$. Hence, inserting~(\ref{eq:L2T_aux}) into~(\ref{eq:delv2_TA}) yields
  \begin{align}\label{eq:delv2_Timp}
    \langle |\Delta{\mathbfit V}|^2\rangle_{\rm imp}&\simeq t\,R^2 \langle \Lambda^2 T_{sl} \rangle\\ \nonumber
     &\simeq t\,R^2 \frac{4\pi}{5}\frac{(GM)^2}{c^2}n{\sqrt\frac{2\pi}{3 \langle v^2\rangle}}.
\end{align}
\item {\bf Adiabatic regime} applies to tidal fluctuations with a characteristic time-span that is much longer than the orbital period of the tracer particle $(T_{ch}\gg w^{-1})$. In this limit, the adiabatic correction scales as $A\sim (w T)^{-3}\approx (w T_{sl})^{-3}$, and~(\ref{eq:L2TA}) becomes
    \begin{align}\label{eq:L2Ad}
    \langle \Lambda^2 T A\rangle_{\rm ad}&\simeq \frac{1}{w^3}\langle \frac{\Lambda^2}{T_{sl}^2} \big\rangle\\ \nonumber
    &= \frac{1}{w^3}\frac{4 C q^2}{\pi T_0^2 }\int_0^{1/\chi}\d\xi\frac{\xi^{14/3}}{(1+\xi^2)^2} \\ \nonumber
      &= \frac{1}{w^3}\frac{4 C q^2}{\pi T_0^2}\big[\frac{3}{5 \chi^{5/3}}-\frac{11\pi}{12}+6\chi^{1/3}+\mathcal{O}(\chi^{2/3})\big].
    \end{align}
Inserting~(\ref{eq:L2Ad}) into~(\ref{eq:delv2_TA}) and taking $\chi\ll 1$ yields
    \begin{align}\label{eq:delv2_Tad}
      \langle |\Delta{\mathbfit V}|^2\rangle_{\rm ad}&\simeq  t\,R^2 \langle \Lambda^2 T A\rangle_{\rm ad} \\ \nonumber
      &\simeq t\,\frac{R^5}{V^3}\frac{24\pi}{5} \frac{(GM)^2}{c^5}n\langle v^2\rangle.
    \end{align}
\end{itemize}
Comparison of~(\ref{eq:delv2_Timp}) and~(\ref{eq:delv2_Tad}) reveals key differences between the impulsive and adiabatic regimes. First, the impulsive velocity variance scales as $\langle |\Delta{\mathbfit V}|^2\rangle_{\rm imp}\sim \langle v^2\rangle^{-1/2}$, whereas in the adiabatic regime $\langle |\Delta{\mathbfit V}|^2\rangle_{\rm ad}\sim \langle v^2\rangle$. The opposite behaviour means that while the outskirts of self-gravitating objects are predominantly heated by nearby substructures with small relative velocities, 
only fast encounters can perturb the internal regions of the potential. Furthermore, we find that $\langle |\Delta{\mathbfit V}|^2\rangle_{\rm ad}\sim R^5/V^{3}$, which implies that particles close to the centre of the potential moving with high peculiar velocities will be barely affected by fluctuations of the external tidal field.

Fig.~\ref{fig:L2TA} shows numerical solutions to Equation~(\ref{eq:L2TA}) for different substructure size-to-average separation ratios, $c/D$. Blue-dotted and orange-dashed lines show the impulsive and adiabatic limits derived from Equations~(\ref{eq:L2T}) and~(\ref{eq:L2Ad}), respectively. Note first that accounting for the force dependence of $T(\Lambda)$ is particularly important for `rarefied` distributions of substructures ($D\gg c$). Indeed, combination of~(\ref{eq:L2T}),~(\ref{eq:Tch}) and~(\ref{eq:lamvar2}) shows that adopting a constant value for the duration of tidal fluctuations, $T\approx T_{ch}$, overestimates the magnitude of tidal heating by a factor $\langle \Lambda^2 \rangle T_{ch}/\langle \Lambda^2 T A\rangle_{\rm imp}=  (6^{1/3}2/9)(D/c)\simeq 0.4 D/c$. Importantly, Fig.~\ref{fig:L2TA} shows that the impulsive range of frequencies extends towards systematically higher frequencies as the size ratio $c/D$ decreases. The accuracy of the straight-line approximation also improves in this limit, giving a progressively better match between the numerical values of~(\ref{eq:L2TA}) and its asymptotic limits $\langle \Lambda^2 T A\rangle_{\rm imp}$ and $\langle \Lambda^2 T A\rangle_{\rm ad}$ as $c/D\to 0$.

\section{Evolution driven by tidal heating}\label{sec:ensemble}
In this Section we use the probability theory introduced by Pe\~narrubia (2015), hereafter P15, to describe the non-equilibrium state of particle ensembles acted on by a stochastic tidal field. 
For simplicity, we assume that the potential $\Phi_s$ is time-independent and has spherical symmetry, such that $\Phi_s({\mathbfit R},t)=\Phi_s(R)$. Extending the analysis to self-gravitating, time-dependent potentials is more involved and will be presented in a separate contribution.

\subsection{Diffusion in a static potential}\label{sec:static}
The presence of a stochastic tidal field in the equations of motion~(\ref{eq:eqmots}) induces random fluctuations of the integrals $\{E,{\mathbfit L}\}$.
In an impulse regime (see \S\ref{sec:heating}), the location of a tracer particle is assumed to remain constant. Hence, the variation of orbital energy is equal to the change of kinetic energy
\begin{align}\label{eq:delE}
   \Delta E=\frac{1}{2}({\mathbfit V}+\Delta {\mathbfit V})^2-\frac{1}{2}{\mathbfit V}^2={\mathbfit V}\cdot\Delta{\mathbfit V}+\frac{1}{2}(\Delta {\mathbfit V})^2,
\end{align}
while the angular momentum varies by an amount
\begin{align}\label{eq:delL}
  \Delta {\mathbfit L}={\mathbfit R}\times({\mathbfit V}+\Delta {\mathbfit V})- {\mathbfit R}\times{\mathbfit V}={\mathbfit R}\times\Delta {\mathbfit V}.
\end{align}
If one assumes that the population of substructures are isotropically distributed around the tracer particle, and that the velocity impulses are small, $|\Delta {\mathbfit V}|\ll V$, then
it is straightforward to show that the average variation of these quantities can be written as 
\begin{align}\label{eq:delEL}
  \langle \Delta E\rangle &=\langle{\mathbfit V}\cdot\Delta{\mathbfit V}\rangle+ \frac{1}{2}\langle|\Delta {\mathbfit V}|^2\rangle=\frac{1}{2}\langle|\Delta {\mathbfit V}|^2\rangle \\ \nonumber 
   \langle \Delta {\mathbfit L}\rangle &=\langle{\mathbfit R}\times\Delta {\mathbfit V}\rangle=0,
\end{align}
whereas at leading order $\mathcal{O}(|\Delta {\mathbfit V}|/V)$ the variance of the integral fluctuations can be approximately written as
\begin{align}\label{eq:delEL2}
  \langle (\Delta E)^2\rangle &\simeq  \langle ({\mathbfit V}\cdot\Delta {\mathbfit V})^2\rangle =\frac{1}{3}V^2\langle |\Delta{\mathbfit V}|^2\rangle, \\ \nonumber 
   \langle |\Delta {\mathbfit L}|^2\rangle &= \langle |{\mathbfit R}\times\Delta {\mathbfit V}|^2\rangle=\frac{2}{3}R^2\langle |\Delta{\mathbfit V}|^2\rangle.
\end{align}
Recall that in our notation brackets denote averages over the spectrum of tidal fluctuations, i.e. $\langle X\rangle=\int \d^3 \Lambda\, p(\Lambda) \,X$, see \S\ref{sec:forces}.

\subsubsection{Free-diffusion in an infinite domain}\label{sec:free}
Following P15, let us define a {\it statistical ensemble of tracer particles} as the collection of a large number of individual mass-less particles with energy $E=E_0$ and angular momentum ${\mathbfit L}={\mathbfit L}_0$ at the time $t_0=0$. The probability that these particles have integrals in the range $E,E+\d E$ and ${\mathbfit L}, {\mathbfit L}+\d {\mathbfit L}$ at a later time $t>t_0$ is given by the function $p(E,{\mathbfit L},t|E_0,{\mathbfit L}_0,t_0)$, which a solution to a 4-dimensional diffusion equation (see P15 for details)
\begin{align}\label{eq:pEL}
    t\frac{\partial p}{\partial t}\approx \tilde C_E\frac{\partial p}{\partial E}\bigg |_{(E_0,{\mathbfit L}_0)} + \tilde C_L \frac{\partial p}{\partial{\mathbfit L}}\bigg |_{(E_0,{\mathbfit L}_0)}+ \tilde D_E\frac{\partial^2 p}{\partial E^2}\bigg |_{(E_0,{\mathbfit L}_0)} 
    +\tilde D_L\frac{\partial^2 p}{\partial L^2}\bigg |_{(E_0,{\mathbfit L}_0)}\\\nonumber
    + \tilde D_{EL}\frac{\partial^2 p}{\partial E \partial{\mathbfit L}}\bigg |_{(E_0,{\mathbfit L}_0)},
\end{align}
with initial conditions $p=\delta(E-E_0)\delta({\mathbfit L}-{\mathbfit L}_0)$ at $t=t_0$.
The simplest solution corresponds to particles that diffuse freely (i.e. in a domain with no boundaries) in the 4 dimensions of the integral-of-motion space, and with a probability density that is separable in energy and angular momentum, such that
\begin{align}\label{eq:p}
  p(E,{\mathbfit L},t|E_0,{\mathbfit L}_0,t_0)=p(E,t|E_0,{\mathbfit L}_0,t_0)\, p({\mathbfit L},t|E_0,{\mathbfit L}_0,t_0).
 \end{align}
Under these conditions, P15 shows that the probability functions $p$ are Green's functions (or propagators) with a Gaussian form
\begin{align}\label{eq:pE}
  p(E,t|E_0,{\mathbfit L}_0,t_0)= \frac{1}{(4\pi\tilde D_E)^{1/2}}\exp\bigg\{-\frac{ (E-E_0+ \tilde C_E)^2}{4 \tilde D_E} \bigg\},
\end{align}
and
\begin{align}\label{eq:pL}
   p({\mathbfit L},t|E_0,{\mathbfit L}_0,t_0)= \frac{1}{(4\pi\tilde D_L)^{3/2}}\exp\bigg\{-\frac{ ({\mathbfit L}-{\mathbfit L}_0+ \tilde {\mathbfit C}_L)^2}{4 \tilde D_L}\bigg\},
\end{align}
where $\{\tilde C_E, \tilde C_L,\tilde D_E, \tilde D_L\}$ are coefficients evaluated at $(E_0,{\mathbfit L}_0,t-t_0)$. If the fluctuations of the external tidal field are much weaker than the binding forces  all coefficients approach zero, and the probability functions~(\ref{eq:pE}) and~(\ref{eq:pL}) become sharply peaked about $E=E_0$ and ${\mathbfit L}={\mathbfit L}_0$, respectively. As expected, in the limit $\langle |\Delta{\mathbfit V}|^2\rangle\to 0$, one has that $p(E,t|E_0,{\mathbfit L}_0,t_0)\to \delta (E-E_0)$, and $p({\mathbfit L},t|E_0,{\mathbfit L}_0,t_0)\to \delta ({\mathbfit L}-{\mathbfit L}_0)$, which recovers the initial conditions.

The diffusion coefficients are derived from phase-space averages of~(\ref{eq:delEL}) and~(\ref{eq:delEL2}) over a collection of tracer particles with the same combination of energy and angular momentum at $t_0=0$. Using the adiabatically-corrected velocity impulses~(\ref{eq:delv2_TA}) leads to {\it drift coefficients} 
\begin{align}\label{eq:CEL}
  \tilde C_E(E,L,t)&= -\overline{\langle \Delta E\rangle }= -\frac{1}{2}\overline{\langle |\Delta {\mathbfit V}|^2\rangle}=-t\frac{1}{2}\overline{R^2\langle \Lambda^2 T A\rangle}, \\ \nonumber
  \tilde {\mathbfit C}_L(E,L,t)&=-\overline{\langle \Delta {\mathbfit L}\rangle}=0,
  \end{align}
and {\it diffusion coefficients}
\begin{align}\label{eq:DEL}
  2\tilde D_E(E,L,t)&= \overline{\langle (\Delta E)^2\rangle}-\overline{\langle \Delta E \rangle}^2 \simeq t\frac{1}{3}\overline{R^2V^2\langle \Lambda^2 TA\rangle},\\ \nonumber
  2\tilde D_L(E,L,t)&= \overline{\langle|\Delta {\mathbfit L}|^2\rangle}-\overline{\langle\Delta {\mathbfit L}\rangle}^2=t\frac{2}{3}\overline{R^4\langle \Lambda^2 TA\rangle},
\end{align}
with upper bars denoting a phase-space average over tracer particle ensembles (see Appendix~\ref{sec:aver}). Note that the coefficients are isotropic in the angular momentum space, i.e. $C_i(E,{\mathbfit L})=C_i(E,L)$ and $D_i(E,{\mathbfit L})=D_i(E,L)$, with a subindex $i=\{E,L\}$. In addition, the cross-coefficients vanish by symmetry, $\overline{\Delta E\Delta {\mathbfit L}}=\tilde D_{EL}=0$, which justifies the separability of $p(E,{\mathbfit L},t|E_0,{\mathbfit L}_0,t_0)$ implicitly assumed in Equation~(\ref{eq:p}).

In an impulsive regime,  averages over tracer-particle (upper bars) and substructure (brackets) ensembles in~(\ref{eq:CEL}) and~(\ref{eq:DEL}) become statistically independent, which simplifies our mathematical treatment greatly. Indeed, for tidal fluctuations with a characteristic duration~(\ref{eq:Tch}) that is much shorter than the orbital frequency, i.e. $w T_{ch}\ll 1$, one can approximate $\langle \Lambda^2 T A\rangle\approx \langle \Lambda^2 T\rangle$, which is independent of the phase-space distribution of tracer particles. Hence, Equations~(\ref{eq:CEL}) and~(\ref{eq:DEL}) become
\begin{align}\label{eq:CELimp}
    \tilde C_{E,{\rm imp}}(E,L,t)&= -t\frac{\overline{R^2}}{2}\langle \Lambda^2 T \rangle, \\ \nonumber
  \tilde C_{L,{\rm imp}}(E,L,t)&=0, \\ \nonumber
  \tilde D_{E,{\rm imp}}(E,L,t)&=
  t\frac{\overline{R^2 V^2}}{6}\langle \Lambda^2 T\rangle,\\ \nonumber
  \tilde D_{L,{\rm imp}}(E,L,t)&= t\frac{\overline{R^4}}{3}\langle \Lambda^2 T\rangle.
\end{align}

\subsubsection{Diffusion in a confined region}\label{sec:diff_conf}
In a self-gravitating potential $\Phi_s$, gravitationally-bound particles can only diffuse in a limited volume of the integral-of-motion space. In particular, energies must be negative definite, $E<0$, while the angular momentum is confined within the interval $0\le L\le L_c(E)$, where $L_c(E)$ corresponds to the angular momentum of a circular orbit with energy $E$. In order to confine particles within a given domain we must solve the diffusion equation~(\ref{eq:pEL}) with appropriate boundary conditions. To this aim, first we set an {\it absorbing} boundary at $E=0$, such that $p(E\ge 0)=0$, which implies that particles with $E\ge 0$ escape from a gravitating system to never return\footnote{This approximation is only approximately correct. In reality, particles can diffuse in and out of the boundary $E=0$. Although this calls for introducing a {\it permeable} boundary, the mathematical treatment significantly more involved (e.g. Carslaw \& Jaeger 1986) and must be left for follow-up work.}. 
The angular momentum space is treated as a cubic box with {\it reflecting} surfaces at the limits of each dimension\footnote{Note that diffusion processes depend on the space geometry. Here we adopt Cartesian symmetry for mathematical convenience, see Appendix~\ref{sec:diff} for details.}, such that $\partial_L p(L_i=\pm L_c)=0$, where $i=x,y,z$, which limits the values of the components within $-L_c\le L_i\le L_c$.
In order to obey these boundary conditions, Appendix~\ref{sec:diff} shows that Equations~(\ref{eq:pE}) and~(\ref{eq:pL}) must be respectively replaced by
\begin{align}\label{eq:pE_con}
  p(E,t|E_0,{\mathbfit L}_0,t_0)=\frac{1}{\sqrt{4\pi \tilde D_E}}\exp\bigg[-\frac{(E-E_0+\tilde C_E)^2}{4 \tilde D_E}\bigg] \\ \nonumber
-\,\frac{1}{\sqrt{4\pi \tilde D_E}}\exp\bigg(E_0\frac{\tilde C_E}{\tilde D_E}\bigg)\exp\bigg[-\frac{(E+E_0+\tilde C_E)^2}{4 \tilde D_E}\bigg],
\end{align}
and
\begin{align}\label{eq:pL_con}
  p({\mathbfit L},t|E_0,{\mathbfit L}_0,t_0)&=\sum_{n,m,l=0}^\infty \frac{\alpha_{nml}}{L_c^3}\cos\big[\frac{n\pi (L_{x,0}+L_c)}{2 L_c}\big]\cos\big[\frac{n\pi (L_x+L_c)}{2 L_c}\big] \\ \nonumber
  \times&  \cos\big[\frac{m\pi (L_{y,0}+L_c)}{2 L_c}\big]\cos\big[\frac{m\pi (L_y+L_c)}{2 L_c} \big]\\ \nonumber
    \times& \cos\big[\frac{l\pi (L_{z,0}+L_c)}{2 L_c}\big]\cos\big[\frac{l\pi (L_z+L_c)}{2 L_c}\big]\exp[-\lambda_{nml}\tilde D_L],
\end{align}
where $L_c=L_c(E_0)$, $\lambda_{nml}=\pi^2(n^2+m^2+l^2)/(4 L_c^2)$ is a separation constant, and 
$\alpha_{000}=1/8$, $\alpha_{n00}=\alpha_{0m0}=\alpha_{00l}=1/4$, $\alpha_{nm0}=\alpha_{n0l}=\alpha_{0ml}=1/2$, and $\alpha_{nml}=1$~ for $n,m,l\ge 1$.
As demonstrated in Appendix~\ref{sec:diff}, the function $p({\mathbfit L},t|E_0,{\mathbfit L}_0,t_0)$ recovers the Gaussian propagator~(\ref{eq:pL}) on short time-scales, $\pi^2 D_L/(4 L_c^2) \,t\ll 1$, where $D_L=\tilde D_L/t$ is a `static' coefficient defined in \S\ref{sec:flux}.
As the time progresses, however, all modes with $n,m,l\ge 1$ decay exponentially, and the probability function converges asymptotically towards a constant value $p({\mathbfit L},t|E_0,{\mathbfit L}_0,t_0)\to 1/(8L_c^3)$ in the limit $t\to\infty$. This is an important result, as it shows that stochastic tidal heating tends to isotropize the initial angular momentum distribution on a time-scale $\pi^2 D_L/(4 L_c^2) \,t\gtrsim 1$. This calls for defining the isotropization time-scale as
\begin{align}\label{eq:tiso}
  t_{\rm iso}=\frac{4L_c^2}{\pi^2 D_L}=\frac{12}{\pi^2}\frac{ L_c^2}{\overline{R^4\langle \Lambda^2 TA\rangle}},
\end{align}
where $D_L=\tilde D_L/t$ is given by~(\ref{eq:DEL}).
In \S\ref{sec:kepler} we discuss the properties of $t_{\rm iso}$ in some detail.

According to the Jeans theorem, the initial equilibrium state of a gravitational system is fully defined by the distribution of particles in the integral-of-motion space $N(E_0,{\mathbfit L}_0,t_0)$, which determines the probability to find a particle in the 4-dimensional volume element $\d E_0\d^3L_0$ centred at a specific energy $E_0=V^2/2+\Phi_s(R)$ and an angular momentum vector ${\mathbfit L}_0={\mathbfit R}\times{\mathbfit V}$ at the time $t_0=0$. The (non-equilibrium) state of a gravitating system at a later time $t>t_0$ is found by convolving the initial distribution $N(E_0,{\mathbfit L}_0,t_0)$ with~(\ref{eq:pE_con}) and~(\ref{eq:pL_con}), which yields
\begin{align}\label{eq:Nt}
  N(E,{\mathbfit L},t)&=\int\int \d E_0 \d^3L_0 \,p(E,{\mathbfit L},t|E_0,{\mathbfit L}_0,t_0)N(E_0,{\mathbfit L}_0,t_0) \\ \nonumber
  =&\int \d E_0\,p(E,t|E_0,{\mathbfit L}_0,t_0)\int\d^3L_0\,p({\mathbfit L},t|E_0,{\mathbfit L}_0,t_0) N(E_0,{\mathbfit L}_0,t_0) 
\end{align}
 hence exploiting the fact that $p(E,{\mathbfit L},t|E_0,{\mathbfit L}_0,t_0)$ is a Green's function (see P15 for details).
In Section~\ref{sec:appl} we discuss the evolution of planetary orbits driven by stochastic tidal heating as an illustration of the Green's convolution~(\ref{eq:Nt}).

\subsection{Fokker-Planck equations}\label{sec:flux}
Let us write down the Fokker-Planck equations in the integral-of-motion space in order to gain insight onto the physical meaning on the diffusion coefficients. To this end let us assume that the transformation $N(E_0,L_0,t_0)\to N(E,L,t)$ takes place during a short, but non-zero, time interval $t-t_0$, and that the variation of energy and angular momentum is small, such that $|\Delta E|/|E|\ll 1$ and $|\Delta {\mathbfit L}|/L\ll 1$.  Thus, expanding the left-hand side of Equation~(\ref{eq:Nt}) up to the first order in $t$, and the right-hand side as a Taylor series to the second order in $\Delta E$ and $\Delta {\mathbfit L}$ yields
(Spitzer 1987; P15)
\begin{align}\label{eq:FP}
    t\frac{\partial N}{\partial t}\approx -\frac{\partial }{\partial E}[N\overline{\langle \Delta E\rangle } ] - \frac{\partial }{\partial{\mathbfit L}}[N\overline{\langle \Delta {\mathbfit L}\rangle }]+ \frac{1}{2}\frac{\partial^2 }{\partial E^2}[N \overline{\langle (\Delta E)^2\rangle }]  \\ \nonumber
+\frac{1}{2}\frac{\partial^2 }{\partial L^2}[N\overline{\langle |\Delta {\mathbfit L}|^2}\rangle  ] + \frac{\partial^2 }{\partial E \partial{\mathbfit L}}[N\overline{\langle \Delta E \Delta {\mathbfit L}\rangle }].
\end{align}
The Fokker-Planck equations~(\ref{eq:FP}) simplify considerably in an isotropic tidal field, where $\overline{\langle \Delta {\mathbfit L}\rangle }=\overline{\langle \Delta E \Delta {\mathbfit L}\rangle }=0$ and $2\tilde D_L=\overline{\langle |\Delta {\mathbfit L}|^2}\rangle$ by symmetry. Also, from Equations~(\ref{eq:delEL}) and~(\ref{eq:delEL2}) it is clear that $\langle (\Delta E)^2\rangle\sim V^2\langle |\Delta\mathbfit{V}|^2\rangle$, whereas $\langle \Delta E\rangle^2\sim \langle |\Delta\mathbfit{V}|^2\rangle^2$. Since the Brownian approach requires $\langle |\Delta\mathbfit{V}|^2\rangle \ll V^2$, one can safely approximate $2\tilde D_E=\overline{\langle (\Delta E)^2\rangle}-\overline{\langle \Delta E \rangle}^2\approx \overline{\langle (\Delta E)^2\rangle}$, such that 
\begin{align}\label{eq:FP2}
  t\frac{\partial N}{\partial t}\approx\frac{\partial }{\partial E}[N\tilde C_E ]+ \frac{\partial^2 }{\partial E^2}[N \tilde D_E]+\frac{\partial^2 }{\partial L^2}[N\tilde D_L].
\end{align}

The linear dependence of the coefficients $\tilde C$ and $\tilde D$ with the length of the time-interval $t$ in~(\ref{eq:CEL}) and~(\ref{eq:DEL}), respectively, calls for the definition of {\it static} drift and diffusion coefficients
\begin{align}\label{eq:CDEL2}
  C_E(E,L)&=\frac{\tilde C_E(E,L,t)}{t}=-\frac{1}{2}\overline{R^2\langle \Lambda^2 T A\rangle}\\ \nonumber
  C_L(E,L)&=\frac{\tilde C_L(E,L,t)}{t}=0,\\ \nonumber
  D_E(E,L)&=\frac{\tilde D_E(E,L,t)}{t}=\frac{1}{6}\overline{R^2V^2\langle \Lambda^2 TA\rangle},\\ \nonumber
  D_L(E,L)&=\frac{\tilde D_L(E,L,t)}{t}=\frac{1}{3}\overline{R^4\langle \Lambda^2 TA\rangle}.
\end{align}
such that Equation~(\ref{eq:FP2}) becomes
\begin{align}\label{eq:FP3}
  \frac{\partial N}{\partial t}\approx\frac{\partial }{\partial E}[N C_E ]+ \frac{\partial^2 }{\partial E^2}[N  D_E]+\frac{\partial^2 }{\partial  L^2}[N D_L].
\end{align}
Random fluctuations of the tidal field induce a bulk motion of tracer particles in the integral-of-motion space.  At leading order, the main variation corresponds to a flux of particles crossing the energy layer $E$ at a fixed angular momentum ${\mathbfit L}$, which can be estimated from Equations~(\ref{eq:FP3}) and~(\ref{eq:CDEL2}) as
\begin{align}
\label{eq:JE}
J(E,{\mathbfit L})&=\int_E\frac{\partial N}{\partial t}\d E' \approx  N(E,{\mathbfit L},t_0)C_E(E,L) \\ \nonumber
&=-\frac{1}{2}\overline{R^2\langle \Lambda^2 T A\rangle}N(E,{\mathbfit L},t_0),
\end{align}
the negative sign of $J$ implies that stochastic tidal heating leads to a {\it steady} flow of particles drifting from bound energies $E< 0$ towards $E\to 0$. This realization has important consequences for the survival of self-gravitating objects in a fluctuating tidal field, as discussed below. Note also that in an impulsive regime Equation~(\ref{eq:JE}) has a simple form
\begin{align}
\label{eq:JE_imp}
J_{\rm imp}(E,{\mathbfit L})=-\frac{\overline{R^2}}{2}\langle \Lambda^2 T\rangle N(E,{\mathbfit L},t_0).
\end{align}
The fact that $J_{\rm imp}\propto \overline{R^2}$ shows that the particle flow is particularly strong in the outskirts of the potential $\Phi_s(R)$.


Over a sufficiently long interval of time, particles with a combination of integrals $(E_0,{\mathbfit L}_0)$ at $t_0=0$ will gain sufficient kinetic energy as to escape the potential $\Phi_s(R)$, which is known as ``tidal evaporation'' (Spitzer 1958 and references therein). The unbinding time, $t_{\rm esc}(E_0,L_0)$, can be estimated from Equation~(\ref{eq:CDEL2}) by expressing $\overline{\langle \Delta E\rangle }=-C_E(E_0,L_0) t=\frac{1}{2}\overline{R^2\langle \Lambda^2 T A\rangle}t$. Hence, setting $E(t_{\rm esc})=0$ yields an average {\it escape time}
\begin{align}\label{eq:tesc}
  t_{\rm esc}(E_0,L_0)=\frac{(0-E_0)}{-C_E(E_0,L_0)}=\frac{2|E_0|}{\overline{R^2\langle \Lambda^2 T A\rangle}}.
\end{align}
In the impulse approximation ($A\approx 1$), Equation~(\ref{eq:tesc}) reduces to
\begin{align}\label{eq:tesc_imp}
t_{\rm esc,imp}(E_0,L_0)=\frac{2|E_0|}{\overline{R^2}}\frac{1}{\langle \Lambda^2 T \rangle},
\end{align}
which is independent of the velocity distribution of the tracer particles.

\subsection{Example: Keplerian potential}\label{sec:kepler}
It is worth illustrating the above results with an analytical example. To this end, consider an ensemble of
tracer particles orbiting in a Keplerian potential
\begin{align}\label{eq:phis_kep}
\Phi_s(R)=-\frac{G m}{R}.
\end{align}
As in previous Sections, we assume that the system is surrounded by an homogeneous distribution of Hernquist (1990) spheres with a mass $M$, scale-length $c$ and number density $n$ (see \S\ref{sec:prob} for details).

The drift and diffusion coefficients appearing in~(\ref{eq:CEL}) and~(\ref{eq:DEL}), respectively, can be calculated analytically under the assumption that (i) random tidal interactions occur in an impulsive regime ($A\approx 1$), and (ii) substructures do not spatially overlap with each other ($c\ll D$). Hence, combination of~(\ref{eq:CELimp}),~(\ref{eq:r2_kep}),~(\ref{eq:r4_kep}),~(\ref{eq:r2v2_kep}) and~(\ref{eq:L2T}) yields
\begin{align}\label{eq:CELimp_kep}
    \tilde C_{E,{\rm imp}}(a,e,t)&= -t\, a^2\big(1+\frac{3}{2}e^2\big)\frac{2\pi}{5}\bigg(\frac{GM}{c}\bigg)^2 n{\sqrt\frac{2\pi}{3 \langle v^2\rangle}}, \\ \nonumber
  \tilde C_{L,{\rm imp}}(a,e,t)&=0, \\ \nonumber
  \tilde D_{E,{\rm imp}}(a,e,t)&=
  t\, a\big(1-\frac{e^2}{2}\big)\frac{2\pi}{15}(G m)\bigg(\frac{GM}{c}\bigg)^2  n{\sqrt\frac{2\pi}{3 \langle v^2\rangle}},\\ \nonumber
  \tilde D_{L,{\rm imp}}(a,e,t)&= t\, a^4\big(1+5e^2 +\frac{15}{8}e^4\big)\frac{4\pi}{15}\bigg(\frac{GM}{c}\bigg)^2 n{\sqrt\frac{2\pi}{3 \langle v^2\rangle}}.
\end{align}
Here, $a$ and $e$ are the semi-major axis and eccentricity of and orbit with energy $E$ and angular momentum $L$
\begin{align}\label{eq:ae}
  E&=-\frac{G m}{2a}~~~~~~~~~~~~~~~~~~{\rm for}~~~~~ E\le 0, \\ \nonumber
  L^2&=G m\,a(1-e^2)~~~~~~~{\rm with}~~~ 0\le e\le 1,
\end{align}
where $e=0$ corresponds to a circular orbit with angular momentum
\begin{align}\label{eq:LcE}
L_c(E)=\sqrt{G m\,a}=\frac{G m}{(-2 E)^{1/2}}.
\end{align}

Combination of~(\ref{eq:tesc}), ~(\ref{eq:CELimp_kep}) and~(\ref{eq:ae}) yields the mean impulsive time-scale required to unbind a tracer particle as a function of semi-major axis and orbital eccentricity
\begin{align}\label{eq:tesc_kep}
  t_{\rm esc,imp}(a,e)&=\frac{G m}{a^3(1+\frac{3}{2}e^2)}\frac{1}{\langle \Lambda^2 T \rangle}\\ \nonumber
&=\frac{5}{2}\frac{c^2}{(1+\frac{3}{2}e^2)}\frac{G m}{(GM)^2}\sqrt{\frac{3\langle v^2\rangle}{2 \pi}}\bigg(\frac{D}{a}\bigg)^3.
\end{align}
This expression reveals a number of interesting features. Notice first that the unbinding time-scale is mainly determined by the ratio between the semi-major axis of the orbit and the average separation between substructures, $t_{\rm esc}\sim (D/a)^{3}$, and depends only weakly on eccentricity. In particular circular orbits ($e=0$) remain bound a factor $5/2$ longer than radial orbits $(e=1)$.
In addition, Equation~(\ref{eq:tesc_kep}) shows that the escape time-scale decreases linearly with the velocity dispersion of the substructure population, $t_{\rm esc}\sim \langle v^2\rangle^{1/2}$, which highlights the fact that impulsive energy injections are dominated by nearby substructures moving with a small relative velocity (see \S\ref{sec:adiab}).

\begin{figure*}
\begin{center}
\includegraphics[width=168mm]{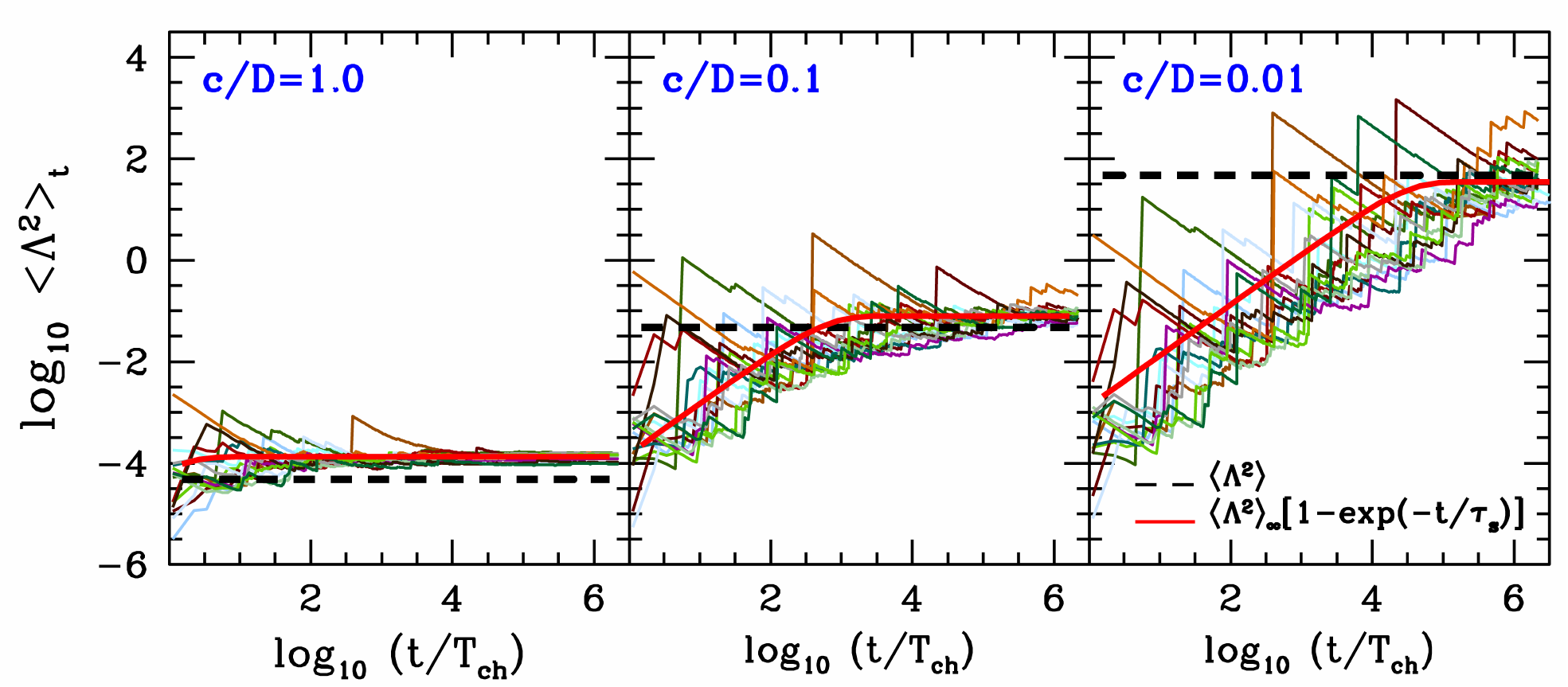}
\end{center}
\caption{Time-averaged variance of tidal fluctuations, $\langle \Lambda^2\rangle_t$, Equation~(\ref{eq:lam_tvar}), as a function of the length of the time interval $t$ measured in units of the characteristic duration of tidal fluctuations, $T_{\rm ch}$, Equation~(\ref{eq:Tch}). Thin curves correspond to ensembles of $N=5000$ Hernquist spheres with a mass $M=10^{-7}$ and a size $c$ given in units of the mean separation between substructures, $D=(2\pi n_0)^{-1/3}$. Red lines show best-fitting Equation~(\ref{eq:lam_fit}). Note that if substructures are not spatially overlapping ($c/D\ll 1$) the averaged value of $\langle \Lambda^2\rangle_t$ converges exponentially towards the analytical function~(\ref{eq:lamvar2}) (horizontal black dashed lines) on a time-scale $t\gg \tau_s$, where $\tau_s$ is the so-called sampling time (see text). }
\label{fig:lt_3}
\end{figure*}

It is interesting to compare the time-scale required to isotropize the angular momentum distribution, Equation~(\ref{eq:tiso}), against the unbinding time-scale~(\ref{eq:tesc}). Adopting the impulse approximation, and inserting~(\ref{eq:LcE}),~(\ref{eq:r4_kep}) and~(\ref{eq:L2T}) into~(\ref{eq:tiso}) yields
\begin{align}\label{eq:tiso_kep}
  t_{\rm iso,imp}(a,e)&=\frac{12}{\pi^2}\frac{G m}{a^3(1+5e^2+\frac{15}{8}e^4)}\frac{1}{\langle \Lambda^2 T \rangle}\\ \nonumber
&=\frac{30}{\pi^2}\frac{c^2}{(1+5e^2+\frac{15}{8}e^4)}\frac{G m}{(GM)^2}\sqrt{\frac{3\langle v^2\rangle}{2 \pi}}\bigg(\frac{D}{a}\bigg)^3,
\end{align}
which indicates that isotropization and unbinding time-scales share the same power-law  behaviour, albeit with the former being more sensitive to orbital eccentricity than the latter. In particular, dividing~(\ref{eq:tiso_kep}) by~(\ref{eq:tesc_kep}) yields a ratio
\begin{align}\label{eq:tisoesc}
  \bigg(\frac{t_{\rm iso}}{t_{\rm esc}}\bigg)_{\rm imp}=\frac{12}{\pi^2}\frac{1+\frac{3}{2}e^2}{1+5e^2+\frac{15}{8}e^4}.
\end{align}
which has values between $12/\pi^2\simeq 1.22$ and $80/(21\pi^2)\simeq 0.386$ for $e=0$ and $e=1$, respectively, with $t_{\rm iso}=t_{\rm esc}$ occurring at $e\simeq 0.256$. This result implies that orbits with low orbital eccentricity tend to escape from the potential $\Phi_s$ before the momentum distribution is fully randomized. We will return to this issue below.

\section{$N$-body tests}\label{sec:nbody}
In this Section, we run a number of restricted N-body experiments that follow the dynamical evolution of tracer particles subject to random tidal interactions with a large population of extended substructures. Our goal is to test the main assumptions introduced in previous Sections in order to describe tidal heating as a random walk (a.k.a. Brownian motion) of velocities that leads to a diffusion process a 4-dimensional integral-of-motion space.

\subsection{Numerical set-up}\label{sec:setup}
To simplify our numerical analysis and speed up $N$-body computations we adopt an analytical Dehnen (1993) model for the host galaxy potential, where
\begin{align}\label{eq:phi_dehn}
\Phi_g(r)=\frac{4\pi G\rho_0}{3-\gamma}\times\begin{cases}
-\frac{1}{2-\gamma}\big[1-\big(\frac{r}{r+r_0}\big)^{2-\gamma}\big] & ,\gamma\ne 2 \\ 
\ln\big(\frac{r}{r+r_0}\big) & , \gamma=2.
\end{cases}
\end{align}
In what follows, we use $N$-body units with $G=\rho_0=r_0=1$.

Substructures are assumed to be spherically distributed about the host centre following a number density profile 
\begin{align}\label{eq:n_dehn}
n(r)=\frac{n_0}{[1+(r/r_0)^\alpha]^{\beta/\alpha}},
\end{align}
which is normalized to $4\pi\int_0^\infty\d r\, r^2\, n(r)=N$. The scale on which the number density profile varies can be calculated analytically from~(\ref{eq:n_dehn}) as
\begin{align}\label{eq:d_n_dehn}
d(r)\equiv \bigg|\frac{\nabla n}{n}\bigg|^{-1}=\frac{r}{\beta}\bigg[1+\bigg(\frac{r_0}{r}\bigg)^\alpha\bigg].
\end{align}
The local approximation is justified at radii where $d$ is larger than the mean separation between substructures, i.e. $d\gg D=(2\pi n(r)]^{-1/3}$ (see Appendix A of Paper I). For $\alpha>1$, the distance $d$ diverges as $r\to 0$, hence one can safely approximate $n\approx n_0$ and $D\approx (2\pi n_0)^{-1/3}$ at $r\ll r_0$.
  
If one further assumes that substructures move isotropically about the centre of the potential, then the distribution function can be written as (Eddington 1916)
\begin{align}\label{eq:eddin}
  f(E)=\frac{1}{\sqrt{8}\pi^2}\bigg[\int_E^0\frac{\d\Phi}{\sqrt{\Phi-E}}\frac{\d^2 n}{\d\Phi^2}+\frac{1}{\sqrt{-E}}\bigg(\frac{\d n}{\d \Phi}\bigg)_{\Phi=0}\bigg],
\end{align}
where $n[r(\Phi_g)]$ corresponds to the profile~(\ref{eq:n_dehn}) expressed as a function of the potential~(\ref{eq:phi_dehn}). Dynamical equilibrium is guaranteed if and only if the distribution function is definite positive, $f\ge 0$, within the energy range sampled by the orbital distribution. For the models described above this condition requires a steep slope at large radii. In our experiments we set $\gamma=0$, $\alpha=3$ and $\beta=60$. Although not shown here, we have explicitly checked that the particular choice of indices $\gamma$, $\alpha$ and $\beta$ does not significantly change the results shown below.
The initial radii and velocities of $N$ substructures are drawn randomly from the distribution function $f(E)$ using a rejection algorithm, while the directions of the position and velocity vectors are isotropically distributed over the surface of a sphere. The velocity dispersion $\langle v^2\rangle$ is measured from the relative velocities of the substructure ensemble. For the experiments shown below we find $\langle v^2\rangle^{1/2}\simeq 0.45$.
The orbits of individual substructures in the potential $\Phi_g$ are integrated forward in time using a leap-frog algorithm whose time-step is chosen such that the energy is conserved at least at a $10^{-9}$ level in isolation.

For simplicity, tracer particles orbit an isolated Keplerian potential~(\ref{eq:phis_kep}) and follow circular orbits at $t=0$, such that the initial tangential velocity is set to $V=V_c(R)=|R\nabla\Phi_s(R)|^{1/2}=(G m/R)^{1/2}$, and the orbital frequency to $w=V/R=(Gm/R^3)^{1/2}$. The initial energy and angular momentum are therefore $E_0=V_c^2/2+\Phi_s(R)$ and $L_0=RV_c(R)$, while the semi-major axis and eccentricity are $a_0=-Gm/(2E_0)$ and $e_0=0$, respectively.

Fluctuations of the external tidal field are added into the equations of motion following two different prescriptions
\begin{enumerate}
\item {\bf Direct-force computation} of the tidal field. Here the combined tidal tensor induced by a set of $N$-Hernquist (1993) spheres is calculated as a direct summation of the tidal forced induced by individual substructures
  \begin{align}\label{eq:tmatrix}
T^{jk}= \sum_{i=1}^N t_i^{jk}\equiv \sum_{i=1}^N\frac{G M}{r'_i(r'_i+c)^2}\delta_{jk}-\bigg(\frac{2 GM}{(r'_i+c)^3}+\frac{G M}{r'_i(r'_i+c)^2}\bigg)\frac{x^j x^k}{r'^2},
  \end{align}
  where $\delta_{jk}$ is the Kronecker delta and ${\mathbfit r}'_i=\{x_i^j\}_{j=1,2,3}={\mathbfit r}_s-{\mathbfit r}_i$ is the relative distance between the self-gravitating object and the $i^{th}$ substructure. 
In the models shown below, the tidal tensor~(\ref{eq:tmatrix}) is evaluated at a radius, $r_s=0.1 r_0$, where the number density of substructures~(\ref{eq:n_dehn}) is approximately constant, $n\approx n_0$. It is worth stressing that this approach comes at a large computational cost.
  \vskip0.2cm
\item {\bf Monte-Carlo sampling} the distribution of velocity increments $\Psi({\mathbfit V}, \Delta {\mathbfit V},\Delta t)$ given by Equation~(\ref{eq:Psi}). Here we use adiabatic-corrected coefficients $\langle |\Delta{\mathbfit V}|^2\rangle$ computed from equations~(\ref{eq:delv2_TA}) and~(\ref{eq:L2TA}), with a time interval $t$ set equal to the time-step of the leap-frog scheme, $\Delta t$.
 In order to incorporate the effects of a fluctuating tidal field, first we compute the orbit of the test particle in the potential $\Phi_s$ from $\mathbfit{R}(t)\to \mathbfit{R}(t+\Delta t)$ by solving the equations of motion
 \begin{align}\label{eq:mc}
   \frac{{\d^2 \mathbfit R}}{\d t^2}=-\nabla\Phi_s({\mathbfit R}).
 \end{align}
 Subsequently, we add a random velocity `kick' to the velocity vector computed from~(\ref{eq:mc}) in isolation, ${\mathbfit V}(t+\Delta t)$, such that
 $${\mathbfit V}(t)\to {\mathbfit V}(t+\Delta t) + {\it Ran}(\Delta{\mathbfit V})$$
 where ${\it Ran}(\Delta{\mathbfit V})$ correspond to random velocity increments drawn from the isotropic Gaussian distribution $\Psi({\mathbfit V}, \Delta {\mathbfit V},\Delta t)$.
\end{enumerate}

The potential $\Phi_s$ is assumed to be in isolation, i.e. the host galaxy potential is set to $\Phi_g=0$, which implies that particles with a specific energy $E\ge 0$ will escape from the system.
This choice helps to isolate the effect of stochastic tidal fluctuations by removing the smooth component of the tidal tensor from the equations of motion ($T_g=0$). 
Equations~(\ref{eq:eqmots}) and~(\ref{eq:mc}) are solved using a
Runge-Kutta scheme (e.g. Press et al. 1992) with a fixed time-step $\Delta t=0.01 w^{-1}$. For the models shown below, this yields an energy conservation at a $\sim 10^{-9}$ level in the absence of stochastic tidal fluctuations. One should bear in mind is that at a fixed time-step direct calculations of the force induced by a population of $N-$substructures consumes approximately a factor $N$-times more CPU-power than the Monte-Carlo method. This difference becomes astronomical in the case of Galactic substructures, as discussed in Section~\ref{sec:micro}.


\subsection{Tidal force fluctuations}\label{sec:num_fluctu}
We start by testing the analytical expressions for the distribution of tidal force fluctuations, $p(\bb \Lambda)$, derived in Section~\ref{sec:forces}. Recall that our analytical framework assumes that (i) the Euler and Coriolis terms appearing in the non-inertial rest frame of the test particle can be neglected, and that (ii) individual substructures induce tidal forces that point in random directions. 
In order to quantify the amplitude of tidal fluctuations, we first measure the modulus of the combined tidal vector $\bb\Lambda$ acting on a single tracer particle from a {\it direct summation} of the tidal forces~(\ref{eq:Flambda}) induced by individual subtructures
\begin{align}\label{eq:lam_t}
\Lambda=\frac{|{\mathbfit F}_t|}{R}=\frac{1}{R}\bigg|\sum_{i=1}^N t_i\cdot {\mathbfit R}\bigg|,
\end{align}
The average of the variance within an interval of time $t$ is calculated as
\begin{align}\label{eq:lam_tvar}
\langle \Lambda^2\rangle_t=\frac{1}{t}\int_0^t\d \tau \Lambda^2(\tau).
\end{align}
Fig.~\ref{fig:lt_3} shows the time evolution of $\langle \Lambda^2\rangle_t$ computed from $N_{\rm ens}=16$ independent ensembles of $N=5000$ substructures with a mass $M=10^{-7}$ and three different sizes ($c$) given in units of the average separation between substructures ($D$). 
 The `saw'-like shape of the curves is caused by impulse variations of $\Lambda^2(t)$ over reduced periods of time. 
As expected, on time-scales $t\gtrsim T_{\rm ch}$, where $T_{\rm ch}$ is the average duration of tidal fluctuations given by~(\ref{eq:Tch}), {\it ensemble-averaged} values of the variance, $\langle \Lambda^2\rangle_{\rm ens}(t)=N_{\rm ens}^{-1}\sum_{k=1}^{N_{\rm ens}}\langle \Lambda^2\rangle_{t,k}$, converge towards the analytical expression for $\langle \Lambda^2\rangle$ derived in the limit $N\to \infty$ (black-dashed lines, Equation~\ref{eq:lamvar2}). However, on time-intervals $t\lesssim T_{\rm ch}$ Equation~(\ref{eq:lamvar2}) systematically overestimates the value of $\langle \Lambda^2\rangle_t$. The reason for this mismatch is simple to understand: as discussed in \S\ref{sec:forces} (see also Paper~I), the variance $\langle \Lambda^2 \rangle$ is completely dominated by the contribution of the nearest object. If the time interval is very short, $t\lesssim T_{\rm ch}$, the positions of substructures do not change appreciably, and the spectrum of tidal fluctuations $p(\Lambda)$ is sharply peaked about the mean, $\langle \Lambda \rangle$. As the time progresses, $t\gg T_{\rm ch}$, stochastic fluctuations begin to sample the large-force tail of the distribution, $\Lambda\sim\lambda_0\gg \langle \Lambda \rangle$, which leads to an exponential convergence of $\langle \Lambda^2\rangle_t$ towards the asymptotic value given by Equation~(\ref{eq:lamvar2}).

\begin{figure}
\begin{center}
\includegraphics[width=82mm]{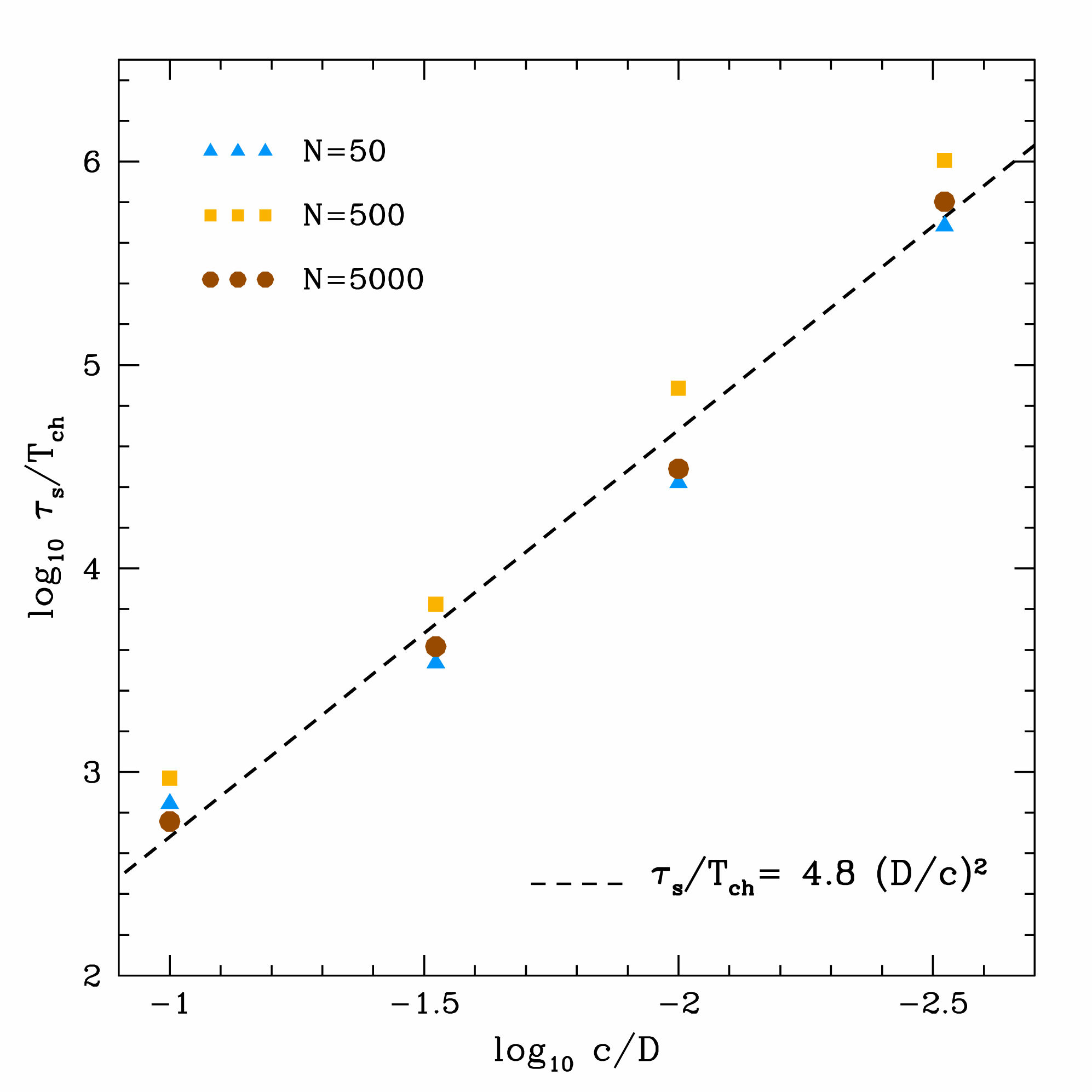}
\end{center}
\caption{Sampling time-scale ($\tau_s$) measured in units of the characteristic duration of tidal fluctuations ($T_{\rm ch}$) as a function of the ratio between the size of substructures ($c$) and their mean separation ($D$). The average {\it number} of tidal fluctuations required to sample the large-force tail of the probability function $p(\bb \Lambda$) scales as $\tau_s\approx 4.8 T_{\rm ch}(D/c)^{2}$ for $c/D\ll 1$ and $N\gg 1$. }
\label{fig:fit}
\end{figure}
Comparison between the three panels of Fig.~\ref{fig:lt_3} highlights the strong dependence of the magnitude of the tidal fluctuations on the size of substructures. Indeed, according to the analytical expression given by Equation~(\ref{eq:lamvar2}), the force variance scales as 
$\langle \Lambda^2\rangle \sim (D/c)^{3}$ for $c/D\ll 1$.
The scatter about the ensemble-average value also increases significantly as the size-to-separation ratio $c/D$ decreases, which can be traced back to the fact that $\langle \Lambda^2\rangle_t$ is largely governed by the few substructures that come closest to the tracer particle within the time interval $t$. Note that for spatially-overlapping populations ($c\gtrsim D$) there is a systematic mismatch between the numerical and analytical values. However, such deviations are to be expected given the analytical expression~(\ref{eq:lamvar2}) is only strictly valid for 'rarefied' populations where $c/D\ll 1$ (for details, see Paper I). 

Fig.~\ref{fig:lt_3} also reveals that the convergence of $\langle \Lambda^2\rangle_t$ towards Equation~(\ref{eq:lamvar2}) shifts to systematically later times as the size of individual substructures shrinks with respect to their average separation. Given that the time interval $t$ is given in units of the average duration of tidal fluctuations, $T_{\rm ch}$, the x-axis represents an average {\it number} of fluctuations within a given time span. Therefore, one may conclude that the average number of fluctuations required to reach the asymptotic behaviour of $\langle \Lambda^2\rangle_{\rm ens}(t)$ increases as the ratio $c/D$ decreases.
To quantify the convergence time we fit the numerical curves with an exponential function (red lines) 
\begin{align}\label{eq:lam_fit}
\langle \Lambda^2\rangle_{\rm ens}(t)=\langle \Lambda^2\rangle_\infty[1-\exp(-t/\tau_s)],
\end{align}
where $\langle \Lambda^2\rangle_\infty$ corresponds to the asymptotic value of the variance in the limit $t\to \infty$, $\delta_s(t)=1-\exp(-t/\tau_s)$ is the so-called {\it sampling delay} function, and $\tau_s$ is defined as the {\it sampling time-scale}. On long time-scales, $t\gtrsim\tau_s$, the variance of tidal fluctuations converges asymptotically towards $\langle \Lambda^2\rangle_\infty\approx \langle \Lambda^2\rangle$, which suggests that $\tau_s$ measures the average time required to sample the large-force tail of $p(\Lambda$).
Fig.~\ref{fig:fit} shows that the sampling time can be approximately written as $\tau_s\approx 4.8 T_{\rm ch} (D/c)^2$, and that this relation is largely independent of the number of substructures, $N$. According to Equation~(\ref{eq:pr_closest}), the quadratic slope reflects the fact that the sampling time-scale is inversely proportional to the probability to find the closest substructure at a distance $r\sim c$. Indeed, from Equation~(\ref{eq:pr_closest}) we find that $p(r)\sim r^{2}$ for $c/D\ll 1$ ($ 4\pi c^3 n/3\ll 1$).
For reasons that will become evident in \S\ref{sec:micro}, it is useful to define the {\it sampling frequency}
\begin{align}\label{eq:omega_s}
w_s\equiv \frac{1}{\tau_s}\approx 1.49 \,c^2 n\langle v^2\rangle^{1/2}.
\end{align}
Note that the sampling time becomes arbitrary long in the particle limit, i.e. $\tau_s\to \infty$ ($w_s\to 0$) as $c\to 0$, which means that the amplitude of tidal fluctuations induced by a background of point-masses never converges, and will therefore reach arbitrarily-large values in the limit $t\to\infty$.

\subsection{Diffusion in the integral-of-motion space}\label{sec:num_diff}
In this Section we follow the evolution of binary 'stars'  subject to the tidal fluctuations computed in \S\ref{sec:num_fluctu}. For simplicity, tracer particles are set on circular orbits at $t_0=0$, which reduces the dimensionality of the problem to the initial (specific) energy, $E_0=-G m/(2 a_0)$.
\begin{figure}
\begin{center}
\includegraphics[width=84mm]{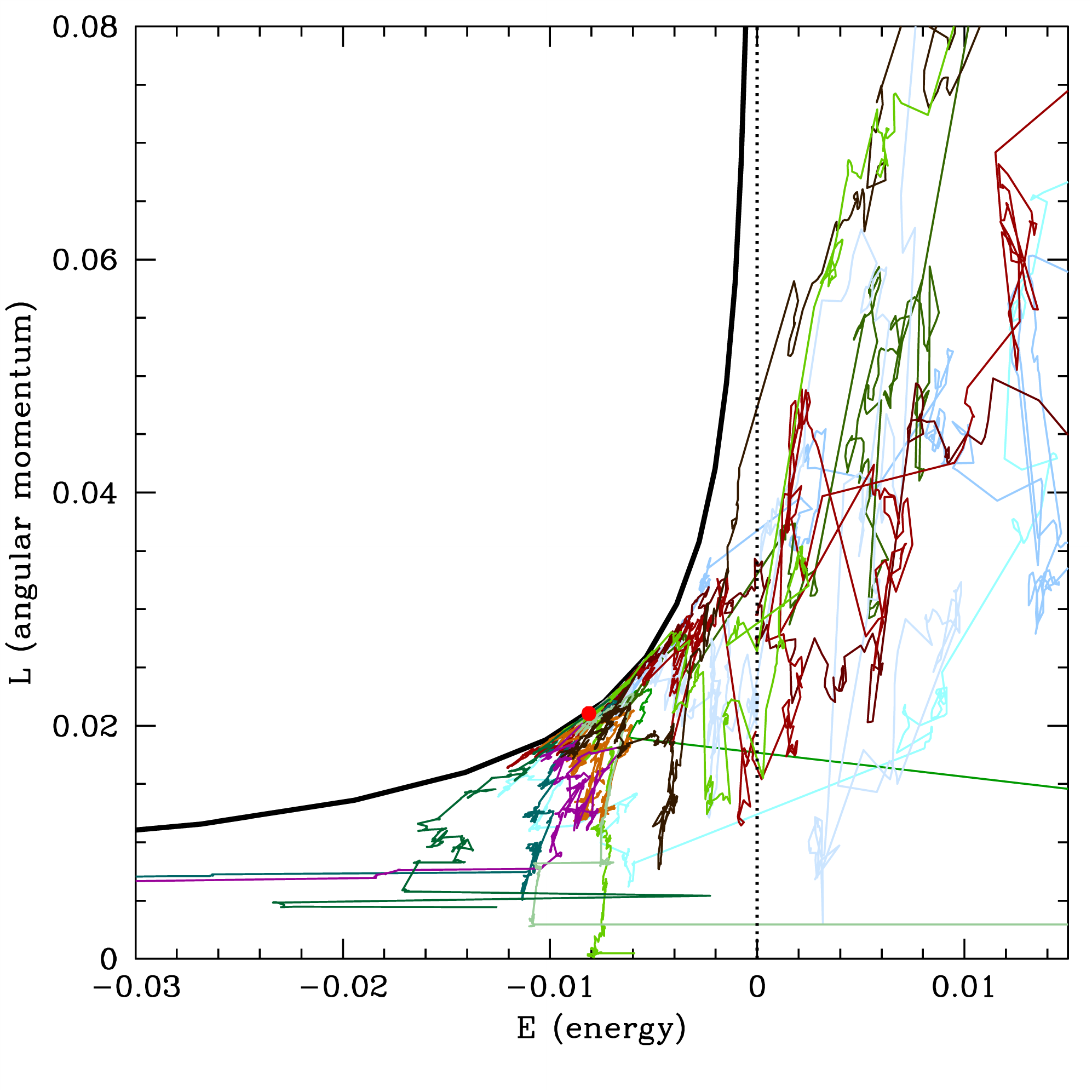}
\end{center}
\caption{Brownian motion of single tracer particles in the integral-of-motion space of a Dehnen potential with $\gamma=0$ and $G=\rho_0=r_0=1$. Initially, particles move on a circular orbit with a semi-major axis $a=10 D$, where $D=(2\pi n_0)^{-1/3}$ in the mean distance between substructures. For ease of reference, the initial location is marked with a red dot, and the zero-energy threshold ($E=0$) with a vertical dotted line. Thin lines show the response to the tidal fluctuations generated by independent ensembles of $N=5000$ substructures with a mass $M=10^{-7}$ and a size-to-separation ratio $c/D=0.1$. Note (i) the role of a reflecting barrier played by the circular angular momentum $L_c(E)$ (thick black line), and (ii) the net flux of particles travelling from negative towards positive energies.}
\label{fig:brown}
\end{figure}

\begin{figure*}
\begin{center}
\includegraphics[width=152mm]{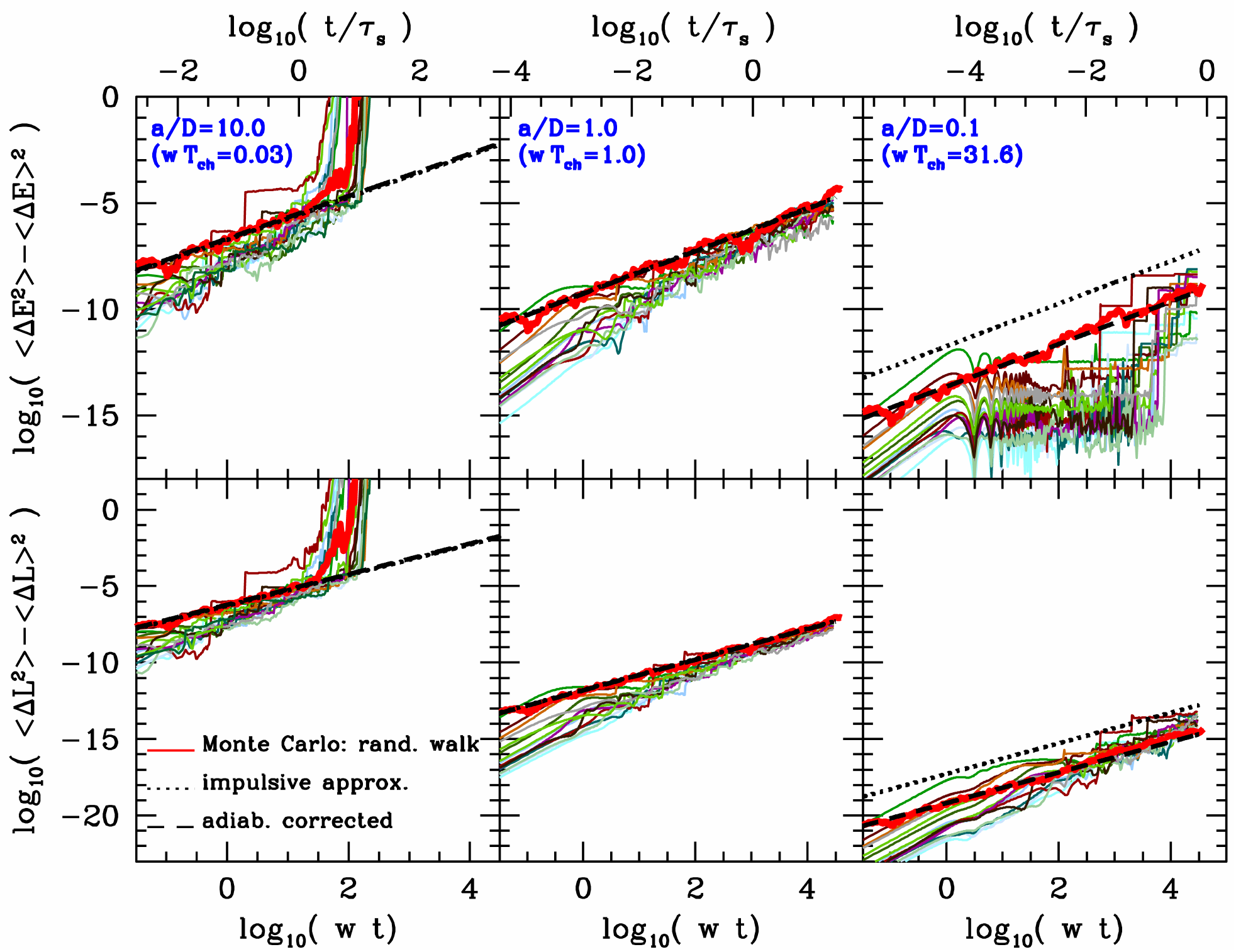}
\end{center}
\caption{ Ensemble-averaged energy (upper panels) and angular momentum (lower panels) variance of tracer particles moving on circular orbits in a Keplerian potential~(\ref{eq:phis_kep}) surrounded by $N=5000$ Hernquist spheres with a mass $M=10^{-7}$, and a scale-radius $c=0.1D$, where $D$ is the average separation between substructures. Left, middle and right panels show orbits with semi-major axes $a/D=10, 1.0$ and 0.1, and orbital frequencies $w T_{\rm ch}=0.03$ (impulsive), 1.0 (intermediate) and 31.6 (adiabatic), where $T_{\rm ch}$ is the average duration of tidal fluctuations~(\ref{eq:Tch}).  Thin coloured curves correspond to models where the tidal force is computed directly from the forces induced by individual substructures. Black-dotted lines show the statistical relation $\sigma_E^2=2 D_E\,t$ and $\sigma_L^2=2 D_L\,t$ derived from diffusion equations under the impulse approximation, Equation~(\ref{eq:CELimp_kep}), whereas black-dashed lines show the adiabatic-corrected values~(\ref{eq:DEL}).
Thick-red lines correspond to Monte-Carlo models that sample velocity 'kicks' from the distribution $\Psi({\mathbfit V}, \Delta {\mathbfit V},\Delta t)$ given by Equation~(\ref{eq:Psi}) (see \S\ref{sec:setup} for details). Note that diffusion and Monte-Carlo curves approach the direct-force results on time-intervals $t\gtrsim \tau_s$, where $\tau_s$ is the sampling time-scale~(\ref{eq:lam_fit}).
  }
\label{fig:ens_6}
\end{figure*}
Fig.~\ref{fig:brown} illustrates how random velocity increments acquired by tracer particles over short intervals of time lead to a Brownian motion in the integral-of-motion space. The energy and angular momentum of single tracer particle varies randomly as a response to the fluctuating tidal field generated by ensembles of $N=5000$ Hernquist spheres with a mass $M=10^{-7}$ and a size $c=0.1D$, separated by an average distance $D=(2\pi n_0)^{-1/3}$. Thin lines correspond to different random realizations of the substructure population. At $t_0=0$ tracer particles have a semi-major axis $a_0=10 D$ and an angular momentum $L_0=L_c(E_0)=\sqrt{G m a_0}$, which we mark with a red dot for ease of reference. 
As a response to the fluctuating tidal field, on average particles tend to gain kinetic energy, thus `drifting' towards the energy threshold $E=0$. 
Simultaneously, a random walk in angular momentum takes the particles away from the initial location on the circular-velocity curve $L_c(E)$ (solid black line). Thus, scattered orbits gradually become more radial and less energetically bound.

In order to test the impulse approximation and the adiabatic correction explored in Section~\ref{sec:adiab}, we integrate the evolution of tracer particles for a time-interval $w\,t_{\rm max}=50000$, where $w=2\pi/P=\sqrt{Gm/a_0^{3}}$ is the frequency of a circular orbit.
Left, middle and right panels of Fig.~\ref{fig:ens_6} correspond to orbits with initial semi-major axes, $a_0/D=10, 1.0$ and 0.1. These models cover $\sim 3$ orders of magnitude in orbital frequencies, $w T_{\rm ch}\simeq 0.03, 1.0$ and 31.6, respectively, where $T_{\rm ch}$ is the typical duration of tidal fluctuations~(\ref{eq:Tch}).
With this choice of parameters, particles with the largest/smallest semi-major axis react impulsively/adiabatically to the fluctuations of the tidal field. In units of the (impulsive) escape time~(\ref{eq:tesc_imp}), our models cover a much larger range of values, $w\,t_{\rm esc}\simeq 1.3\times 10^3, 4.0\times 10^7 $ and $1.3\times 10^{12}$ for $a_0/D=10.0, 1.0$ and 0.1, respectively. Hence, of the three orbits explored here, only particles with $a_0=10 D$ are evolved for a sufficiently long time to escape from the system (i.e. $t_{\rm esc}<t_{\rm max}$).
Recall that on time-intervals $t\ll t_{\rm esc}\simeq t_{\rm iso}$ particles are not ``aware'' of the finite boundaries of the integral-of-motion space (see Appendix~\ref{sec:diff}), hence the evolution of energy and angular momentum is expected to follow the relation exhibited by free-diffusion processes in an unconfined volume, $\sigma^2=2Dt$. To test this theoretical expectation, we plot the time-evolution of the energy variance, $\sigma_E^2=\langle \Delta E^2\rangle - \langle \Delta E\rangle^2$, and angular momentum variance, $\sigma_L^2=\langle |\Delta \mathbfit{L}|^2\rangle-\langle |\Delta \mathbfit{L}|\rangle^2$. Values derived from a direct-force summation (thin coloured lines) are compared against the theoretical predictions derived under the impulse approximation (black-dotted lines) and the adiabatically-corrected values (black-dashed lines).

The first noteworthy result is the convergence of $\sigma_E^2(t)$ and $\sigma_L^2(t)$ computed from direct-force $N$-body models towards the free-diffusing relation $\sigma^2=2 D t$. As expected from Section~\ref{sec:num_fluctu}, numerical curves approach Equation~(\ref{eq:DEL})
on time-scales $t\gtrsim \tau_s$, where $\tau_s$ is the sampling time-scale plotted in Fig.~\ref{fig:fit}. This can be better seen by looking at the upper axis plots, which measures the time interval in units of the sampling time, $\tau_s$.
Notice that on very short time-scales, $t\ll \tau_s$, the time-evolution of $\sigma_E^2$ and $\sigma_L^2$ undergoes periodic oscillations rather than a monotonic increase. These oscillations are particularly prominent in the upper-right panel, which corresponds to orbits with such a high orbital frequency that particles perform $\sim 10^4$ revolutions around the potential $\Phi_s$ before the location of the closest substructures changes appreciably. Hence, for these objects the external tidal field undergoes a ``parallax'' oscillation in phase with their orbital motion, which causes cyclic variations of energy and, to a lesser extent angular momentum.
Once the spectrum of tidal fluctuations $p(\Lambda)$ is thoroughly sampled and convergence has been reached, we find that the adiabatic-corrected values of $\sigma_E^2$ and $\sigma_L^2$ derived from Weinberg (1994a,b,c) formula (black-dashed lines) show an excellent match to all the models explored here. In contrast, the values derived under the impulse approximation largely overestimate the increments of energy and angular momentum. This is particularly evident for 'adiabatic' models with $w\,T_{\rm ch}\gg 1$ (right panels).

Finally, it is worth highlighting the accelerated increase of energy and angular momentum variance exhibited by the `impulsive' ($w T_{\rm ch}\ll 1$) models at late times. In particular, particles with a large semi-major axis $a_0=10 D$ show a rapid growth of $\sigma_E^2(t)$ and $\sigma_L^2(t)$ at $w \,t\gtrsim 10^2$ that cannot be reproduced by Equation~(\ref{eq:DEL}). Although not shown here, we find that the regime of rapid heating coincides with the unbinding of test particles as the orbital energy approaches the boundary $E=0$. 
Interestingly, Monte-Carlo models show that sampling velocity 'kicks' from the probability function~(\ref{eq:Psi}) (red solid lines) successfully describe the average evolution of $\sigma_E^2(t)$ and $\sigma_L^2(t)$ of all our models. Remarkably, these models also reproduce the accelerated heating experienced by test particles as they escape from the system.

\section{Application to planetary orbits}\label{sec:appl}
This Section illustrates the theoretical framework described in \S\ref{sec:ensemble} by 
following the evolution of planetary discs composed of test particles (or `comets') orbiting around a Keplerian potential subject to stochastic fluctuations of an external tidal field. Tidal heating is modelled following two different approaches: (i) Monte-Carlo sampling velocity impulses as described in \S\ref{sec:nbody}, and (ii) convolving the initial energy--angular momentum distribution with Green's propagators. For convenience, we choose units where $G=m=D=\langle v^2\rangle =1$. In these units, the mass and size of substructures are set to $M=10^{-3}$ and $c=0.1$, respectively. 

For simplicity, we adopt idealized conditions at $t_0=0$ where particles move on a razor-thin disc with energies homogeneously distributed over an interval $E_0\in(E_1,E_2)$, and an angular momentum vector that points along the (positive) $z$-axis, $\mathbfit{L}=(0,0,L_c[E])$. The initial state of the ensemble is thus fully specified by the probability function 
\begin{align}\label{eq:N0}
N(E_0,{\mathbfit L}_0,t_0)=N_0 \delta(L_{x,0})\delta(L_{y,0})\delta[L_{z,0}-L_c(E_0)],
\end{align}
where $N_0=1/(E_2-E_1)$ is a normalization constant which guarantees that $\int_{E_1}^{E_2} \d E_0\int \d^3L\, N(E_0,{\mathbfit L}_0,t_0)=1$. Equation~(\ref{eq:N0}) implies that at $t=t_0$ comets move on circular orbits on the plane $z=0$ within a radial range $R\in [R_1,R_2]=[Gm/(-2 E_1),Gm/(-2 E_2)]$.

\subsection{Evolution in the integral-of-motion space}\label{sec:dist}
The non-equilibrium distribution of integrals at $t>t_0$ can be directly derived from the convolution of the initial distribution~(\ref{eq:N0}) with the Green's functions~(\ref{eq:pE_con}) and~(\ref{eq:pL_con}), which yields
\begin{align}\label{eq:Ntpl}
  N(E,{\mathbfit L},t)
  &=N_0 \int_{E_1}^{E_2}\d E_0 \int_{-L_c}^{+Lc}\d L_{x,0}\int_{-L_c}^{+Lc}\d L_{y,0}\int_{-L_c}^{+Lc}\d L_{z,0}\\ \nonumber
 &\qquad {} \times \delta(L_{x,0})\delta(L_{y,0})\delta(L_{z,0}-L_c)p(E,{\mathbfit L},t|E_0,{\mathbfit L}_0,t_0)  \\ \nonumber
&= N_0 \int_{E_1}^{E_2} \d E_0 \frac{1}{\sqrt{4\pi \tilde D_E}} \bigg\{ \exp\bigg[-\frac{(E-E_0+\tilde C_E)^2}{4 \tilde D_E}\bigg] \\ \nonumber
&\qquad {}-\exp\bigg(E_0\frac{\tilde C_E}{\tilde D_E}\bigg)\exp\bigg[-\frac{(E+E_0+\tilde C_E)^2}{4 \tilde D_E}\bigg] \bigg\} \\ \nonumber
  &\qquad {}\times \sum_{n=0}^\infty\sum_{m=0}^\infty\sum_{l=0}^\infty \frac{\alpha_{nml}}{L_c^3}\exp[-\lambda_{nml}\tilde D_L]\\ \nonumber
 &\qquad {}\times \cos\bigg(\frac{n \pi}{2}\bigg)\cos\big[\frac{n\pi (L_x+L_c)}{2 L_c}\big] \\ \nonumber
  &\qquad {}\times\cos\bigg(\frac{m\pi}{2}\bigg)\cos\big[\frac{m\pi (L_y+L_c)}{2 L_c} \big]\\ \nonumber
  &\qquad {}\times \cos(l \pi)\cos\big[\frac{l\pi (L_z+L_c)}{2 L_c}\big] ,
\end{align}
with coefficients $\tilde C_E=\tilde C_E(E_0,L_c,t-t_0)$, $\tilde D_E=\tilde D_E(E_0,L_c,t-t_0)$ and $\tilde D_L=\tilde D_L(E_0,L_c,t-t_0)$ and an integration constant $\lambda_{nml}=\pi^2 (n^2+m^2+l^2)/(4 L_c^2)$.
Equation~(\ref{eq:Ntpl}) simplifies considerably if we assume that the distribution of angular momentum in the $x$ and $y-$directions are tightly peaked around the initial null value, and thus far from the reflecting barriers at $\pm L_c$. Under this approximation, propagators in the $L_x$ and $L_y$ dimensions have a Gaussian form~(\ref{eq:pL}) (see Appendix~\ref{sec:diff}), and Equation~(\ref{eq:Ntpl}) reduces to
\begin{align}\label{eq:Nt2}
  N(E,{\mathbfit L},t)&= N_0 \int_{E_1}^{E_2} \d E_0 \frac{1}{\sqrt{4\pi \tilde D_E}}\bigg\{ \exp\bigg[-\frac{(E-E_0+\tilde C_E)^2}{4 \tilde D_E}\bigg] \\ \nonumber 
  &\qquad {}-\exp\bigg(E_0\frac{\tilde C_E}{\tilde D_E}\bigg)\exp\bigg[-\frac{(E+E_0+\tilde C_E)^2}{4 \tilde D_E}\bigg] \bigg\}\\ \nonumber
  &\qquad {}\times \frac{1}{4\pi \tilde D_L}\exp\bigg[-\frac{L_x^2+L_y^2}{4 \tilde D_L}\bigg] \\ \nonumber
  &\qquad {}\times\frac{1}{L_c}\bigg\{\frac{1}{2}+\sum_{\ell=1}^\infty(-1)^\ell\cos\big[\frac{\ell\pi (L_z+L_c)}{2 L_c}\big] e^{-\lambda_\ell\tilde D_L}\bigg\}
\end{align}
where $\lambda_\ell=\pi^2 \ell^2/(4 L_c^2)$ and $\cos(\ell\pi)=(-1)^\ell$. In order to gain physical intuition into the physical behaviour of $N(E,{\mathbfit L},t)$, it is useful to reduce the dimensionality of the probability function by marginalizing over each integral dimension.

\subsubsection{Energy space}\label{sec:en_sp}
Let us first calculate the distribution of energy $N(E,t)$ by integrating each side of 
 Equation~(\ref{eq:Nt2}) over $\d^3 L=2\pi L_R \,\d L_R\,\d L_z$, where $L_R=\sqrt{L_x^2+L_y^2}$ is the planar component of the angular momentum, which yields
\begin{align}\label{eq:NE}
  N(E,t)&=\int \d^3 L\, N(E,{\mathbfit L},t) \\ \nonumber
  &= N_0 \int_{E_1}^{E_2} \d E_0 \frac{1}{\sqrt{4\pi \tilde D_E}}\bigg\{ \exp\bigg[-\frac{(E-E_0+\tilde C_E)^2}{4 \tilde D_E}\bigg] \\ \nonumber 
  &\qquad {}-\exp\bigg(E_0\frac{\tilde C_E}{\tilde D_E}\bigg)\exp\bigg[-\frac{(E+E_0+\tilde C_E)^2}{4 \tilde D_E}\bigg] \bigg\}\\ \nonumber
  &\qquad {}\times 2\pi \int_0^{L_c}\d L_R \frac{L_R}{4\pi \tilde D_L}\exp\bigg[-\frac{L_R^2}{4 \tilde D_L}\bigg] \\ \nonumber
  &\qquad {}\times\frac{1}{L_c}\int_{-L_c}^{+L_c}\d L_z\bigg\{\frac{1}{2}+\sum_{\ell=1}^\infty(-1)^\ell\cos\big[\frac{\ell\pi (L_z+L_c)}{2 L_c}\big] e^{-\lambda_\ell \tilde D_L}\bigg\}.
\end{align}
If the variation of the radial component of the angular momentum is small, $|\Delta L_R|\ll L_c$, the integral over $\d L_R$ takes the value of unity
$$2\pi \int_0^{L_c}\d L_R\frac{L_R}{4\pi \tilde D_L}e^{-L_R^2/(4 \tilde D_L)}\approx 2\pi \int_0^{\infty}\d L_R \frac{L_R}{4\pi \tilde D_L}e^{-L_R^2/(4 \tilde D_L)} =1,$$
whereas, by construction, the integral over $\d L_z$ is normalized to unity
\begin{align}
\frac{1}{L_c}\int_{-L_c}^{+L_c}\d L_z\bigg\{\frac{1}{2}+\sum_{\ell=1}^\infty(-1)^\ell\cos\big[\frac{\ell\pi (L_z+L_c)}{2 L_c}\big] e^{-\lambda_\ell \tilde D_L}\bigg\} &\\ \nonumber
=1+\sum_{\ell=1}^\infty(-1)^\ell \frac{2 L_c}{\ell\pi}\sin(\ell\pi)e^{-\lambda_\ell \tilde D_L}=1.& \nonumber
\end{align}
Thus, Equation~(\ref{eq:NE}) reduces to a single integral
\begin{align}\label{eq:NE2}
  N(E,t)&= N_0 \int_{E_1}^{E_2} \d E_0 \frac{1}{\sqrt{4\pi \tilde D_E}}\bigg\{ \exp\bigg[-\frac{(E-E_0+\tilde C_E)^2}{4 \tilde D_E}\bigg] \\ \nonumber 
  &\qquad {}-\exp\bigg(E_0\frac{\tilde C_E}{\tilde D_E}\bigg)\exp\bigg[-\frac{(E+E_0+\tilde C_E)^2}{4 \tilde D_E}\bigg] \bigg\}, \\ \nonumber
  &= N_0 \int_{E_1}^{E_2} \d E_0\, p(E,t|E_0,L_c,t_0),
\end{align}
where $p(E,t|E_0,L_c,t_0)$ is the energy propagator~(\ref{eq:pE_con}) with coefficients computed for circular orbits, $\tilde C_E=\tilde C_E(E_0,L_c,t-t_0)$ and $\tilde D_E=\tilde D_E(E_0,L_c,t-t_0)$. 
The role of the negative right-hand term of~(\ref{eq:NE2}) must be highlighted, as it causes the removal of comets that cross the absorbing barrier at $E=0$ (see Appendix~\ref{sec:diff}). This means that the total number of bound comets $\int_{-\infty}^0 \d E\, N(E,t)$ decreases steadily with time, as it is shown in more detail below. 

\subsubsection{Angular momentum space}\label{sec:angmom_sp}
Similarly, the probability to find comets with an angular momentum component between $L_R, L_R+\d L_R$ at the time $t$ can be calculated as
\begin{align}\label{eq:NLR}
  N(L_R,t)&=\int_{-\infty}^0\d E \int_{-L_z}^{+L_z} \d L_z \,N(E,L_R,L_z,t) \\ \nonumber
  &=N_0\int_{E_1}^{E_2} \d E_0 \frac{1}{4 \pi \tilde D_L}\exp\bigg[-\frac{L_R^2}{4 \tilde D_L}\bigg]\\ \nonumber
&\qquad {}\times  \int_{-\infty}^0\d E\, p(E,t|E_0,L_c,t_0)\\ \nonumber 
  &=\frac{N_0}{2}\int_{E_1}^{E_2} \d E_0 \frac{1}{4 \pi \tilde D_L}\exp\bigg[-\frac{L_R^2}{4 \tilde D_L}\bigg]\bigg\{ 1+\erf\bigg(\frac{\tilde C-E_0}{2\sqrt{\tilde D_E}}\bigg)\\ \nonumber 
  &\qquad {}-\exp\bigg(E_0\frac{\tilde C_E}{\tilde D_E}\bigg)\bigg[1+\erf\bigg(\frac{\tilde C+E_0}{2\sqrt{\tilde D_E}}\bigg)\bigg] \bigg\},
\end{align}
whereas the vertical component of angular momentum has a distribution
\begin{align}\label{eq:NLz}
  N(L_z,t)&\approx\int_{-\infty}^0\d E \times 2\pi \int_{0}^{\infty} \d L_R\, L_R \,N(E,L_R,L_z,t) \\ \nonumber
 &= N_0  \int_{E_1}^{E_2}  \d E_0\, p(L_z,t|E_0,L_c,t_0) \int_{-\infty}^0\d E p(E,t|E_0,L_c,t_0) \\ \nonumber
 &\qquad {}\times 2\pi \int_0^\infty \d L_R \frac{L_R}{4 \pi \tilde D_L}\exp\bigg[-\frac{L_R^2}{4 \tilde D_L}\bigg]\\ \nonumber
  &= \frac{N_0}{2}  \int_{E_1}^{E_2} \d E_0\frac{1}{L_c}\bigg\{\frac{1}{2}+\sum_{\ell=1}^\infty(-1)^\ell\cos\big[\frac{\ell\pi (L_z+L_c)}{2 L_c}\big] e^{-\lambda_\ell \tilde D_L}\bigg\} \\ \nonumber
 &\qquad {}\times \bigg\{ 1+\erf\bigg(\frac{\tilde C_E-E_0}{2\sqrt{\tilde D_E}}\bigg) \\ \nonumber
  &\qquad {}-\exp\bigg(E_0\frac{\tilde C_E}{\tilde D_E}\bigg)\bigg[1+\erf\bigg(\frac{\tilde C_E+E_0}{2\sqrt{\tilde D_E}}\bigg)\bigg] \bigg\},
\end{align}
here
\begin{align}\label{eq:pLz}
p(L_z,t|E_0,L_c,t_0)=  \frac{1}{L_c}\bigg\{\frac{1}{2}+\sum_{\ell=1}^\infty(-1)^\ell\cos\big[\frac{\ell\pi (L_z+L_c)}{2 L_c}\big] e^{-\lambda_\ell \tilde D_L}\bigg\},
\end{align}
is the one-dimensional angular momentum propagator~(\ref{eq:pL1D_sol}) evaluated at $L_{z,0}=L_c$ with a coefficient $\tilde D_L=\tilde D_L(E_0,L_c,t_0)$.

Unfortunately, the distributions $N(E,t), N(L_R,t)$ and $N(L_z,t)$ must be solved numerically owing to the non-trivial dependence of the coefficients $\tilde C_E, \tilde D_E$ and $\tilde D_L$ on $E_0$. Below we illustrate the time-evolution of angular momentum and energy via comparison against Monte-Carlo $N$-body models.

\begin{figure*}
\begin{center}
\includegraphics[width=176mm]{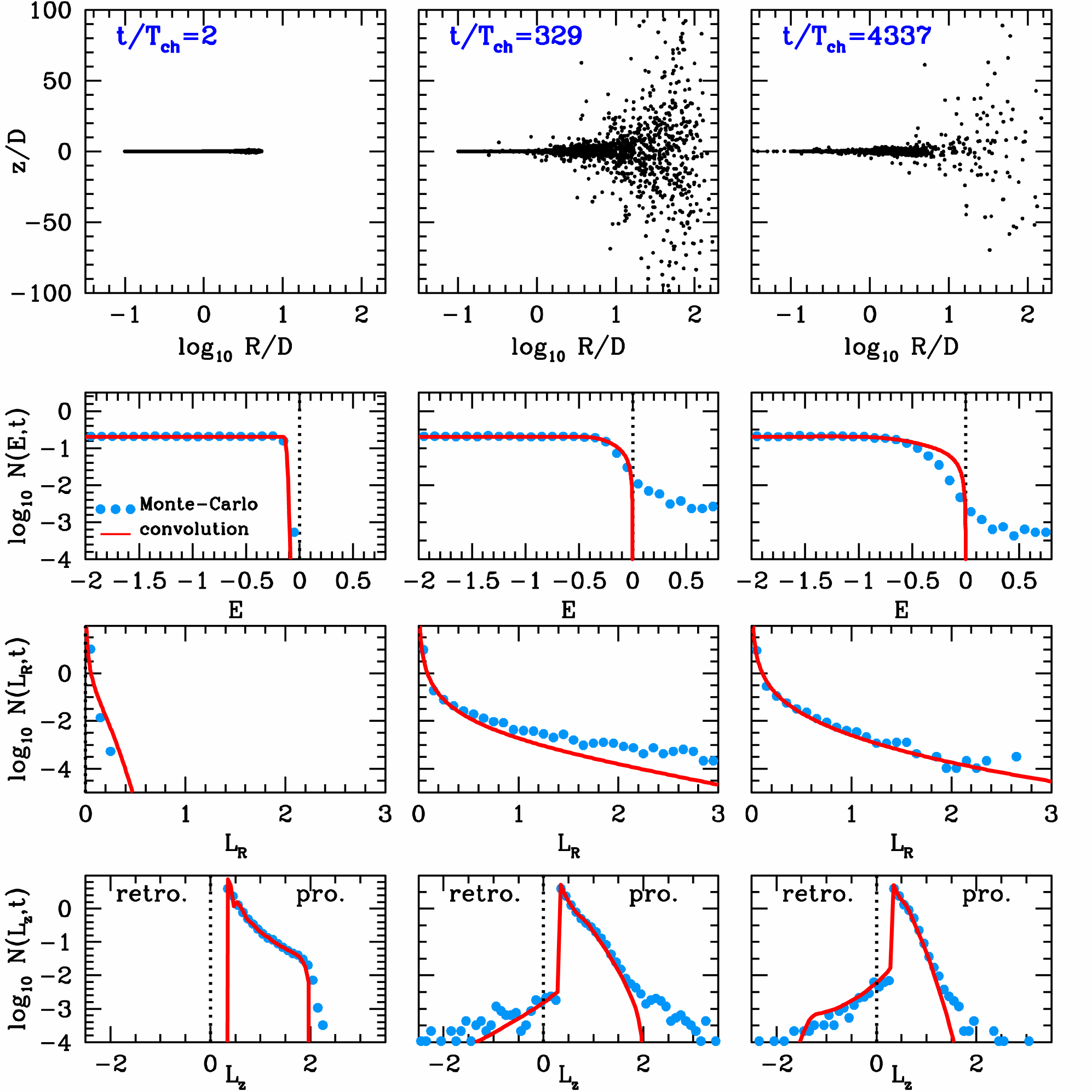}
\end{center}
\caption{Three snapshots of a Monte-Carlo $N$-body experiment that follows the tidal heating of a planetary disc subject to external tidal fluctuations. Middle panels coincide with the time at which the evaporation rate peaks.
  At $t_0=0$ all tracer particles move on circular orbits in the plane $z=0$ within a radial interval $R/D\in (R_1,R_2)=(0.1,4)$. Background substructures have a mass $M=10^{-3}$ and size $c=0.1$ in units where $G=m=D=\langle v^2\rangle=1$.  {\it Upper panels:} Locations in cylindrical coordinates $(R,z)$ of tracer particles. As a result of orbital scattering, the initial disc configuration heats up in the outer regions, leading to the formation of Oort-like clouds at large distances.  {\it Second row} plot the energy distribution, $N(E,t)$, at the same snapshots. Blue dots show the distributions measured from Monte-Carlo models, while red curves correspond to the analytical convolution~(\ref{eq:NE2}). For ease of reference, we mark the boundary between bound and unbound orbits ($E=0$) with vertical dotted lines. Note that by construction the analytical curves cannot reproduce the probability tail extending at $E>0$. {\it Third row} plot the distribution of the radial angular momentum component, $N(L_R,t)$, of bound particles. Red curves show the analytical result~(\ref{eq:NLR}). Recall that the initial distribution is a delta function centred at zero, $N(L_R,t_0=0)=\delta(L_R)$. Off-plane motions translate into broader distributions as time goes by. {\it Fourth row} show the distribution of the vertical component of the angular momentum, $N(L_z,t)$, of energetically-bound particles. At $t_0=0$ all particles move on prograde, circular orbits ($L_z>0$). At later times a significant fraction of comets scatter across the boundary $L_z=0$ (marked with vertical dotted lines), thus acquiring a retrograde motion.}
\label{fig:MC}
\end{figure*}

\subsubsection{Monte-Carlo $N$-body experiments}\label{sec:MC}
Fig.~\ref{fig:MC} plots three snap-shots in the evolution of $10^5$ tracer particles initially distributed on a disc at $z=0$ within a radial interval $R\in(R_1,R_s)=(0.1,4)$. As time progresses, the effects of tidal fluctuations become particularly strong in the outskirts of the disc, where orbits can be scattered off to large distances from the disc plane, $|z|/D\gtrsim 100$. Interestingly, the resulting spatial distribution at late times is reminiscent to that of the Oort cloud in the Solar system (e.g. Oort 1950; Hills 1981). Note also that the number of test particles at small radii $R\ll R_1$ also increases with time. As we will see below, these comets have been scattered onto radial, or even retrograde motions.

A sharp drop in the number of particles beyond $R\gtrsim 100$ is also noticeable at late times. Indeed, at such large distances most comets are gravitationally unbound from the central potential, thus moving on parabolic trajectories away from the system. This is clearly visible in the distribution of orbital energies, $N(E,t)$, plotted in the second row of Fig.~\ref{fig:MC}, which shows a long probability tail that extends across $E=0$ (marked with vertical dotted lines) towards positive energies at intermediate and late times. The fraction of particles crossing the energy layer $E=0$ increases very rapidly on short time scales, $t\lesssim 100$, and slowly decline thereafter. A detailed study of the time-evolution of the escape rate will be presented in Section~\ref{sec:rate}. Comparison of the energy distribution with the analytical expression~(\ref{eq:NE2}) (red lines) shows good agreement at low energies ($E\lesssim 0$), which progressively worsens in the proximity of the barrier at $E=0$, a region dubbed the `fringe' by Spitzer \& Shapiro (1972). As pointed out by these authors, ``Fokker-Plank equations are valid throughout most of the [system], but not in the fringe''. Indeed, both Fokker-Planck and diffusion equations rely on a Taylor-expansion of Einstein's master equation for small increments of energy $|\Delta E|\ll |E|$ within a finite time interval $t>0$ (e.g. see Sections~2.2 and 3.3 of P15), a condition that does not hold at $E\sim 0$. For the same reason, diffusion equations do not reproduce the `accelerated' unbinding of particles found in Monte-Carlo models (see left panel of Fig.~\ref{fig:ens_6}), which in turn leads to an overprediction of the number of comets in the fringe. Yet, the largest discrepancy between diffusion and Monte-Carlo models is found in the region of positive energies. Here, Monte-Carlo models show a non-zero probability to find particles at $E\gtrsim 0$ at intermediate and late times. By construction, the tail at positive energies cannot be reproduced by diffusion models with an absorbing barrier at $E=0$, which sets $N(E,t)=0$ for $E\ge 0$ (see Appendix~\ref{sec:diff}). 
  
The third row of Fig.~\ref{fig:MC} shows that the radial angular momentum distributions $N(L_R,t)$ widens progressively as the time interval increases. Given that at $t_0=0$ all particles move on a disc $z=0$, such that $N(L_R,t_0=0)=\delta (L_R)$, non-zero values of $L_R$ must be caused by particles scattered off the initial plane. To see this more clearly, let us define the angle $\cos\theta=\hat{\mathbfit L}\cdot \hat z$, such that $\theta_0=0$ at $t=t_0$. Given that the maximum vertical offset from the disc plane goes as $z_{\rm max}= a\sin\theta$ for $e\approx 0$, we find that $z_{\rm max}/a=L_R/L$, hence showing that the thickening of the outer disc is related to the widening of the distribution $N(L_R,t)$. It is also interesting to observe that the excess of particles above the analytical red curve coincides with highest evaporation rates. Indeed, particles that experience large velocity kicks also tend to gain angular momentum, which explains the correlation between the fringe population and the excess of probability in the large-angular momentum tail of the distribution.

The distribution of the rotational angular momentum, $N(L_z,t)$, is plotted in the bottom panels of Fig.~\ref{fig:MC}, which show a number of remarkable features. Note first that the initial truncations at $L_{z,min}=\sqrt{G m R_1}\approx 0.32$ and $L_{z,max}=\sqrt{G m R_2}= 2$ are quickly erased by tidal perturbations. In particular, the tail extending towards positive angular momentum corresponds to particles scattered into the fringe region, which is poorly described by the analytical expression~(\ref{eq:NLz}).
More interestingly, at intermediate and late times we find a second tail at low, or even negative values of $L_z$. These are comets that move on nearly radial or counter-rotating orbits which periodically plummet into the inner regions of the potential from the outskirts of the planetary system, a process discussed in detail by Hills (1981). The fraction of comets on retrograde orbits correlates with the number of particles escaping from the system. We return to this issue in \S\ref{sec:retro}. 
Finally, the experiments shown in Fig.~\ref{fig:MC} confirm that particles become energetically unbound before the isotropization of the angular momentum distribution is complete. This is in agreement with Equation~(\ref{eq:tisoesc}), which shows that the isotropization of circular orbits occurs on longer time-scales than the average time that these particles remain bound to the Keplerian potential, i.e. $t_{\rm iso}(e=0)>t_{\rm esc}(e=0)$.

\subsection{Evaporation rate}\label{sec:rate}
Given enough time, the fraction of energetically-bound comets decays to zero. This is a consequence of the ergodic theorem, which states that one-dimensional Brownian motion will visit every point of space at least once, including the absorbing wall at $E=0$. 
We can use the energy distribution~(\ref{eq:NE2}) to derive an analytical expression for the loss of comets as a function of time. As a first step, it is useful to calculate the fraction of bound comets as
\begin{align}\label{eq:fb}
  f_b(t)&= \int_{-\infty}^0\d E\, N(E,t) \\ \nonumber
&=  \int_{E_1}^{E_2} \d E_0\, N_0\int_{-\infty}^0\d E\,p(E,t|E_0,L_c,t_0)\\ \nonumber
&=\frac{N_0}{2}\int_{E_1}^{E_2} \d E_0 \bigg\{ 1+\erf\bigg(\frac{\tilde C_E-E_0}{2\sqrt{\tilde D_E}}\bigg)\\ \nonumber 
 &\qquad {}-\exp\bigg(E_0\frac{\tilde C_E}{\tilde D_E}\bigg)\bigg[1+\erf\bigg(\frac{\tilde C_E+E_0}{2\sqrt{\tilde D_E}}\bigg)\bigg] \bigg\}.
\end{align}


The escape, or ``evaporation'' rate is defined as the time-derivative of the fraction of comets that have escaped from the potential $\Phi_s$ in a time interval $t$, i.e. $f_{\rm esc}(t)\equiv 1-f_b(t)$. From~(\ref{eq:fb})
\begin{align}\label{eq:esc_rate}
 \mathcal{R}_{\rm esc}(t)= \frac{\d f_{\rm esc}}{\d t}&=-\frac{\d f_b}{\d t}=-\int_{-\infty}^0\d E\frac{\partial N}{\partial t}\\ \nonumber
  &= -\int_{-\infty}^0\d E \frac{\partial}{\partial t}\int_{E_1}^{E_2} \d E_0\, N_0 p(E,t|E_0,L_c,t_0)\\ \nonumber 
  &= -\int_{E_1}^{E_2} \d E_0 \,N_0\int_{-\infty}^0\d E \frac{\partial}{\partial t}p(E,t|E_0,L_c,t_0).
\end{align}
The right-hand integral in Equation~(\ref{eq:esc_rate}) represents a flux of particles crossing the energy layer $E=0$ at the time $t$ which started from $\delta(E-E_0)\delta(L_{x,0})\delta(L_{y,0})\delta(L_{z,0}-L_c)$ at $t_0=0$. One can thus use the energy term of the diffusion equation~(\ref{eq:pEL}) and the propagator~(\ref{eq:pE_con}) to measure the flux through the barrier $E=0$ as 
\begin{align}\label{eq:flux_E}
  j_E(0,t|E_0,L_c,t_0)&\equiv \int_{-\infty}^0\d E \frac{\partial}{\partial t}p(E,t|E_0,L_c,t_0)\\ \nonumber
  &=C_E \,p(0,t|E_0,L_c,t_0)+D_E\,\frac{\partial p}{\partial E}\bigg|_{E=0} \\ \nonumber
  &=\frac{1}{2t}\frac{1}{\sqrt{4\pi \tilde D_E}}\bigg\{\exp\bigg[-\frac{(-E_0+\tilde C_E)^2}{4 \tilde D_E}\bigg](\tilde C_E+E_0) \\ \nonumber
 &\qquad {}-\,\exp\bigg(E_0\frac{\tilde C_E}{\tilde D_E}\bigg)\exp\bigg[-\frac{(+E_0+\tilde C_E)^2}{4 \tilde D_E}\bigg](\tilde C_E-E_0)\bigg\},
\end{align}
where we have used the relation between static and time-dependent coefficients $\tilde C_E=C_E t$ and $\tilde D_E=D_E t$ (see \S\ref{sec:flux}).

Combination of~(\ref{eq:esc_rate}) and~(\ref{eq:flux_E}) shows that the escape rate corresponds to the average flux of probability crossing $E=0$ at the time $t$ which originates from the energy interval $(E_1,E_2)$ at $t_0=0$
\begin{align}\label{eq:esc_rate2}
  \mathcal{R}_{\rm esc}(t)&= -N_0\int_{E_1}^{E_2} \d E_0\,j_E(0,t|E_0,L_c,t_0)\\ \nonumber
  &=-\frac{N_0}{2t}\int_{E_1}^{E_2} \d E_0 \frac{1}{\sqrt{4\pi \tilde D_E}}\bigg\{\exp\bigg[-\frac{(-E_0+\tilde C_E)^2}{4 \tilde D_E}\bigg](\tilde C_E+E_0) \\ \nonumber
&\qquad {}-\,\exp\bigg(E_0\frac{\tilde C_E}{\tilde D_E}\bigg)\exp\bigg[-\frac{(+E_0+\tilde C_E)^2}{4 \tilde D_E}\bigg](\tilde C_E-E_0)\bigg\}.
\end{align}

\begin{figure}
\begin{center}
\includegraphics[width=84mm]{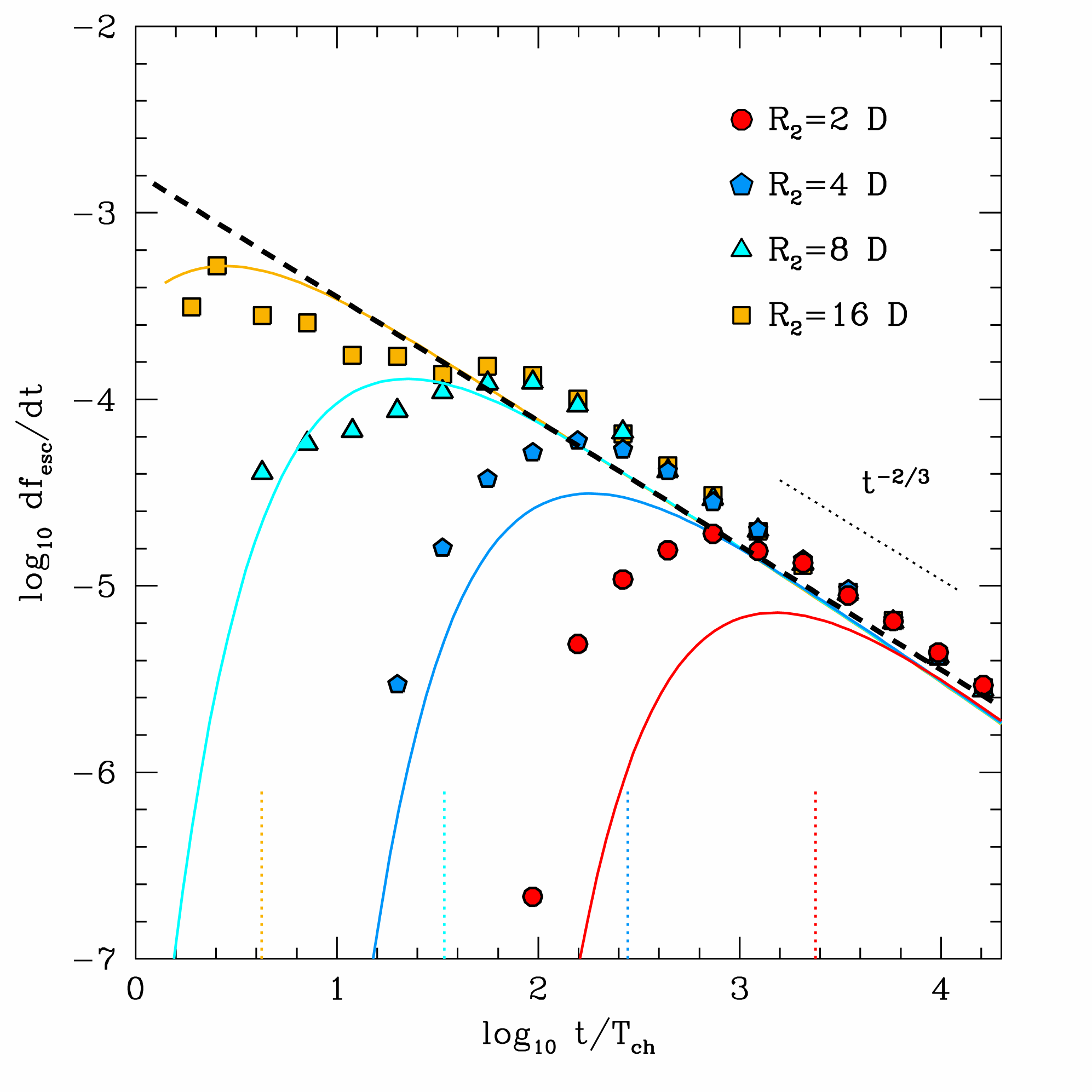}
\end{center}
\caption{Escape (or 'evaporation') rate of disc particles as a function of time. Initially, disc particles move on circular orbits within a radial interval $R\in(0.1,R_2)$, with $R_2$ being the outer-most radius. As in Fig.~\ref{fig:MC}, the mass and size of substructures are $M=10^{-3}$ and $c=0.1$, respectively, with units $G=m=D=\langle v^2\rangle=1$. Coloured symbols show the results from Monte-Carlo $N$-body models, while solid lines denote analytical rates given by Equation~(\ref{eq:esc_rate2}). The power-law asymptotic behaviour~(\ref{eq:esc_rate_pl}) is plotted with a black-dashed line. Note that all escape rates roughly peak at the escape time-scale associated with the outer-most circular orbit, $t_{\rm esc}(E_2)$ (Equation~(\ref{eq:tesc_kep}), marked with vertical dotted lines).}
\label{fig:rate_E}
\end{figure}

\begin{figure}
\begin{center}
\includegraphics[width=84mm]{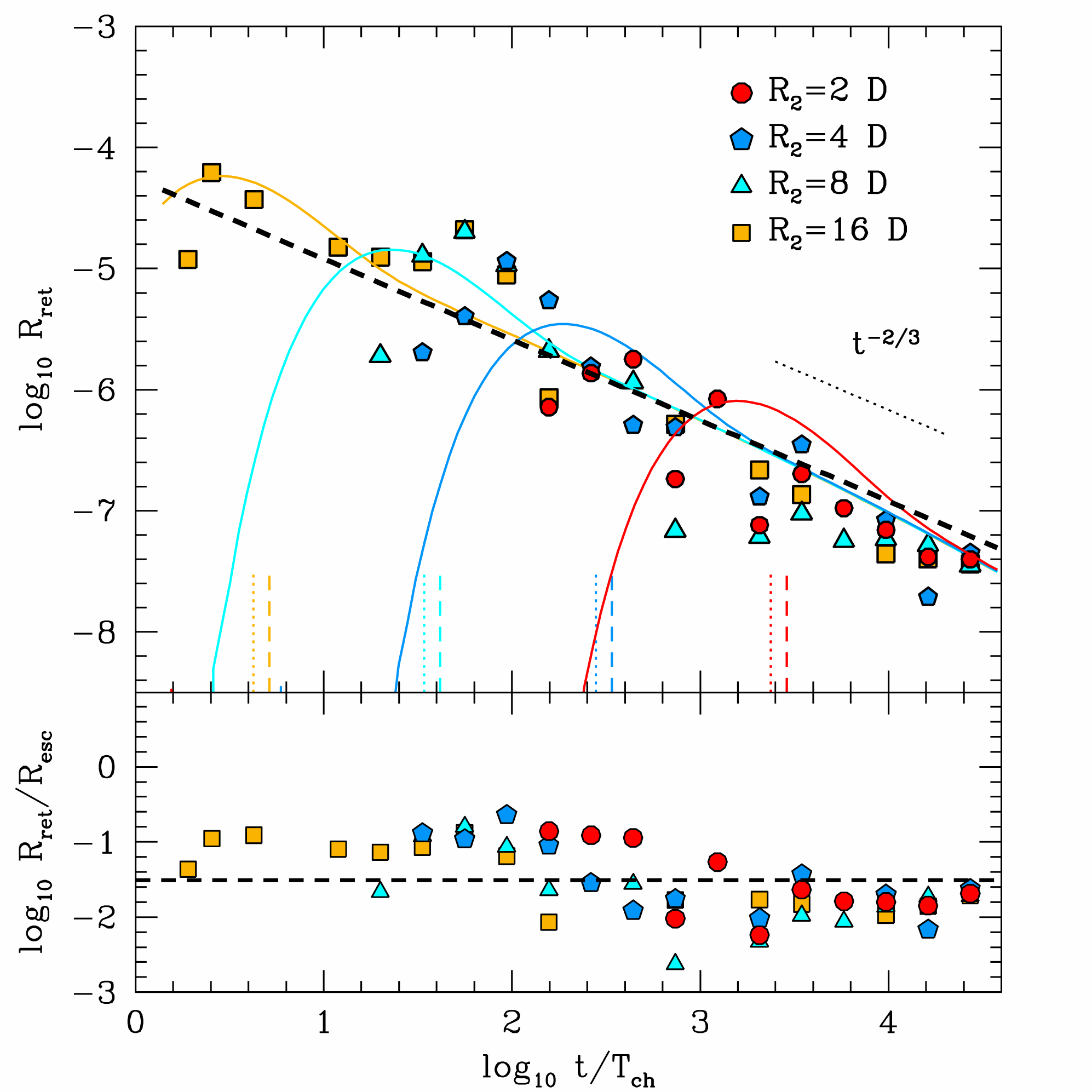}
\end{center}
\caption{{\it Upper panel:} Rate of production of retrograde orbits as a function of time for the models shown in Fig.~\ref{fig:rate_E}. Solid lines denote analytical rates given by Equation~(\ref{eq:rate_ret}). The power-law asymptotic behaviour~(\ref{eq:rate_ret_pl}) is plotted with a black-dashed line. Escape and isotropization time-scales at the outer-most radii, $t_{\rm esc}(E_2)$ and $t_{\rm iso}(E_2)$, Equations~(\ref{eq:tesc_kep}) and~(\ref{eq:tiso_kep}), are marked with vertical dotted and dashed lines, respectively. {\it Lower panel:} Ratio between escape and retrograde rates as a function of time. Note the slow convergence towards the power-law limit $\mathcal{R}_{\rm ret}/\mathcal{R}_{\rm esc}\sim 0.031$ predicted by Equation~(\ref{eq:rate_ret_pl}), marked here with a horizontal dashed line. }
\label{fig:rate_Lz}
\end{figure}
One can show that the rate at which comets leave a Keplerian potential approaches a power-law function $\d f_{\rm esc}/\d t\sim t^{-2/3}$  on time-scales $t_{\rm esc}(E_2) \ll t \ll t_{\rm esc}(E_1)$, where $t_{\rm esc}(E_0)=E_0/C_E(E_0,L_c,t_0)$ corresponds to the average unbinding time of particles moving on circular orbits with an energy $E_0$ at $t=t_0$. To demonstrate this property, it is useful to introduce a dimension-less variable $\beta = t_{\rm esc}/t$. As shown in \S\ref{sec:kepler}, the impulsive approximation is accurate on time-scales $t\sim t_{\rm esc}$, hence one can use the analytical coefficients~(\ref{eq:CELimp_kep}) to solve Equation~(\ref{eq:esc_rate2}), which yields
$$E_0\frac{\tilde C}{\tilde D}=E_0\frac{ C}{ D}=\frac{3}{2}$$
and 
$$\frac{(-E_0+\tilde C_E)^2}{4\tilde D_E}=\frac{(-E_0+ C_E \,t)^2}{4 D_E\,t}=\frac{3}{8}\frac{(1-\beta)^2}{\beta}.$$
After some algebra, Equation~(\ref{eq:esc_rate2}) can be written as
\begin{align}\label{eq:esc_rate3}
  \mathcal{R}_{\rm esc}(t)=\frac{N_0}{ 2^{5/6}3^{1/2}5^{1/3}4\pi^{1/6}}\frac{K(E_1,E_2,t)}{t^{2/3}}\bigg(\frac{ G^2 M m}{c}\bigg)^{2/3}n^{1/3}\bigg(\frac{2 \pi}{3\langle v^2\rangle}\bigg)^{1/6},
  \end{align}
with
\begin{align}\label{eq:Int_cons}
  K(E_1,E_2,t)&=
  \int_{\beta_2}^{\beta_1}\d\beta\,\frac{1}{\beta^{7/6}}\bigg\{\exp\bigg[-\frac{3}{8}\frac{(1-\beta)^2}{\beta}\bigg](1+\beta) \\ \nonumber
  &\qquad {}-\exp\bigg(\frac{3}{2}\bigg)\exp\bigg[-\frac{3}{8}\frac{(1+\beta)^2}{\beta}\bigg](1-\beta)\bigg\}.
  \end{align}
On a time-scale $\beta_2 \ll \beta \ll \beta_1$ Equation~(\ref{eq:Int_cons}) becomes approximately constant, $K(E_1,E_2,t)\simeq K$. To calculate the value of $K$ one must replace the lower and upper limits of the integral by zero and infinity, respectively, such that
\begin{align}\label{eq:Int_cons2}
  K&\approx \int_{0}^{\infty}\d\beta\,\frac{1}{\beta^{7/6}}\bigg\{\exp\bigg[-\frac{3}{8}\frac{(1-\beta)^2}{\beta}\bigg](1+\beta)\\ \nonumber
  &\qquad{}-\exp\bigg(\frac{3}{2}\bigg)\exp\bigg[-\frac{3}{8}\frac{(1+\beta)^2}{\beta}\bigg](1-\beta)\bigg\} \\ \nonumber
  &=4 e^{3/4}{\rm K_{-5/6}}(3/4)\simeq 7.046,
   \end{align}
where $K_n(z)$ corresponds to the modified Bessel function\footnote{This function satisfies the equation $z^2K_n''+z K_n' -(z^2 +n^2)K_n=0$ (e.g. Press et al. 1992)}. Hence, inserting~(\ref{eq:Int_cons2}) into~(\ref{eq:esc_rate3}) yields a escape rate that approaches asymptotically a power-law 
\begin{align}\label{eq:esc_rate_pl}
 \mathcal{R}_{\rm esc}(t)\simeq N_0\frac{0.275 }{t^{2/3}}\bigg(\frac{ G^2 M m}{c}\bigg)^{2/3}n^{1/3}\bigg(\frac{2 \pi}{3\langle v^2\rangle}\bigg)^{1/6}.
\end{align}

Fig.~\ref{fig:rate_E} plots the escape rate measured from Monte-Carlo runs (symbols) and the analytical values derived from diffusion equations~(\ref{eq:esc_rate2}) (solid lines). Lines and symbols are colour-coded according to the energy of the outer-most circular orbit at $t_0=0$, $E_2=-Gm/(2 R_2)$. This Figure shows several noteworthy results. Note first that the rate at which particles leave the potential roughly peaks at the escape time of the outer-most radius, $t_{\rm esc}(E_2)$, which is marked with vertical dotted lines for reference. Within short time intervals, $t\ll t_{\rm esc}(E_2)$ the flux of particles that become energetically unbound is negligible in all models, except for very extended discs ($R_2\gtrsim 10D$) at $t_0$. On longer time-scales, $t\gtrsim t_{\rm esc}(E_2)$, the escape rate converges asymptotically to the power-law curve~(\ref{eq:esc_rate_pl}) (black-dashed line), which scales as $\mathcal{R}_{\rm esc}\sim t^{-2/3}$. 
Comparison between Monte-Carlo models and analytical curves shows that diffusion equations systematically underestimate the fraction of particles crossing $E=0$. This is particularly visible at peak time, $t\sim t_{\rm esc}(E_2)$. The origin of this mismatch can be traced back to the accelerated unbinding of particles as they escape from the system (see left panels of Fig.~\ref{fig:ens_6}), which cannot be properly described by diffusion or Fokker-Planck equations, as discussed in \S\ref{sec:MC}. The agreement between Monte-Carlo and diffusion models improves as the fraction of particles in the fringe decreases and the escape rate approaches $\mathcal{R}_{\rm esc}\to 0$.

\subsection{Production of retrograde orbits}\label{sec:retro}
As demonstrated in Appendix~\ref{sec:diff}, stochastic tidal fluctuations tend to isotropize the initial distribution of angular momentum. This implies a non-zero probability that at any given time a tracer particle will scatter onto a retrograde motion. Here we use Equation~(\ref{eq:NLz}) to calculate the fraction of comets whose rotational direction is reversed as a function of time. Recall that by convection $L_{z,0}>0$ at $t=t_0$, hence the fraction of comets on retrograde orbits can be written as
\begin{align}\label{eq:fret}
  f_{\rm ret}(t)&= \int_{-L_c}^0\d L_z\, N(L_z,t) \\ \nonumber
    &= N_0 \int_{E_1}^{E_2} \d E_0 \int_{-L_c}^0 \d L_z \,p(L_z,t|E_0,L_c,t_0)\int_{-\infty}^0\d E \,p(E,t|E_0,L_c,t_0)\\ \nonumber
    &= N_0 \int_{E_1}^{E_2} \d E_0 \bigg\{\frac{1}{2}+\sum_{\ell=1}^\infty(-1)^\ell\frac{\sin(\ell\pi/2)}{\ell\pi/2} e^{-\lambda_\ell D_L \,t}\bigg\}\\ \nonumber
  &\qquad {}\times \frac{1}{2}\bigg\{ 1+\erf\bigg(\frac{\tilde C_E-E_0}{2\sqrt{\tilde D_E}}\bigg)\\ \nonumber
  &\qquad {}-\exp\bigg(E_0\frac{\tilde C_E}{\tilde D_E}\bigg)\bigg[1+\erf\bigg(\frac{\tilde C_E+E_0}{2\sqrt{\tilde D_E}}\bigg)\bigg] \bigg\} .
\end{align}

By definition, the rate at which retrograde orbits are produced corresponds to the time-derivative of~(\ref{eq:fret})
\begin{align}\label{eq:rate_ret}
\mathcal{R}_{\rm ret}(t)=  \frac{\d f_{\rm ret}}{\d t}&=
  N_0 \int_{E_1}^{E_2} \d E_0 \int_{-L_c}^0 \d L_z \,p(L_z,t|E_0,L_c,t_0)\\ \nonumber
 &\qquad {}\times \int_{-\infty}^0\d E \frac{\partial}{\partial t}p(E,t|E_0,L_c,t_0)\\ \nonumber 
  &+N_0 \int_{E_1}^{E_2} \d E_0 \int_{-\infty}^0\d E p(E,t|E_0,L_c,t_0) \\ \nonumber
 &\qquad {}\times \int_{-L_c}^0 \d L_z \frac{\partial}{\partial t}p(L_z,t|E_0,L_c,t_0)\\ \nonumber 
  &=   N_0 \int_{E_1}^{E_2} \d E_0 \bigg\{\frac{1}{2}+\sum_{\ell=1}^\infty(-1)^\ell\frac{\sin(\ell\pi/2)}{\ell\pi/2} e^{-\lambda_\ell \tilde D_L}\bigg\}  \\ \nonumber
  &\qquad {}\times j_E(0,t|E_0,L_c,t_0)\\ \nonumber
  &+  \frac{N_0}{2} \int_{E_1}^{E_2} \d E_0 \bigg\{ 1+\erf\bigg(\frac{\tilde C_E-E_0}{2\sqrt{\tilde D_E}}\bigg)\\ \nonumber
 &\qquad {}-\exp\bigg(E_0\frac{\tilde C_E}{\tilde D_E}\bigg)\bigg[1+\erf\bigg(\frac{\tilde C_E+E_0}{2\sqrt{\tilde D_E}}\bigg)\bigg] \bigg\}\\ \nonumber
 &\qquad {}\times j_L(0,t|E_0,L_c,t_0),
\end{align}
where $j_E(0,t|E_0,L_c,t_0)$ is the flux of particles crossing the energy boundary $E=0$, Equation~(\ref{eq:flux_E}), and
\begin{align}\label{eq:flux_J}
  j_L(0,t|E_0,L_c,t_0)&=\int_{-L_c}^0\d L_z \frac{\partial}{\partial t}p(L_z,t|E_0,L_c,t_0)\\ \nonumber
  &= D_L \frac{\partial }{\partial L_z}p(L_z,t|E_0,L_c,t_0)\bigg|_{L_z=0}\\ \nonumber
&=  \sum_{\ell=1}^\infty(-1)^{\ell}\frac{\sin(\ell\pi/2)}{\ell\pi/2} (-\lambda_\ell D_L)e^{-\lambda_\ell \tilde D_L},
\end{align}
is the flux of probability crossing $L_z=0$ at the time $t$ which originated from $\delta(E-E_0)\delta(L_R)\delta(L_z-L_c)$ at the time $t_0=0$. With these definitions in place, it is straightforward to derive a physical interpretation of the two terms appearing in Equation~(\ref{eq:rate_ret}). In particular, the left-hand term has a negative value and accounts for the loss of retrograde orbits as they cross the absorbing barrier at $E=0$, while the right-hand one is positive and describes a flux of particles that diffuse into the retrograde interval $L_z<0$ at a fixed bound fraction.

One can follow similar steps as in Section~\ref{sec:rate} to find analytical solutions for $\d f_{\rm ret}/\d t$ on time-scales $t_{\rm esc}(E_2)\ll t\ll t_{\rm esc}(E_1)$. 
Let us first re-write Equation~(\ref{eq:rate_ret}) as
\begin{align}\label{eq:rate_ret2}
  \mathcal{R}_{\rm ret}(t)=-\frac{1}{2}\mathcal{R}_{\rm esc}(t)+\sum_{\ell=1}^\infty(-1)^\ell\frac{\sin(\ell\pi/2)}{\ell\pi/2}I_\ell,
\end{align}
where $\mathcal{R}_{\rm esc}$ is the escape rate defined by Equation~(\ref{eq:esc_rate2}), and
\begin{align}\label{eq:Iell}
I_\ell&=N_0\int_{E_1}^{E_2}\d E_0\,e^{-\lambda_\ell\tilde D_L}\bigg[j_E(0,t)-\frac{\lambda_\ell D_L}{2}\bigg\{ 1+\erf\bigg(\frac{\tilde C_E-E_0}{2\sqrt{\tilde D_E}}\bigg)\\ \nonumber
 &\qquad {}-\exp\bigg(E_0\frac{\tilde C_E}{\tilde D_E}\bigg)\bigg[1+\erf\bigg(\frac{\tilde C_E+E_0}{2\sqrt{\tilde D_E}}\bigg)\bigg] \bigg\}\bigg].
\end{align}
After some algebra, one can show that the series in Equation~(\ref{eq:rate_ret2}) converges to a power-law function
\begin{align}\label{eq:series}
 \sum_{\ell=1}^\infty(-1)^\ell\frac{\sin(\ell\pi/2)}{\ell\pi/2} I_\ell\simeq  0.146 \frac{N_0}{t^{2/3}}\bigg(\frac{ G^2 M m}{c}\bigg)^{2/3}n^{1/3}\bigg(\frac{2 \pi}{3\langle v^2\rangle}\bigg)^{1/6}.
  \end{align}
Inserting~(\ref{eq:series}) into~(\ref{eq:rate_ret2}) leads to a scale-free rate
\begin{align}\label{eq:rate_ret_pl}
  \mathcal{R}_{\rm ret}(t)&=0.0086\frac{N_0}{t^{2/3}}\bigg(\frac{ G^2 M m}{c}\bigg)^{2/3}n^{1/3}\bigg(\frac{2 \pi}{3\langle v^2\rangle}\bigg)^{1/6} \\ \nonumber
  &\approx 0.031 \mathcal{R}_{\rm esc}(t),
\end{align}
where $\mathcal{R}_{\rm esc}(t)$ is the escape rate is given by~(\ref{eq:esc_rate_pl}).
Thus, we find that on time-scales $t_{\rm esc}(E_2)\ll t\ll t_{\rm esc}(E_1)$ the fraction of bound comets scattered onto retrograde orbits per unit of time is approximately $\sim 3\%$ of those escaping from the central potential.

Upper panel of Fig.~\ref{fig:rate_Lz} shows the rate of production of retrograde orbits ($\mathcal{R}_{\rm ret}$) as a function of time for the models plotted in Fig.~\ref{fig:rate_E}. 
There is an overall good agreement between Monte-Carlo simulations and the analytical expression~(\ref{eq:rate_ret}) (thin-colour lines), which improves as the time interval increases. As in Fig.~\ref{fig:rate_E}, we find that the maximum rate $\mathcal{R}_{\rm ret}(t)$ approximately occurs at the average escape time of the initial outer-most disc radius, $t_{\rm esc}(E_2)$ (marked with vertical dotted lines), which is similar to the isotropization time-scale (vertical dashed lines), as expected from~(\ref{eq:tisoesc}). On longer time intervals, $t \gg t_{\rm esc}(E_2)$, the scale rate approaches the power-law behaviour $\mathcal{R}_{\rm ret}(t)\sim t^{-2/3}$ given by Equation~(\ref{eq:rate_ret_pl}).
The ratio $\mathcal{R}_{\rm ret}/\mathcal{R}_{\rm esc}$ is plotted in the lower panel of Fig.~\ref{fig:rate_Lz}. We find that the rate of comets attaining a retrograde motion reaches a maximum $\mathcal{R}_{\rm ret}/\mathcal{R}_{\rm esc}\sim 0.1$ shortly before the escape rate peaks. Thereafter, the rate ratio slowly converges towards the value predicted by Equation~(\ref{eq:rate_ret_pl}), i.e. $\mathcal{R}_{\rm ret}/\mathcal{R}_{\rm esc}\sim 0.03$ (marked with a horizontal dashed line).

\begin{figure}
\begin{center}
\includegraphics[width=84mm]{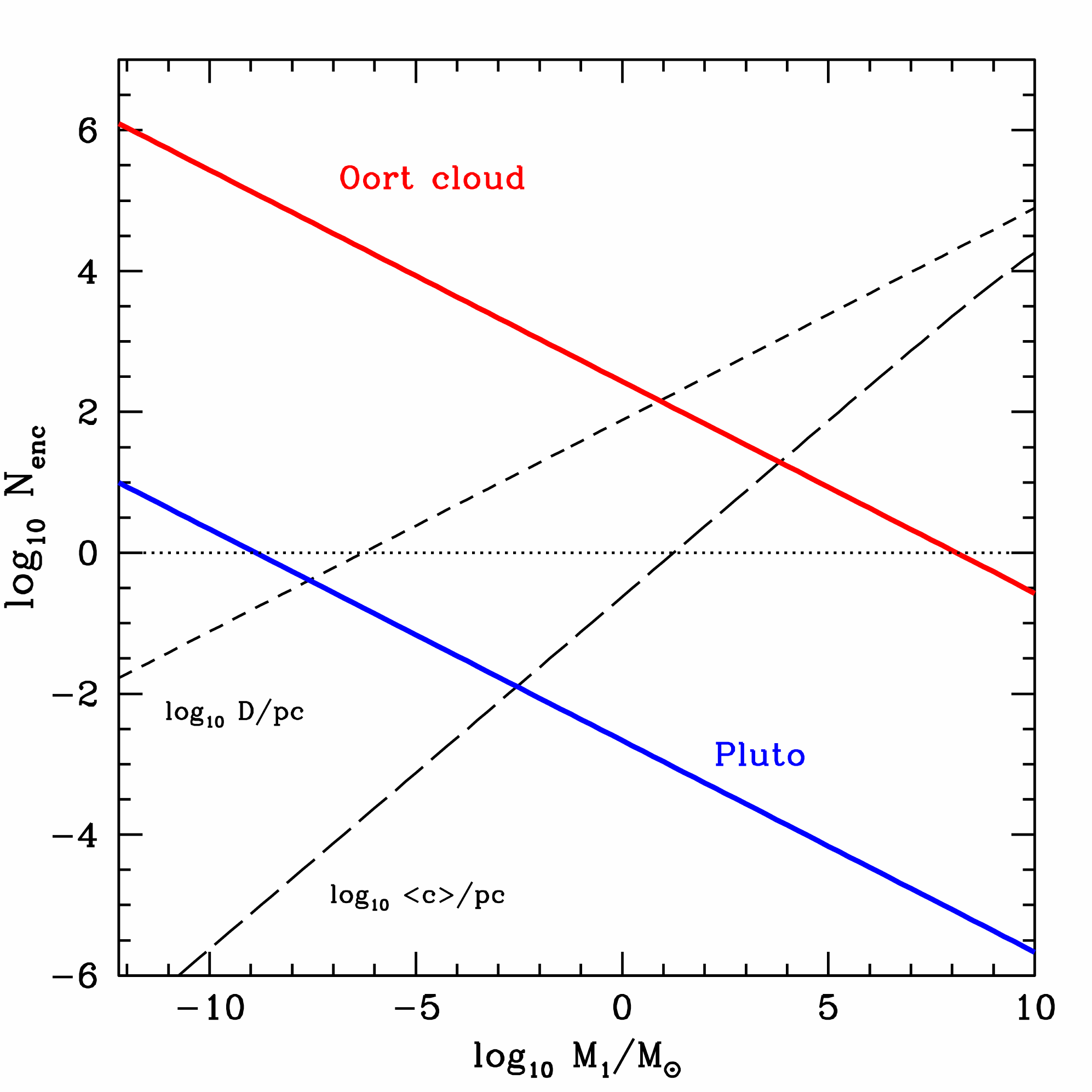}
\end{center}
\caption{Averaged number of tidal fluctuations experienced by comets in the Solar system during a single orbital revolution, $N_{\rm enc}=P/T_{\rm ch}(r_\odot)$, as a function of the minimum mass of dark matter substructures, $M_1$ (see text). Red and blue lines correspond to cometary orbits with semi-major $a=10^5\,{\rm AU}$ (Oort cloud), and $a=40\,{\rm AU}$ (Pluto), respectively. For reference, we plot the average separation between subhaloes $D(r_\odot)=[2\pi n(r_\odot)]^{-1/3}$ (black long-dashed lines) as well as their average scale radius $\langle c\rangle $ (black dashed lines). Notice that the outer-most regions of the Solar system may be perturbed by dark matter substructures with $M\ll 10^{6}M_\odot$.}
\label{fig:aq}
\end{figure}
\section{Discussion: Tidal heating by dark microhaloes}\label{sec:micro}
Previous Sections assume that substructures have a fixed mass $(M)$ and scale-radius $(c)$. Here, we extend our statistical formalism to ensembles of objects that cover a wide range of masses and sizes. For illustration, we discuss the effect of dark matter subhaloes on weakly-bound objects in the Solar System, as this provides an interesting, and rather special case of study.


Following Paper I (see \S4.1 for details), let us consider {\em statistical ensembles} of dark matter subhaloes whose mass function and number density profile match those found in the Aquarius simulations of Milky Way-sized haloes
\begin{align}
  \frac{\d n}{\d M}(r,M)=B_0\bigg(\frac{M}{M_0}\bigg)^\alpha g(r),
  \label{eq:nrm}
  \end{align}
with $B_0=2.02\times 10^{-13}M_\odot^{-1}\kpc^{-3}$ and $M_{0}=2.52\times 10^{7}M_\odot$. The function $g(r)=\exp\{-(2/\gamma)[ (r/r_{-2})^\gamma-1]\}$ is an Einasto profile with parameters $\gamma=0.678$, $r_{-2}=199\kpc$ and $\alpha=-1.9$ (Springel et al. 2008; see also Han et al. 2016, Erkal et al. 2016).
The subhalo mass function is believed to be truncated from below on scales comparable to the free-streaming length of the DM-particle candidates. For DM made of WIMPs with a mass $\sim 1-1000$ GeV the cut-off of the mass spectrum lies on sub-solar mass scales, $M_1/M_\odot\sim 10^{-12}$--$10^{-3}$ (e.g. Schmid et al. 1999; Hofmann et al. 2001; Green et al. 2005; Loeb \& Zaldarriaga 2005; Diemand et al. 2005). As a result, the total number of galactic substructures predicted by Equation~(\ref{eq:nrm}) is extremely large
\begin{align}  \label{eq:Ntot}
  N =\int \d^3r  \int_{M_1}^{M_2}\d M\,\frac{\d n}{\d M} ~\sim ~10^{15}\,\bigg(\frac{M_1}{10^{-6}M_\odot}\bigg)^{-0.9},
\end{align}
and {\it diverges} in the perfect-fluid limit $M_1\to 0$.

The velocity distribution of DM substructures in the Solar neighbourhood is assumed to be Maxwellian, Equation~(\ref{eq:vgauss}), with a dispersion $\sigma$ that can be computed from the isotropic Jeans equations as
\begin{align}
  \sigma^2(r)=\frac{1}{g(r)}\int_r^{\infty}\d r'\,g(r') \frac{\d \Phi_g(r')}{\d r},
  \label{eq:sig}
\end{align}
here $\Phi_g(r)$ is the potential of the Aquarius halo, which can be described by a Navarro, Frenk \& White (1997) model with a virial mass and radius $M_{200}=1.84\times 10^{12}M_\odot$, $r_{200}=246\kpc$, respectively, and a concentration $c_{200}=16.11$ (Springel et al. 2008). Inserting the number density profile~(\ref{eq:nrm}) into~(\ref{eq:sig}) yields $\sigma(r_\odot)=296\kms$. 
The mean relative velocity between the Sun and the dark matter microhaloes in Equation~(\ref{eq:vgauss}) corresponds to the circular velocity of the Milky Way at the solar radius $V_\odot=V(r_\odot)=230\kms$ at $r_\odot\approx 8\kpc$ (Eilers et al. 2018). Hence, we find that the averaged velocity~(\ref{eq:v2av}) is
\begin{align}
\langle v^2 \rangle^{1/2}_\odot=\sqrt{ V_\odot^2 +3\sigma^2(r_\odot)}\simeq 562\kms.
  \label{eq:v2a_sun}
\end{align}

As the cutoff of the mass spectrum is shifted to lower values, the average separation between subhaloes decreases. From Equation~(\ref{eq:nrm}) we find
\begin{align}
D(r_\odot)=\bigg[2\pi \int_{M_1}^{M_2}\d M\,\frac{\d n}{\d M}\bigg]^{-1/3}\approx 1.21\pc~\bigg(\frac{M_1}{10^{-6}M_\odot}\bigg)^{0.3},
  \label{eq:Dmr}
\end{align}
whereas the characteristic duration of tidal fluctuations shortens in proportion to $D(r_\odot)$. Combination of~(\ref{eq:Tch}),~(\ref{eq:v2a_sun}) and~(\ref{eq:Dmr}) yields
\begin{align}
 T_{\rm ch}(r_\odot)=0.88\frac{D(r_\odot)}{\langle v^2 \rangle^{1/2}_\odot}\approx 1840~{\rm yr}\bigg(\frac{M_1}{10^{-6}M_\odot}\bigg)^{0.3}.
  \label{eq:Tch_sun}
\end{align}
which is comparable to the orbital period of comets with a semi-major axis
\begin{align}
  a_{\rm ch}=\bigg[\frac{T_{\rm ch}(r_\odot)}{2\pi}\bigg]^{2/3}(GM_\odot)^{1/3}\approx 150\AU ~\bigg(\frac{M_1}{10^{-6}M_\odot}\bigg)^{0.2}.
 \label{eq:ach_sun}
\end{align}  
Given that the mean-life of a tidal fluctuation~(\ref{eq:Tch_sun}) is comparable to the time-span between subsequent fluctuations, comets with a semi-major axis $a\gg a_{\rm ch}$ will experience multiple `encounters' with dark microhaloes during a single orbital revolution. To illustrate this point, Fig.~\ref{fig:aq} plots the average number of fluctuations per orbital period, $N_{\rm enc}=P/T_{\rm ch}(r_\odot)$, as a function of the minimum subhalo mass, $M_1$. The upper-limit of the subhalo mass function is set at $M_2=10^{11}M_\odot$, although the results shown in Fig.~\ref{fig:aq} are largely independent of this choice insofar as $M_1\ll M_2$. As expected, orbits with a semi-major axis comparable or smaller to that of Pluto ($a \sim 40\AU$, blue solid line) must perform several orbital revolutions before experiencing a single fluctuation of the local tidal field, and therefore responds adiabatically to the subhalo background (see \S\ref{sec:adiab}). In contrast, comets in the Oort cloud ($a\sim 10^5\AU$, red solid line) feel a rapidly fluctuating tidal field, and will react impulsively to gravitational interactions with dark matter substructures.

The magnitude of tidal fluctuations depends very strongly on the relation between the mass and size of dark matter substructures. Following Paper I, it is useful to quantify this dependency by introducing a power-law size function
\begin{align}
   c(M)=c_0\bigg(\frac{M}{M_0}\bigg)^\beta,
  \label{eq:cm}
\end{align}
For the subhaloes found in the Via Lactea II simulation at redshift $z=0$ (Diemand et al. 2008) the best-fitting parameters are $c_0\simeq 0.53\kpc$ and $\beta\simeq 0.5$ (Erkal et al. 2016).
 The average size of subhaloes in a mass bin $M\in (M_1,M_2)$ derived from~(\ref{eq:cm}) is
\begin{align}
  \langle c \rangle=\frac{\int_{M_1}^{M_2}\d M\, c(M)(\d n/\d M)}{\int_{M_1}^{M_2}\d M\, (\d n/\d M)} \approx 2\times 10^{-4}\pc ~\bigg(\frac{M_1}{10^{-6}M_\odot}\bigg)^{0.5},
  \label{eq:cma}
\end{align}
 which is grossly consistent with the size of microhaloes expected from an inflation-produced primeval fluctuation spectrum (Berezinsky et al. 2003), as well as with simulations that resolve the size-mass relation of the first structures formed in the Universe (Ishiyama 2014; see also S{\'a}nchez-Conde \& Prada 2014).
Fig.~\ref{fig:aq} shows that the gap between the mean size of dark matter substructures (black long-dashed line) and their average separation (black short-dashed line) widens as the value of $M_1$ decreases. One can use~(\ref{eq:cma}) and~(\ref{eq:Dmr}) to compute the size-to-separation ratio, which goes as $\langle c \rangle/D(r_\odot)\sim 4\times 10^{-4}[M_1/(10^{-6} M_\odot)]^{0.2}$, thus approaching the point-mass behaviour $\langle c \rangle/D(r_\odot)\to 0$ in the perfect-fluid limit $M_1\to 0$.

For dark matter microhaloes with power-law mass and size functions Equations~(\ref{eq:nrm}) and~(\ref{eq:cm}) can be combined as (see Paper~I) 
\begin{align}\label{eq:nmc}
  \frac{\d^2 n}{\d M\d c}&=\frac{\d n}{\d M}\delta[c-c(M)]\\ \nonumber
 &= B_0\bigg(\frac{M}{M_0}\bigg)^\alpha g(r)\,\delta\bigg[c-c_0\bigg(\frac{M}{M_0}\bigg)^\beta\bigg].
\end{align}
Note that~(\ref{eq:nmc}) assumes an exact correspondence between size and mass (i.e. no scatter).

Over a time interval $t$, the average velocity increments experienced by comets located at a fixed distance from the Sun $R$ can be calculated as $\langle |\Delta{\mathbfit V}|^2\rangle=t R^2\langle \Lambda^2 T A\rangle \delta_s(t)$, where $\langle \Lambda^2 T A\rangle$ is given by Equation~(\ref{eq:L2TA}), and $\delta_s(t)=1-\exp(-t/\tau_s)$ is the sampling-delay function defined in \S\ref{sec:num_fluctu}. Since we are mainly interested in the response of comets in the outer-most regions of the Solar system ($a\gg a_{\rm ch}$), one can safely assume that tidal perturbations occur in an impulsive regime, wherein the adiabatic correction~(\ref{eq:A}) is $A\approx 1$. To average both quantities, $\langle \Lambda^2 T\rangle$ and $\tau_s$, over substructures with a known distribution of masses and sizes one can simply write $n\to \int\int \d M\d c \frac{\d^2 n}{\d M\d c}$. Equation~(\ref{eq:L2T}) then becomes
\begin{align}\label{eq:L2T_micro}
  \langle \Lambda^2 T\rangle &=\frac{4\pi}{5}{\sqrt\frac{2\pi}{3 \langle v^2\rangle}} M_0^{2\beta-\alpha}\frac{G^2 B_0}{c_0^2}\\ \nonumber
  &\quad{}\times
  \begin{cases}
    \frac{M_2^{3+\alpha-2\beta}-M_1^{3+\alpha-2\beta}}{3+\alpha-2\beta} &,~ 3+\alpha-2\beta \ne 0 \\
    {\ln(M_2/M_1)} & ,~3+\alpha-2\beta = 0.
    \end{cases}
\end{align}
This Equation exhibits two well-defined behaviours
\begin{itemize}
\item $3+\alpha-2\beta\le 0$. This regime arises when the size function is sufficiently steep, $\beta \ge (3+\alpha)/2$, leading to an average velocity increment that {\it diverges} in the limit $M_1\to 0$.
  \item $3+\alpha-2\beta> 0$. This case corresponds to relatively shallow size functions, $\beta <(3+\alpha)/2$, leading to an asymptotically convergent value of $\langle \Lambda^2 T\rangle$ as $M_1\to 0$.
 \end{itemize}

The second parameter of relevance is the sampling time-scale, $\tau_s$, which measures the average time interval required to sample the large-force tail of probability function $p(\Lambda)$. The ratio $\tau_s/T_{\rm ch}$ informs on the average number of fluctuations that must occur before the tidal force variance $\langle \Lambda^2\rangle$ converges to the analytical value~(\ref{eq:lamvar2}). As shown in Fig.~\ref{fig:lt_3}, the magnitude of tidal fluctuations is exponentially suppressed on time-scales $t\ll \tau_s$ (i.e. $\delta_s\ll 1$). Here, it is convenient to average the time reciprocal of the sampling time-scale, i.e. the sampling frequency $w_s=1/\tau_s$. From Equation~(\ref{eq:omega_s}) we find
\begin{align}\label{eq:ws_micro}
  \langle w_s \rangle &=1.49\langle v^2\rangle^{1/2}\frac{B_0}{M_0^{\alpha+2\beta}}c_0^2\times
  \begin{cases}
    \frac{M_2^{1+\alpha+2\beta}-M_1^{1+\alpha+2\beta}}{1+\alpha+2\beta} &,~ 1+\alpha+2\beta \ne 0 \\
    {\ln(M_2/M_1)} & ,~1+\alpha+2\beta = 0.
    \end{cases}
\end{align}
Again, this expression reveals two possible regimes
\begin{itemize}
\item $1+\alpha+2\beta\le 0$. In this case the sampling frequency {\it diverges} in the perfect fluid limit $M_1\to 0$, which implies a progressively faster convergence of $\langle\Lambda^2 T\rangle$ towards the analytical value~(\ref{eq:L2T_micro}) as the value of $M_1$ decreases. As expected from Figs.~\ref{fig:lt_3} and~\ref{fig:fit}, this regime arises in substructure populations with a relatively shallow mass function $\beta\le -(1+\alpha)/2$.
  \item $1+\alpha+2\beta> 0$. This regime is associated with a relatively steep size function, $\beta>-(1+\alpha)/2$, which leads to a convergent sampling frequency in the limit $M_1\to 0$. 
\end{itemize}

\begin{figure}
\begin{center}
\includegraphics[width=84mm]{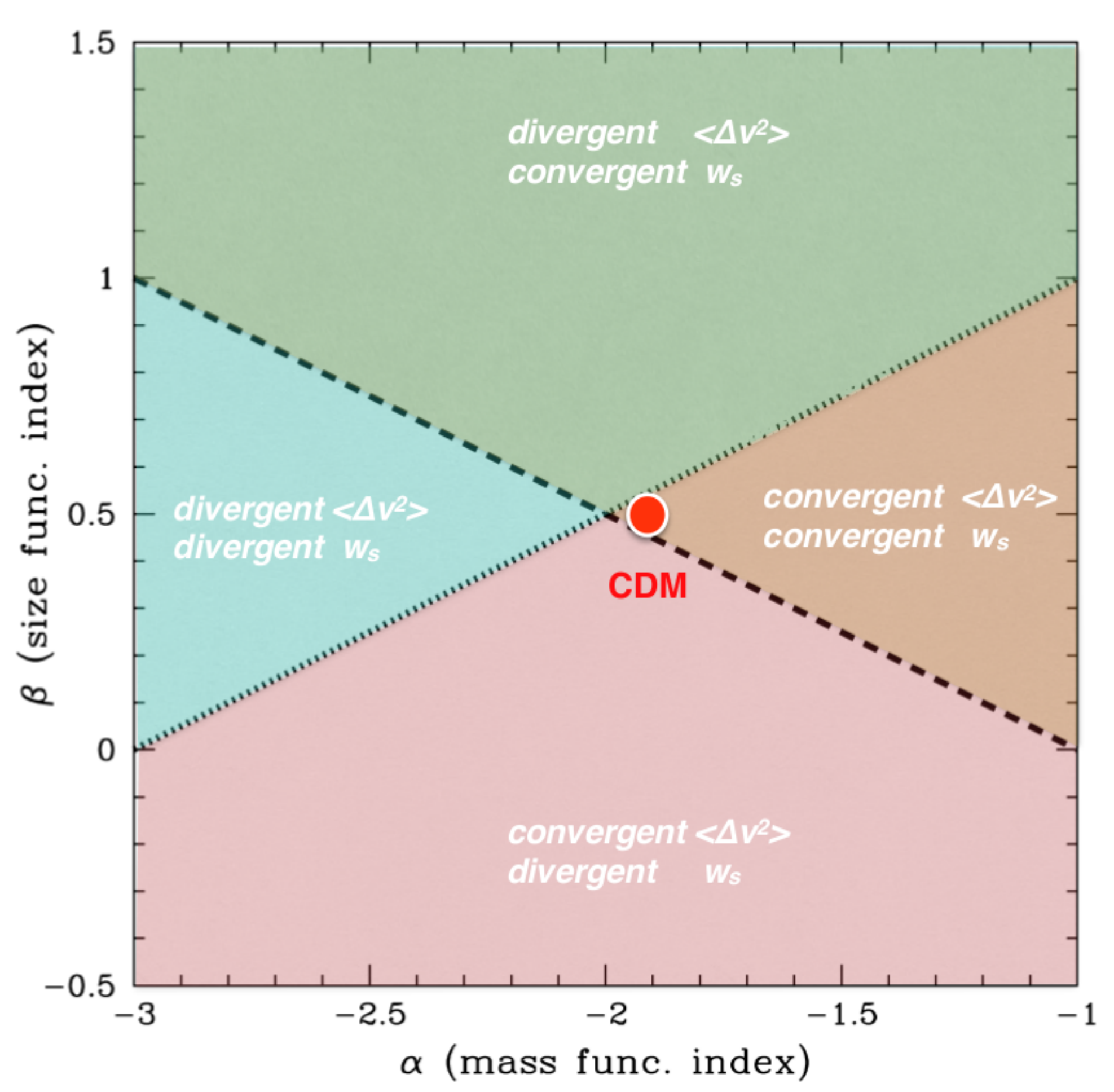}
\end{center}
\caption{Combination of power-law indices of the mass ($\alpha$) and size ($\beta$) functions that determine the amplitude of the tidal force fluctuations~(\ref{eq:L2T_micro}) and the sampling frequency~(\ref{eq:ws_micro}) in the perfect-fluid limit $M_1\to 0$. Dotted and dashed lines represent the transition between divergent/convergent velocity increments ($3+\alpha-2\beta=0$) and sampling frequencies ($1+\alpha+2\beta=0$), respectively.
  Best-fitting indices derived from the Via Lactea II models (Diemand et al. 2007) are marked with a red circle.}
\label{fig:ab}
\end{figure}
It is straightforward to show that the solution to the equations $3+\alpha-2\beta= 0$ and $1+\alpha+2\beta = 0$ is $(\alpha,\beta)=(-2,1/2)$, which corresponds to an ensemble of substructures with a constant mass fraction per logarithmic bin, i.e. $M(\Delta n/\Delta \ln M)\sim {\rm const.}$, and a squared mass-size relation $M\sim c^2$. For this particular combination, both quantities $\langle\Lambda^2 T\rangle$ and $\langle w_s\rangle$ show a mild logarithmic divergence $\sim \ln(M_2/M_1)$ as $M_1\to 0$. Strikingly, Fig.~\ref{fig:ab} shows that substructures in the Via Lactea~II simulation sit closely to this regime. By a narrow margin, however, the best-fitting indices, $(\alpha,\beta)\simeq (-1.9,0.5)$, meet the condition for slowly-converging values of velocity increments suggested by~(\ref{eq:L2T_micro}), i.e. $\beta < (3+\alpha)/2\simeq 0.55$, as well as the condition for a slowly-converging sampling frequency (and thus a finite time-scale $\tau_s$) given by~(\ref{eq:ws_micro}), $\beta >-(1+\alpha)/2\simeq 0.45$.

The above conclusion must be taken with caution, as several issues must be addressed before CDM can make a firm prediction on the convergence of $\langle |\Delta{\mathbfit V}|^2\rangle$ and $\langle w_s\rangle$ at low subhalo masses. 
For example, our statistical ensembles rely on extrapolations of cosmological $N$-body simulations that do not include baryons and have particle-mass resolutions  that lie many orders of magnitude above the mass-scale of dark matter microhaloes.
Indeed, the largest CDM simulations of Milky Way-like haloes to date, i.e. Aquarius (Springel et al. 2008), Via Lactea II (Diemand et al. 2007) and GHALO (Stadel et al. 2009), have particle-masses $\sim10^4$--$10^5M_\odot$, i.e. at least $\sim 10$ orders of magnitude above the WIMP mass-scale. On the other hand, simulations of structure formation that do resolve the free-streaming length of WIMPs (e.g. Ishiyama et al. 2014) typically stop at redshift $z\sim 30$ owing to their large computational cost. To the resolution limitations one must add the formidable numerical challenge of following the dynamical evolution of subhaloes that span $\sim 18$ orders of magnitude in  mass to redshift $z=0$ (van den Bosch 2017, van den Bosch \& Ogiya 2018).

Semi-analytical algorithms offer a flexible alternative approach to reach small low subhalo masses at a relatively low computational cost (e.g. Pe\~narrubia et al. 2010; Stref \& Lavalle 2017; Hiroshima et al. 2018). However, these methods are limited by our incomplete theoretical understanding of the processes of tidal stripping and mass loss (e.g. Daniel et al. 2017 and references therein).

Although often neglected, baryons may play an important role in shaping the mass \& size functions of low-mass subhaloes in the Solar neighbourhood. E.g., fly-by encounters with individual stars (Green \& Goodwin 2007) and the smooth tidal field of the Milky Way disc (D'Onghia et al. 2010; Pe\~narrubia et al. 2010; Errani et al. 2017) will enhance mass loss and modify the size of these objects.
Owing to their low-binding energies, micro-haloes experience a wide variety of mass loss histories depending on their individual orbits and accretion times. Combination of the above effects will necessarily magnify the scatter of the mass-size relation of low-mass subhaloes in the Solar neighbourhood. Note that it is not straightforward to predict the effect of scatter on the behaviour of $\langle \Lambda^2T \rangle$ and $\langle w_s \rangle$ because neither of these quantities scale linearly with subhalo size, $c$. Finally, the suppression of structure formation at low masses leads to a relatively shallow subhalo mass function (Schneider et al. 2012; Angulo et al. 2013; Lovell et al. 2014), which may also affect the magnitude of tidal heating. Addressing these issues goes beyond the main goals of this work and will be studied in a separate contribution.

\section{Summary}\label{sec:sum}
This paper uses stochastic calculus techniques to describe the dynamical response of tracer particles to stochastic variations of the combined tidal field generated by a large population of extended objects.
Over a sufficiently-long interval of time the cumulative effect of tidal fluctuations leads to a random walk of orbital velocities known as ``tidal heating''. Under the assumption that velocity impulses are small, $|\Delta{\mathbfit V}|<<|{\mathbfit V}|$, this paper uses diffusion equations to describe the response of tracer particles to random variations of a gravitational field (e.g. Chandrasekhar 1943; Kandrup 1980), which reduces the problem of tidal heating to the computation of diffusion coefficients, $\langle \Delta {\mathbfit V}\rangle$ and $\langle |\Delta {\mathbfit V}|^2\rangle$, where brackets denote averages over the spectrum of tidal fluctuations, $p({\bb \Lambda})$ (see \S3.2). This approach solves a number of shortcomings in Chandrasekhar's (1941a) classical theory, in which tidal heating is treated as the cumulative effect of isolated encounters between a test particle with an infinite background of individual point-mass particles:
\begin{itemize}
\item Our theory does not require arbitrary truncations of the force spectrum either at large or weak forces.
\item External forces are generated by extended objects (not particles), which allows us to combine tidal heating from baryonic and dark matter substructures covering a wide range of masses and sizes. On the other hand, the theory breaks down when applied to objects with divergent forces (e.g. black holes).
    \item Modelling the {\it combined} tidal force generated by background substructures avoids a detailed analysis of three-body encounters.
  \item Implementation of Weinberg's adiabatic corrections leads to diffusion coefficients that describe perturbations in impulsive as well as adiabatic regimes.
  \item Our analysis accounts for the period of time required to fully sample the spectrum of tidal fluctuations generated by a finite number of substructures.
  \end{itemize}

In Section~4, we treat stochastic tidal heating of self-gravitating systems as a diffusion process in a confined region of the 4-dimensional integral-of-motion space, where the energy is definite negative $E<0$, and the angular momentum components lie in the range $-L_c(E)\le L_i\le L_c(E)$, with $L_c(E)$ being the angular momentum of a circular orbit with fixed energy.
To this aim, Appendix B derives Green's functions with boundary conditions that place an absorbing barrier at $E=0$, and reflecting surfaces at $L_i=\pm L_c(E)$. The coefficients of the Green's functions, $\langle \Delta E\rangle$, $\langle \Delta E^2\rangle$, $\langle \Delta{\mathbfit L}\rangle$ and $\langle |\Delta {\mathbfit L}|^2\rangle$, are calculated from averages over ensembles of tracer particles {\it and} substructures. \S4.3 discusses the analytical case of a Keplerian potential in detail. Our results show that tidal heating leads to (i) a steady flow of probability drifting from bound energies towards $E\to 0$, which causes ``tidal evaporation'' as tracer particles gain sufficient energy as to escape from the system, and (ii) ``isotropization'', as the initial angular momentum distribution is randomized.

Diffusion equations are tested in Section~5 with the aid of $N$-body experiments in which the the local tidal field is computed as a direct summation of the forces generated by a large ($N\gg 1$) population of Hernquist (1990) spheres in dynamical equilibrium. We find that (i) sampling the large-force tail of the distribution $p(\bb \Lambda$) requires of the order of $N_{\rm enc} \approx 4.8 (D/c)^2$ fluctuations, where $D/c\gg 1$ is ratio between the mean separation and the size of individual substructures. This implies that our theory never attains statistical convergence when applied to a background of point-masses ($c\to 0$). (ii) Tracer particles in the ``fringe'' ($E\sim 0$) experience an rapid increase of energy and angular momentum that cannot be explained by our diffusion theory. In contrast, the process of tidal evaporation can be successfully described Monte-Carlo $N$-body simulations that sample random velocity kicks $\Delta {\mathbfit V}$ drawn from a probability function $\Psi({\mathbfit V},\Delta {\mathbfit V}, \Delta t)$, Equation~(\ref{eq:Psi}).

Section~6 follows the evolution of idealized planetary (mass-less) discs with an homogeneous energy distribution at $t_0=0$ subject to stochastic fluctuations of the local tidal field. The tidal evaporation rate ($\mathcal{R}_{\rm esc}$), and the rate of comets moving on retrograde orbits ($\mathcal{R}_{\rm ret}$) are computed analytically from the flux of comets going through the energy layer $E=0$ and the angular momentum boundary $L_z=0$, respectively. We find that on long time-scales both rates approach a scale-free relation $\mathcal{R}\sim t^{-2/3}$, and that in this regime the number of comets acquiring a retrograde motion corresponds to a small fraction of those escaping from the system, $\mathcal{R}_{\rm ret}/\mathcal{R}_{\rm esc}\simeq 0.03$.  
Overall, the energy and angular momentum distributions derived from Monte-Carlo simulations and Green's convolutions are in excellent agreement except for particles in the fringe ($E\sim 0$), where the condition of small increments demanded by the diffusion theory, $|\Delta E/E|\ll 1$, cannot be guaranteed. 

Section~7 inspects the effect of dark matter substructures on the dynamics of weakly-bound objects in the Solar system. To this end, we construct substructure ensembles that mimic the subhalo population found in cosmological $N$-body simulations of Milky Way-sized haloes. Extrapolation of the subhalo mass \& size functions found in the Aquarius (Springel et al. 2008) and Via Lactea II (Diemand et al. 2007) models suggests that comets in the Oort cloud may be sensitive to the presence of dark subhaloes with sub-solar masses. In particular, for a standard WIMP free-streaming mass Equation~(\ref{eq:ach_sun}) indicates that objects with a semi-major axis $a\gtrsim 150\AU$ may experience a large number of tidal interactions with planet-sized microhaloes on a dynamical time-scale. 
Interestingly, recent studies of the dynamics of Trans-Neptunian Objects with semi-major axes $a\gtrsim 250\AU$ find evidence for a non-equilibrium configuration, which may be caused by perturbations from a super-Earth ($M\gtrsim 10M_\oplus$) object known as ``Planet Nine'' (Batygin \& Brown 2016; Becker et al. 2018). The possibility to put constraints on the dark matter clumpiness using objects in the Solar system deserves detailed examination in a separate contribution.


The diffusion and Monte-Carlo techniques presented here provide statistical tools to study the combined effect of {\it baryonic} (e.g. free-floating planets, stars, giant molecular clouds, etc) and {\it dark matter} substructures on the dynamical evolution of a large number of weakly-bound objects (e.g. comets, protoplanetary discs, wide-binary stars, stellar clusters and gas clouds) at a minor computational cost. 
Combining observational constraints from a large variety of visible objects represents one of the best hopes to probe the unknown behaviour of dark matter on subsolar mass scales, a question with important implications for particle physics and cosmology.

{}

\appendix

\section{The autocorrelation function}
Here we calculate the autocorrelation function $W({\bb \Lambda}_0,{\bb \Lambda}_t)$. Our main assumptions are that (i) substructures are randomly distributed over the volume $V'$, and (ii) they move on straight lines.

By definition, $W$ is the convolution of two probability densities
\begin{align}\label{eq:Wap}
W({\bb \Lambda}_0,{\bb \Lambda}_t)=p({\bb \Lambda}_0)\conv p({\bb \Lambda}_t),
\end{align} 
where $p(\bb \Lambda)$ is the probability of experiencing a tidal vector in the interval ${\bb \Lambda}, {\bb \Lambda}+\d{\bb \Lambda}$.
 The computation of $W$ is more straighforward in Fourier space. Application of the convolution theorem to Equation~(\ref{eq:pL}) yields
\begin{align}\label{eq:Wapf}
  \tilde W({\mathbfit k}_0,{\mathbfit k}_t)=\tilde p({\mathbfit k}_0) \tilde p({\mathbfit k}_t),
\end{align} 
where the Fourier transform $\tilde p({\mathbfit k}_0)$ can be written as (Holtsmark 1919; see Paper I)
\begin{align}\label{eq:pf0}
  \tilde p({\mathbfit k}_0)&=\int \d^3\Lambda e^{i {\mathbfit k}_0\cdot \bb\Lambda}p({\bb \Lambda})\\ \nonumber
  &=\frac{1}{V'}\int \d^3 r_1 \times ...\times \frac{1}{V'}\int \d^3 r_N\int \d^3\Lambda e^{i {\mathbfit k}\cdot \bb\Lambda}\delta\big({\bb \Lambda}-\sum_i{\bb \lambda}_{i}\big)\\ \nonumber
  &=\bigg[\frac{1}{V'}\int \d^3 r e^{i {\mathbfit k}_0\cdot {\bb \lambda}({\mathbfit r})}\bigg]^N.
\end{align}
The derivation of $\tilde p({\mathbfit k}_t)$ can be done in a similar way under the assumption that substructures move on straight lines with a relative velocity ${\mathbfit v}$. Hence, if ${\mathbfit r}_i$ is the initial position vector, the probability to find the particle at a later time $t$ at the location ${\mathbfit r}'_i$ is $\delta({\mathbfit r}'_i-{\mathbfit r}_i-{\mathbfit v}t)$. Thus,
\begin{align}\label{eq:pft}
  \tilde p({\mathbfit k}_t)&=\int \d^3\Lambda e^{i {\mathbfit k}_t\cdot \bb\Lambda}p({\bb \Lambda})\\ \nonumber
 &= \int \d^3 r'_1 \delta({\mathbfit r}'_1-{\mathbfit r}_1-{\mathbfit v}t)\times ...\times \int \d^3 r'_N\delta({\mathbfit r}'_N-{\mathbfit r}_N-{\mathbfit v}t)\\ \nonumber
 &\qquad{} \times\int \d^3\Lambda e^{i {\mathbfit k}_t\cdot \bb\Lambda}\delta\big({\bb \Lambda}-\sum_i{\bb \lambda}_{i}\big)\\ \nonumber
  &=\bigg[\int \d^3 r' \delta({\mathbfit r}'-{\mathbfit r}-{\mathbfit v}t) e^{i {\mathbfit k}_t\cdot {\bb \lambda}({\mathbfit r'})}\bigg]^N.
\end{align}
Note that the last equalities in~(\ref{eq:pf0}) and~(\ref{eq:pft}) implicitly assume that the vectors ${\bb \lambda}_i$ are spatially uncorrelated at all times.

The Fourier transform of the autocorrelation function~(\ref{eq:Wap}) becomes
\begin{align}\label{eq:Wapf2}
  \tilde W({\mathbfit k}_0,{\mathbfit k}_t)&=\bigg[\frac{1}{V'}\int \d^3r\, e^{i {\mathbfit k}_0\cdot {\bb \lambda}({\mathbfit r})}\bigg]^N\bigg[\int \d^3 r'\delta({\mathbfit r}'-{\mathbfit r}-{\mathbfit v}t) \exp[i {\mathbfit k}_t\cdot {\bb \lambda}({\mathbfit r}')]\bigg]^N\\ \nonumber
   &= \bigg\{\frac{1}{V'}\int \d^3 r\,\exp\big[i {\mathbfit k}_0\cdot {\bb \lambda}({\mathbfit r}) +  i{\mathbfit k}_t\cdot {\bb \lambda}({\mathbfit r}+{\mathbfit v}t)\big]\bigg\}^N.
\end{align}
The last integral can be re-written as
$$\frac{1}{V'}\int \d^3r e^{i x}=\frac{1}{V'}\int \d^3r\bigg[ 1-\big(1-e^{i x}\big)\bigg]=1-\frac{1}{V'}\int \d^3r\big(1-e^{i x}\big),$$
which elevated to the $N$-th power becomes
$$\bigg[1-\frac{1}{V'}\int \d^3r\big(1-e^{i x}\big)\bigg]^N\approx \exp\bigg[-n\int_{V'} \d^3r\big(1-e^{i x}\big)\bigg]  ~~~~{\rm for}~~N\gg 1,$$
where $n\equiv N/V'$ is the number density of substructures.
Equation~(\ref{eq:Wapf2}) thus becomes
\begin{align}\label{eq:Wapf3}
  \tilde W({\mathbfit k}_0,{\mathbfit k}_t)= \exp[-\phi({\mathbfit k}_0,{\mathbfit k}_t)],
\end{align}
with
\begin{align}\label{eq:phi2}
  \phi({\mathbfit k}_0,{\mathbfit k}_t)\equiv n\int_{V'}\d^3 r \bigg[1-\exp\bigg(i {\mathbfit k}_0\cdot {\bb \lambda}({\mathbfit r}) +  i{\mathbfit k}_t\cdot {\bb \lambda}({\mathbfit r}+{\mathbfit v}t)\bigg) \bigg]. 
\end{align}
For substructures with a velocity distribution $f({\mathbfit v})$, where $\int \d v^3 \,f({\mathbfit v})=1$, Equation~(\ref{eq:phi2}) can be generalized as (Chandrasekhar 1943)
\begin{align}\label{eq:phi3}
  \phi({\mathbfit k}_0,{\mathbfit k}_t)\equiv n\int \d^3 v f({\mathbfit v})\int_{V'}\d^3 r \bigg[1-\exp\bigg(i {\mathbfit k}_0\cdot {\bb \lambda}({\mathbfit r}) +  i{\mathbfit k}_t\cdot {\bb \lambda}({\mathbfit r}+{\mathbfit v}t)\bigg) \bigg]. 
\end{align}
Finally, taking the inverse Fourier transform of Equation~(\ref{eq:Wapf2}) we obtain
\begin{align}\label{eq:Wa}
  W({\bb \Lambda}_0,{\bb \Lambda}_t)=\int\frac{\d^3 k_0}{(2\pi)^3}\int\frac{\d^3 k_t}{(2\pi)^3} \exp\bigg[-i({\mathbfit k}_0\cdot\bb\Lambda_0+{\mathbfit k}_t\cdot\bb\Lambda_t) - \phi({\mathbfit k}_0,{\mathbfit k}_t)\bigg],
\end{align}
thus recovering autocorrelation function derived in Chandrasekhar (1943) under the assumption that the effects of individual interactions can be described as a Markov chain process (see his \S3).

\section{Diffusion in confined regions}\label{sec:diff}
Here we derive Green's propagators in a confined region of the integral-of-motion space. As in previous Sections, we shall assume that diffusion in energy and angular momentum are statistically independent processes, such that solutions to the diffusion equation~(\ref{eq:pEL}) can be assumed to have a separable form~(\ref{eq:p}).

\subsection{Energy space}
Let us assume that particles that attain a positive gravitational energy will leave the system to never come back. This calls for the diffusion equation in energy space
\begin{align}\label{eq:pEb}
  \frac{\partial p}{\partial t}= C_E\nabla_E p\big |_{(E_0,{\mathbfit L}_0)} +  D_E\nabla_E^2p\big |_{(E_0,{\mathbfit L}_0)},
  \end{align}
with an absorbing boundary at $E=0$, such that $p(E=0,t)=0$. To solve the differential equation we use the {\it method of images} originally introduced by Lord Kelvin (see Feller, 1971), where one places an image source (or sink) at $-E_0$ as mirror image of the original source at $+E_0$ with a strength or intensity selected to match the boundary condition at $E=0$. The initial conditions in this case can be written as
\begin{align}\label{eq:im_E}
p=\delta(E-E_0) - \exp(-\eta)\delta(E+E_0),
 \end{align}
where the dimensionless parameter $\eta$ determines the strength of the mirror source. The solution of~(\ref{eq:pEb}) with boundaries~(\ref{eq:im_E}) is a linear combination of two mirrored Gaussians
\begin{align}\label{eq:im_p}
  p(E,t|E_0,{\mathbfit L}_0,t_0)=\frac{1}{\sqrt{4\pi \tilde D}}\exp\bigg[-\frac{(E-E_0+\tilde C)^2}{4 \tilde D}\bigg] \\ \nonumber
-\frac{1}{\sqrt{4\pi \tilde D}}\exp\bigg[-\eta-\frac{(E+E_0+\tilde C)^2}{4 \tilde D}\bigg],
\end{align}
with coefficients evaluated at $\tilde C=\tilde C(E_0,L_0,t-t_0)$ and $\tilde D=\tilde D(E_0,L_0,t-t_0)$. It straightfoward to show that the condition $p(0,t)=0$ implies
\begin{align}\label{eq:nup}
  \eta=-E_0\frac{\tilde C}{\tilde D}.
\end{align}  

\begin{figure}
\begin{center}
\includegraphics[width=86mm]{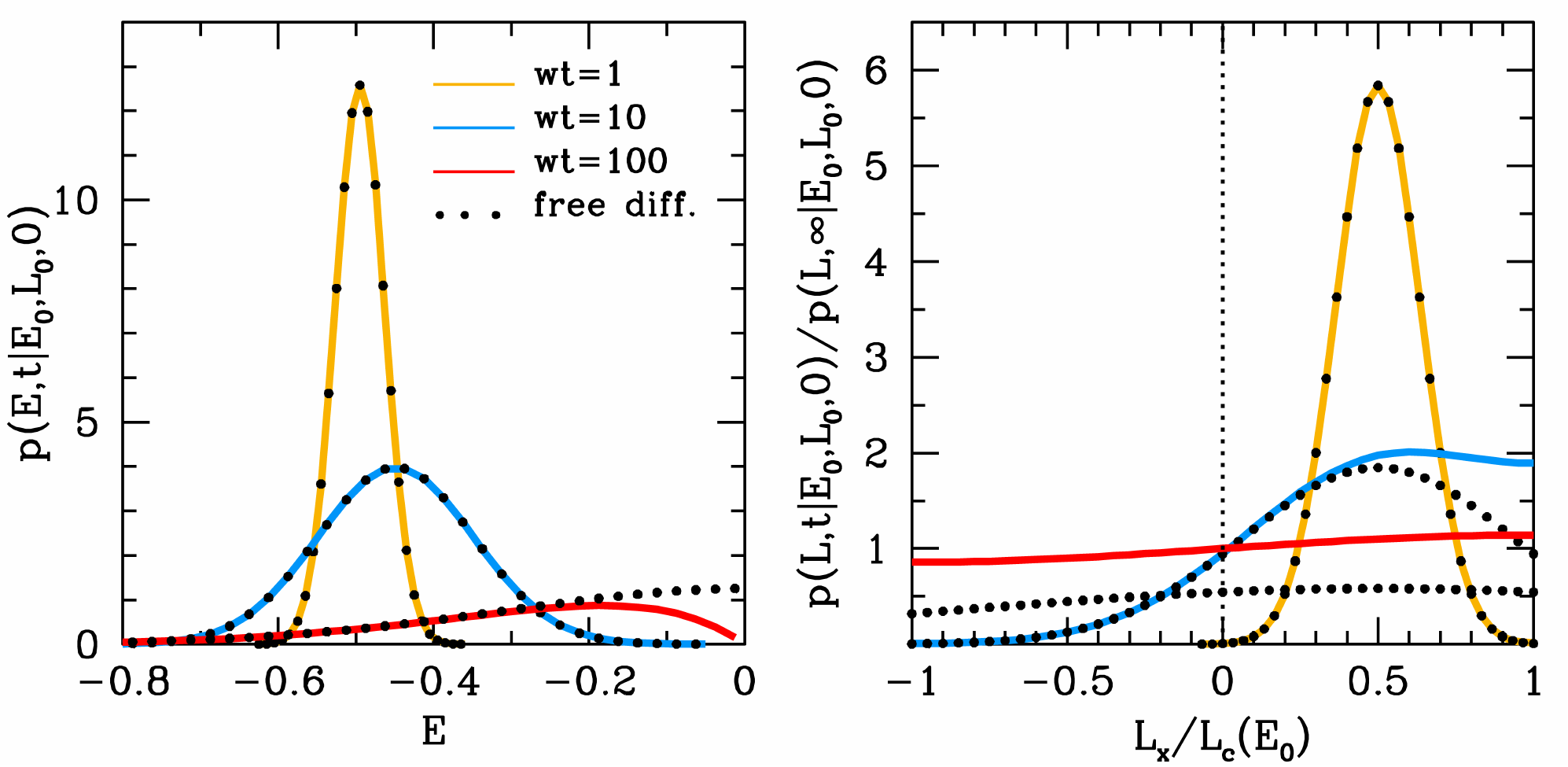}
\end{center}
\caption{Green's propagators in energy $E$ (left panel) and 1D angular momentum $Lx$ (right panel) at three different snapshots. For simplicity, we choose units where $G=m=D=\langle v^2\rangle=1$. In these units substructures have a mass $M=0.01$ and a size $c=0.1$. 
  The initial energy and angular momentum at $t=0$ are $E_0=-0.5$ and $L_x/L_c(E_0)=+0.5$, respectively. Propagators confined in a region of the integral-of-motion volume (solid lines) deviate from a Gaussian function (black dots) in the vicinity of the barriers at $E=0$ and $L_x/L_c=\pm 1$. Note that as time progresses the probability to find particles with counter rotation ($L_x<0$) increases. In the limit $t\to \infty$ the angular momentum progator approaches asymptotically an isotropic distribution $p(L,\infty|E_0,L_0,0)=[2L_c]^{-1}$.}
\label{fig:pr}
\end{figure}

Left panel of Fig.~\ref{fig:pr} plots the Green's propagator~(\ref{eq:im_p}) at three different snaphosts. Far from the boundary $E=0$, the right-hand side of Equation~(\ref{eq:im_p}) has a negligible contribution, and the energy propagator can be approximated by a Gaussian distribution (black dots). However, as the time progresses, the Green's function broadens and shifts systematically towards lower energies. By construction, $p(E,t|E_0,{\mathbfit L}_0,t)\to 0$ as $E\to 0$, which strongly deviates from the behaviour of the free-diffusing Gaussian solution in an infinite volume.
  
\subsection{Angular momentum space}
The angular momentum of gravitationally-bound particles cannot exceed the maximum set by circular orbits. Hence, $L\le L_c(E)$, where $L_c(E)$ is the angular moment of a circular orbit with energy $E$. E.g., for a Keplerian potential~(\ref{eq:phis_kep}) it follows from~(\ref{eq:ae}) that the maximum angular momentum is $L_c(E)=\sqrt{G m a}$, where $a=G m/(-2E)$ is the semi-major axis of orbits with energy $E$. 

Our goal is to solve the diffusion equation
\begin{align}\label{eq:pLb}
  \frac{\partial p}{\partial t}=  D_L\nabla_L^2 p\big|_{(E_0,{\mathbfit L}_0)},
\end{align}
with reflecting boundaries $\nabla_L p=0$ at $L=0, L_c$. Given that diffusion in angular momentum space is assumed to be isotropic, one may attempt to solve~(\ref{eq:pLb}) in spherical coordinates. However, the probability function corresponds to an infinite sum of Bessel functions and spherical harmonics (Carslaw \& Jaeger 1986), which complicates an intuitive interpretation of the result. Instead, here we will solve~(\ref{eq:pLb}) in Cartesian coordinates, where the differential operator $\nabla_L^2=\partial^2/\partial L_x^2 +\partial^2/\partial L_y^2+\partial^2/\partial L_z^2$. The problem at hand, therefore, is that of isotropic diffusion in a cubic box with reflecting boundaries.
To gain insight, first we solve the one-dimensional case, and subsequently provide the general 3D solution.

\subsubsection{One dimension}
Let us re-write the one-dimensional diffusion equation in angular momentum~(\ref{eq:pLb}) using a simplified notation 
\begin{align}\label{eq:pL1D}
 \frac{\partial p}{\partial t}= D\frac{\partial^2 p}{\partial x^2},
\end{align}
where $x=L_x$ is defined within the interval $x\in[-L_c,L_c]$, and $D=D_L(E_0,L_0,t_0)$. To limit the region in which particles can diffuse, we set reflecting boundaries at $x=\pm L_c$, such that $\partial p/\partial x|_{\pm L_c}=0$. Equation~(\ref{eq:pL1D}) admits separable solutions 
\begin{align}\label{eq:pL1D_sep}
 p(x,t)=X(x)T(t).
\end{align}
Inserting~(\ref{eq:pL1D_sep}) into~(\ref{eq:pL1D}) and re-arranging yields
\begin{align}\label{eq:pL1D_sep_eq}
  \frac{1}{X} \frac{\d^2 X}{\d x^2}=\frac{1}{D T} \frac{\d T}{\d t}  =-\lambda,
\end{align}
where $\lambda$ is a separation constant. Since the coefficient $D$ does not contain an explicit dependence on time (see \S\ref{sec:flux}), the function $T(t)$ can be directly integrated from~(\ref{eq:pL1D_sep_eq}), which yields
\begin{align}\label{eq:T}
  T(t)\propto \exp(-\lambda D t).
\end{align}

Let us now define the variable $x'=x+L_c$ and expand $X(x')$ as a Fourier series with a fundamental period $4 L_c$
\begin{align}\label{eq:X}
 X(x')=\frac{a_0}{2}+\sum_{n=1}^\infty a_n\cos(k_n x')+\sum_{n=1}^\infty b_n\sin(k_n x'),
\end{align}
with a wave number $k_n=n\pi /(2L_c)$. 
It is clear from~(\ref{eq:X}) that reflecting boundaries $\d X/\d x'|_{0,2L_c}=0$ require $b_n=0$ for integers $n\ge 1$. Note also that although the Fourier series~(\ref{eq:X}) extends to $[-2L_c,0]$, here we are only interested in the solution inside the interval $[0,2L_c]$.
Inserting~(\ref{eq:X}) into~(\ref{eq:pL1D_sep_eq}) yields a separation constant
\begin{align}\label{eq:lam}
  \lambda_n=k_n^2=\frac{\pi^2 n^2}{4 L_c^2}.
\end{align}
To find the coefficients of the Fourier series we must use the initial condition $p=\delta(x-x_0)=\delta(x'-x_0-L_c)$ at $t=0$. To this aim, we multiply both sides of~(\ref{eq:X}) by $\cos(k_{m}x')$ and integrate from 0 to $2L_c$
\begin{align}\label{eq:X2}
  \int_0^{2L_c}\d x' \,\cos(k_m x')\delta (x'-x_0-L_c)\\ \nonumber
   =\int_0^{2L_c}\d x'\,\cos(k_m x') \big[\frac{a_0}{2}  +\sum_{n=1}^\infty a_n\cos(k_n x')\big].
\end{align}
Using the orthogonality relations
\begin{align}\label{eq:ortho}
  \int_{0}^{2L_c} \d x \; \cos\left(\frac{n\pi x}{2 L_c}\right) & = \left\{\begin{matrix} 0& n\ne 0\\ 2L_c & n=0 \end{matrix} \right.\\ \nonumber
\int_{0}^{2L_c} \d x \; \cos\left(\frac{m\pi x}{2 L_c}\right) \cos\left(\frac{n\pi x}{2L_c}\right) & = \left\{\begin{matrix} 0& m\ne n\\ L_c & m=n \end{matrix} \right.
\end{align}
we find
\begin{align}\label{eq:am}
a_n=\frac{1}{L_c}\cos\bigg[\frac{n\pi (x_0+L_c)}{2 L_c}\bigg]~~~~ {\rm for} ~~~n\ge 0.
\end{align}
Combination of~(\ref{eq:T}) and~(\ref{eq:X}) then yields
\begin{align}\label{eq:pL1D_sol}
  p(x,t)=\frac{1}{L_c}\bigg\{\frac{1}{2}+\sum_{n=1}^\infty\cos\big[\frac{n\pi (x_0+L_c)}{2 L_c}\big]\cos\big[\frac{n\pi (x+L_c)}{2 L_c}\big] \\ \nonumber
  \times \exp[-\lambda_n D t]\bigg\}.
\end{align}

Compare Equation~(\ref{eq:pL1D_sol}) with the Fourier expansion of a Gaussian function
$$p_G(x,t)=\frac{1}{\sqrt{2\pi \sigma^2}}\exp\big[-\frac{(x-x_0)^2}{2\sigma^2}\big],$$
within the interval $x\in[-L_c,L_c]$. Following the same steps as above, we introduce a variable $x'=x+L_c$, and choose a fundamental period $4L_c$, neglecting the even terms of the series $x'\in [-2L_c,0]$

\begin{align}\label{eq:amG}
  a'_n&=\frac{2}{2L_c}\int_0^{2L_c}\d x\,\cos(k_n x')\frac{1}{\sqrt{2\pi \sigma^2}}\exp\big[-\frac{(x'-x_0-L_c)^2}{2\sigma^2}\big]\\ \nonumber
  &=\frac{1}{L_c}\frac{1}{\sqrt{2\pi \sigma^2}}\int_{-x_0-L_c}^{-x_0+L_c}\d y\,\cos\big[\frac{n\pi (y+x_0+L_c)}{2 L_c}\big]\exp\big[-\frac{y^2}{2\sigma^2}\big],
\end{align}
where $y=x'-x_0-L_c$. If the mean of the Gaussian function is located far from the boundaries, $|L_c-x_0|/\sigma \gg 1$, one can safely shift the limits of the integral~(\ref{eq:amG}) to infinity. Hence, Equation~(\ref{eq:amG}) becomes
\begin{align}\label{eq:amG2}
  a'_n &\approx\frac{1}{L_c}\frac{1}{\sqrt{2\pi \sigma^2}}\int_{-\infty}^{+\infty}\d y\,\cos\big[\frac{n\pi (y+x_0+L_c)}{2 L_c}\big]\exp\big[-\frac{y^2}{2\sigma^2}\big]\\ \nonumber
  &=\frac{1}{L_c}\cos\big[\frac{n\pi (x_0+L_c)}{2 L_c}\big]\exp\big[-\frac{\pi^2n^2\sigma^2}{8L_c^2}\big],
\end{align}
which leads to a Fourier series
\begin{align}\label{eq:pG}
  p_G(x,t)=\frac{1}{L_c}\bigg\{\frac{1}{2}+\sum_{n=1}^\infty\cos\big[\frac{n\pi (x_0+L_c)}{2 L_c}\big]\cos\big[\frac{n\pi (x+L_c)}{2 L_c}\big] \\ \nonumber
  \times\exp\big[-\frac{\pi^2n^2\sigma^2}{8L_c^2}\big]\bigg\}.
\end{align}
 It is clear that Equation~(\ref{eq:pG}) recovers~(\ref{eq:pL1D_sol}) if
$$\frac{\pi^2n^2\sigma^2}{8L_c^2}=\lambda_n D t.$$
Applying the separation constant~(\ref{eq:lam}) yields the celebrated behaviour of the free-diffusing Gaussian propagators~(\ref{eq:pL})
$$\sigma^2 = 2 D t,$$
thus demonstrating that far from the boundaries of the confined region particles diffuse as if they were moving in an infinite medium, $p(x,t)\approx p_G(x,t)$.

Right panel of Fig.~\ref{fig:pr} shows the solution~(\ref{eq:pL1D_sol}) at three different snapshots. As expected from the above results, the confined Green's function~(\ref{eq:pL1D_sol}) has a Gaussian form (black dots) on short time scales $\pi^2/(2 L_c)^2 D t\ll 1$. However, as the time progresses all modes with $n>0$ decay exponentially, and the propagator converges asymptotically towards a constant value $p(x,\infty)=1/(2L_c)$. Hence, on long time-scales $\pi^2/(2 L_c)^2D t\gg 1$ the angular momentum distribution progressively becomes isotropic. This behaviour is 
in stark contrast with the free-diffusing solution in an infinite domain, which vanishes $p_G\to 0$ in the limit $t\to \infty$.

\subsubsection{Three dimensions}
The above derivation can be generalized to a three-dimensional space by replacing the diffusion equation~(\ref{eq:pL1D}) with
\begin{align}\label{eq:pL3D}
 \frac{\partial p}{\partial t}= D\bigg(\frac{\partial^2 p}{\partial x^2}+\frac{\partial^2 p}{\partial y^2}+\frac{\partial^2 p}{\partial z^2}\bigg),
\end{align}
and~(\ref{eq:pL1D_sep}) with
\begin{align}\label{eq:pL3D_sep}
 p(x,y,z,t)=X(x)Y(y)Z(z)T(t).
\end{align}
Inserting~(\ref{eq:pL3D_sep}) into~(\ref{eq:pL3D}) and re-arranging yields
\begin{align}\label{eq:pL3D_sep}
  \frac{1}{X} \frac{\d^2 X}{\d x^2}=\frac{1}{Y} \frac{\d^2 Y}{\d y^2}=\frac{1}{Z} \frac{\d^2 Z}{\d z^2}=\frac{1}{D T} \frac{\d T}{\d t}=-\lambda,
\end{align}
with a separation constant 
\begin{align}\label{eq:lam3D}
  \lambda_{nml}=k_n^2+k_m^2+k_l^2=\frac{\pi^2}{4 L_c^2}(n^2+m^2+l^2).
\end{align}
Following similar steps as in the 1D case, we find that the separable solution~(\ref{eq:pL3D_sep}) with reflecting boundaries at $x=y=z=\pm L_c$ can be expressed as an infinite cosine series
\begin{align}\label{eq:pL3D_sol}
  p(x',y',z',t)=\sum_{n,m,l=0}^\infty C_{nml}\cos(k_nx')\cos(k_my')\cos(k_lz') e^{-\lambda_{nml} D t},
\end{align}
with $x'=x+L_c$, $y'=y+L_c$ and $z'=z+L_c$.
Application of the initial conditions $p(x,y,z,t=0)=\delta (x-x_0)\delta (y-y_0)\delta (z-z_0)$ and the orthogonality relations~(\ref{eq:ortho}) yields
\begin{align}\label{eq:pL3D_sol2}
  p(x,y,z,t)&=\frac{1}{L_c^3}\sum_{n,m,l=0}^\infty \alpha_{nml}\exp[-\lambda_{nml} D t] \\ \nonumber
 &\qquad{}\times \cos\big[\frac{n\pi (x_0+L_c)}{2 L_c}\big]\cos\big[\frac{n\pi (x+L_c)}{2 L_c}\big] \\ \nonumber
  &\qquad{}\times  \cos\big[\frac{m\pi (y_0+L_c)}{2 L_c}\big]\cos\big[\frac{m\pi (y+L_c)}{2 L_c} \big]\\ \nonumber
   &\qquad{}\times \cos\big[\frac{l\pi (z_0+L_c)}{2 L_c}\big]\cos\big[\frac{l\pi (z+L_c)}{2 L_c}\big],
\end{align}
with $\alpha_{000}=1/8$, $\alpha_{n00}=\alpha_{0m0}=\alpha_{00l}=1/4$, $\alpha_{nm0}=\alpha_{n0l}=\alpha_{0ml}=1/2$; and $\alpha_{nml}=1$~ for $n,m,l\ge 1$.

\section{Phase-space averages}\label{sec:aver}
Here we briefly discuss how to compute phase-space averages of particle ensembles with a fixed energy and angular momentum (for further details see Appendix A of P15). As a first step, let us define the tangential and radial velocity components as $v_t$ and $v_r$, respectively, such that the angular momentum becomes $L=r v_t$, the velocity volume $\d^3 v=2\pi v_t\d v_t\d v_r =2\pi \d v_r L\d L/r^2$, and the energy $E=v^2/2 +\Phi(r)=v_r^2/2+\Phi(r)+L^2/(2 r^2)$, hence at fixed radius $\d E=v\d v$. The volume accessible to tracer particles with a particular combination of $E$ and $L$ is a spherical shell with inner (peri-centre) and outer (apo-centre) radii, $R_p$ and $R_a$, respectively. At these two radii the radial velocity component vanishes, $v_r=0$, and the kinetic energy is entirely in the tangential direction $v=v_t$. Note that at each radius $v_r$ can be both positive and negative. In what follows we take only positive values of $v_r$ and double the number of particles with this velocity.

In order to express the probability function $N(E,L,t)$ as a function of phase-space coordinates, one needs to map points in the integral-of-motion space onto phase space through the equation
\begin{align}
\label{eq:intspace}
f({\mathbfit r},{\mathbfit v},t)\d^6 \Omega &= f(E,L,t)\bigg|\frac{\partial ({\mathbfit r},{\mathbfit v})}{\partial (E,L)}\bigg|\d E\d L \\ \nonumber
&\equiv N(E,L,t)\d E\d L,
\end{align} 
where $f=N/\omega$ is the probability of finding a particle with integrals $(E,L)$ in the phase-space volume $\d^6\Omega$ centred at the coordinates $({\mathbfit r},{\mathbfit v})$ at the time $t$, and $[\omega]\equiv\partial ({\mathbfit r},{\mathbfit v})/\partial(E,L)$ is the {\it matrix of density of states}. The Jacobian of this matrix, $\omega(E,L,t)$, defines the maximum phase-space volume that particles with a given combination of energy and angular momentum can access. The simplest derivation of $\omega$ is found by integrating the volume $\d^6\Omega=\d^3{\mathbfit r}\d^3{\mathbfit v}=4\pi r^2\d r \times 2\pi v_t\d v_t\d v_r$ over the accessible spherical shell, which yields (Spitzer 1987)
\begin{align}
  \label{eq:intspace}
  N(E,L,t)\d E\d L&=f(E,L,t)\,2\int_{R_p}^{R_a} 4\pi r^2\d r \times 2\pi v_t\d v_t\d v_r \\ \nonumber
  &= f(E,L,t)  \,16\pi^2 L\int_{R_p}^{R_a}\frac{\d r}{v_r} \d E \d L \\ \nonumber
  &\equiv f(E,L,t)\omega(E,L)\d E \d L ,
\end{align}
with a density of states
\begin{align}\label{eq:omega}
  \omega(E,L)=16\pi^2 L\int_{R_p}^{R_a}\frac{\d r}{v_r}=8\pi^2 L \,P(E,L).
   \end{align}
Here $P(E,L)$ denotes to the orbital period
\begin{align}\label{eq:P}
P(E,L)=2\int_{R_p}^{R_a} \frac{\d r}{v_r}=2\int_{R_p}^{R_a} \frac{\d r}{ \{2[E-\Phi(r)-L^2/(2r^2)]\}^{1/2}},
\end{align}
with $R_p$ and $R_a$ being the peri- and apo-centres of the orbit, respectively, which correspond to the radii where the radial velocity vanishes, $v_r=\{2[E-\Phi_s(R)-L^2/(2R^2)]\}^{1/2}=0$.

Tracer particles with a given combination of energy and angular momentum follow a delta function in the integral-of-motion space, $N(E,L)=\delta(E-H)\delta(L-r v_t)$. The distribution function $f=N/\omega$ can be used to compute the average of a generic function $X(r,v)$ in phase-space. Combination of~(\ref{eq:intspace}),~(\ref{eq:omega}) and~(\ref{eq:P}) yields
\begin{align}\label{eq:aver}
  \bar{X}(E,L)&=\frac{1}{\omega}\int \d^3{\mathbfit r}\d^3{\mathbfit v}  \delta(E-H)\delta(L-r v_t)X(r,v) \\ \nonumber\\ \nonumber
 &=\frac{1}{\omega}  16\pi^2 L\int_{R_p}^{R_a}\frac{\d r}{v_r} X(r,v)\\ \nonumber
   &=\frac{2}{P(E,L)}\int_{R_p}^{R_a} \frac{\d r}{v_r}X(r,v) \\ \nonumber
\end{align}
which weights a region of the orbit $\d r$ by the fraction of the orbital period that a particle spends in that region, $(2\d r/v_r)/P$. Hence, Equation~(\ref{eq:aver}) establishes an {\it ergodic correspondence} between ensemble and time averages. It is important to bear in mind that the ergodic property only holds for particle distributions that are fully mixed in phase space (see P15).


The averaged quantities appearing in Equation~(\ref{eq:CELimp}) can be calculated analytically using~(\ref{eq:aver}) as
\begin{align}\label{eq:r2a}
  \overline{R^m}(E,L)=\frac{2}{P(E,L)}\int_{R_p}^{R_a} \d r \frac{ r^m}{ \{2[E-\Phi(r)-L^2/(2r^2)]\}^{1/2}},
\end{align}
for $m=2,4$, and
\begin{align}\label{eq:v2a}
  \overline{R^2 V^2}(E,L)=\frac{2}{P(E,L)}\int_{R_p}^{R_a} \d r \frac{ 2[E-\Phi(r)] r^2}{ \{2[E-\Phi(r)-L^2/(2r^2)]\}^{1/2}},
\end{align}
which in general must be integrated numerically. An exception is the Keplerian potential~(\ref{eq:phis_kep}), which we inspect in some detail below.

\subsection{Keplerian potential}\label{sec:kep}
As a first step, it is useful to express the orbital energy and angular momentum in terms of a semi-major axis ($a$) and eccentricity ($e$) using Equation~(\ref{eq:ae}).
The peri- and apocentres of the orbit correspond to the radii where $v_r=\{2[E-\Phi_s(R)-L^2/(2R^2)]\}^{1/2}=\{2[-G m/(2a)+G m/R-G m\,a(1-e^2)/(2R^2)]\}^{1/2}=0$, which admits two solutions
\begin{align}\label{eq:rpra}
  R_p&=a(1-e)\\\nonumber
   R_a&=a(1+e).
\end{align}
The orbital period can be derived analytically from~(\ref{eq:P}) by changing the integration variable to $\nu=r/a$ and using the roots of the radial velocity~(\ref{eq:rpra}), which yields
\begin{align}\label{eq:P_kep}
  P&=2\int_{a(1-e)}^{a(1+e)} \d r\frac{1}{\{2[-G m/(2a)+G m/R-G m\,a(1-e^2)/(2R^2)]\}^{1/2}}\\ \nonumber
     &= \frac{2 a^{3/2}}{\sqrt{G m}}\int_{1-e}^{1+e}\d \nu\frac{\nu}{\{[(1+e)-\nu][\nu-(1-e)]\}^{1/2}} \\ \nonumber
    &=\frac{2 \pi a^{3/2}}{\sqrt{G m}},
\end{align}
hence, the orbital period is independent of the orbital angular momentum.

Similarly, the averaged second and fourth power of the radius~(\ref{eq:r2a}) can be written as
\begin{align}\label{eq:r2_kep}
  \overline{R^2}&=\frac{2}{P}\int_{a(1-e)}^{a(1+e)} \d r\frac{r^2}{\{2[-G m/(2a)+G m/R-G m\,a(1-e^2)/(2R^2)]\}^{1/2}}\\ \nonumber
     &= \frac{2 a^{3/2}}{\sqrt{G m}}\frac{a^2}{P}\int_{1-e}^{1+e}\d \nu\frac{\nu^3}{\{[(1+e)-\nu][\nu-(1-e)]\}^{1/2}} \\ \nonumber
    &=a^2\big(1+\frac{3}{2}e^2\big),
\end{align}
and
\begin{align}\label{eq:r4_kep}
  \overline{R^4}&=\frac{2}{P}\int_{a(1-e)}^{a(1+e)} \d r\frac{r^4}{\{2[-G m/(2a)+G m/R-G m\,a(1-e^2)/(2R^2)]\}^{1/2}}\\ \nonumber
     &= \frac{2 a^{3/2}}{\sqrt{G m}}\frac{a^4}{P}\int_{1-e}^{1+e}\d \nu\frac{\nu^5}{\{[(1+e)-\nu][\nu-(1-e)]\}^{1/2}} \\ \nonumber
  &=a^4\big(1+5e^2 +\frac{15}{8}e^4\big).
\end{align}
The averaged squared velocity is
\begin{align}\label{eq:v2_kep}
  \overline{V^2}&=\frac{2}{P}\int_{a(1-e)}^{a(1+e)} \d r\frac{2[-G m/(2a)+G m/r]}{\{2[-G m/(2a)+G m/R-G m\,a(1-e^2)/(2R^2)]\}^{1/2}}\\ \nonumber
  &=\frac{4\sqrt{G m}}{P}a^{1/2}\int_{1-e}^{1+e}\d \nu\frac{(1-\nu/2)}{\{[(1+e)-\nu][\nu-(1-e)]\}^{1/2}}\\ \nonumber
  &=\frac{G m}{a},
 \end{align} 
while the integral~(\ref{eq:v2a}) has an analytical solution
\begin{align}\label{eq:r2v2_kep}
  \overline{R^2 V^2}&=\frac{2}{P}\int_{a(1-e)}^{a(1+e)} \d r\frac{2[-G m/(2a)+G m/r]r^2}{\{2[-G m/(2a)+G m/R-G m\,a(1-e^2)/(2R^2)]\}^{1/2}}\\ \nonumber
  &=\frac{4\sqrt{G m}}{P}a^{5/2}\int_{1-e}^{1+e}\d \nu\frac{\nu^2(1-\nu/2)}{\{[(1+e)-\nu][\nu-(1-e)]\}^{1/2}}\\ \nonumber
  &=G m\,a\big(1-\frac{e^2}{2}\big).
\end{align}
Note that $\overline{R^2 V^2}\ne \overline{R^2}\times\overline{V^2}$ for $e\ne 0$ (non-circular orbits).


\begin{thebibliography}{}
  
\bibitem[Angulo et al.(2013)]{2013MNRAS.434.3337A} Angulo, R.~E., Hahn, O., \& Abel, T.\ 2013, \mnras, 434, 3337 


\bibitem[Batygin \& Brown(2016)]{2016AJ....151...22B} Batygin, K., \& Brown, M.~E.\ 2016, \aj, 151, 22 

\bibitem[Becker et al.(2018)]{2018AJ....156...81B} Becker, J.~C., Khain, T., Hamilton, S.~J., et al.\ 2018, \aj, 156, 81 

\bibitem[Benson(2017)]{2017MNRAS.467.3454B} Benson, A.~J.\ 2017, \mnras, 467, 3454

  \bibitem[Berezinsky et al.(2003)]{2003PhRvD..68j3003B} Berezinsky, V., Dokuchaev, V., \& Eroshenko, Y.\ 2003, \prd, 68, 103003 

  
\bibitem[Carslaw \& Jaeger(1986)]{1986chs..book.....C} Carslaw, H.~S., \& Jaeger, J.~C.\ 1986, Conduction of Heat in Solids, by H S Carslaw and J C Jaeger, pp.~520.~Oxford University Press, Apr 1986.~ISBN-10: 0198533683.~ISBN-13: 9780198533689, 520


\bibitem[Chandrasekhar(1941)]{1941ApJ....93..285C} Chandrasekhar, I.~S.\ 1941a, \apj, 93, 285 

  
\bibitem[Chandrasekhar(1941)]{1941ApJ....94..511C} Chandrasekhar, S.\ 1941b, \apj, 94, 511 

\bibitem[Chandrasekhar \& von Neumann(1942)]{1942ApJ....95..489C} Chandrasekhar, S., \& von Neumann, J.\ 1942, \apj, 95, 489 


\bibitem[Chandrasekhar \& von Neumann(1943)]{1943ApJ....97....1C} Chandrasekhar, S., \& von Neumann, J.\ 1943, \apj, 97, 1 

\bibitem[Chandrasekhar(1943)]{1943RvMP...15....1C} Chandrasekhar, S.\ 1943, Reviews of Modern Physics, 15, 1

  
\bibitem[Chandrasekhar(1944)]{1944ApJ....99...25C} Chandrasekhar, S.\ 1944, \apj, 99, 25 



\bibitem[Chavanis(2009)]{2009EPJB...70..413C} Chavanis, P.~H.\ 2009, European Physical Journal B, 70, 413


\bibitem[Cohen et al.(1950)]{1950PhRv...80..230C} Cohen, R.~S., Spitzer, L., \& Routly, P.~M.\ 1950, Physical Review, 80, 230 



\bibitem[Daniel et al.(2017)]{2017MNRAS.468.1453D} Daniel, K.~J., Heggie, D.~C., \& Varri, A.~L.\ 2017, \mnras, 468, 1453 


\bibitem[Diemand et al.(2005)]{2005Natur.433..389D} Diemand, J., Moore, B., \& Stadel, J.\ 2005, \nat, 433, 389 



\bibitem[Diemand et al.(2007)]{2007ApJ...667..859D} Diemand, J., Kuhlen, M., \& Madau, P.\ 2007, \apj, 667, 859 

\bibitem[D'Onghia et al.(2010)]{2010ApJ...709.1138D} D'Onghia, E., Springel, V., Hernquist, L., \& Keres, D.\ 2010, \apj, 709, 1138 

\bibitem[Eddington(1916)]{1916MNRAS..76..572E} Eddington, A.~S.\ 1916, \mnras, 76, 572 

\bibitem[Eilers et al.(2018)]{2018arXiv181009466E} Eilers, A.-C., Hogg, D.~W., Rix, H.-W., \& Ness, M.\ 2018, arXiv:1810.09466 

\bibitem[Erkal et al.(2016)]{2016MNRAS.463..102E} Erkal, D., Belokurov, V., Bovy, J., \& Sanders, J.~L.\ 2016, \mnras, 463, 102 

\bibitem[Errani et al.(2017)]{2017MNRAS.465L..59E} Errani, R., Pe{\~n}arrubia, J., Laporte, C.~F.~P., \& G{\'o}mez, F.~A.\ 2017, \mnras, 465, L59 

  
\bibitem[Feller(1971)]{1971aitp.book.....F} Feller, W.\ 1971, Wiley Series in Probability and Mathematical Statistics, New York: Wiley, 1971, 3rd ed.,  
  
\bibitem[Gnedin \& Ostriker(1999)]{1999ApJ...513..626G} Gnedin, O.~Y., \& Ostriker, J.~P.\ 1999, \apj, 513, 626
  
\bibitem[Green et al.(2005)]{2005JCAP...08..003G} Green, A.~M., Hofmann, S., \& Schwarz, D.~J.\ 2005, JCAP, 8, 003 


\bibitem[Green \& Goodwin(2007)]{2007MNRAS.375.1111G} Green, A.~M., \& Goodwin, S.~P.\ 2007, \mnras, 375, 1111
  
\bibitem[Han et al.(2016)]{2016MNRAS.457.1208H} Han, J., Cole, S., Frenk, C.~S., \& Jing, Y.\ 2016, \mnras, 457, 1208
  
  
\bibitem[Heggie \& Rasio(1996)]{1996MNRAS.282.1064H} Heggie, D.~C., \& Rasio, F.~A.\ 1996, \mnras, 282, 1064


\bibitem[Hernquist(1990)]{1990ApJ...356..359H} Hernquist, L.\ 1990, \apj, 356, 359 

\bibitem[Hills(1981)]{1981AJ.....86.1730H} Hills, J.~G.\ 1981, \aj, 86, 1730

\bibitem[Hiroshima et al.(2018)]{2018PhRvD..97l3002H} Hiroshima, N., Ando, S., \& Ishiyama, T.\ 2018, \prd, 97, 123002 


  
\bibitem[Hofmann et al.(2001)]{2001PhRvD..64h3507H} Hofmann, S., Schwarz, D.~J., \& St{\"o}cker, H.\ 2001, \prd, 64, 083507 

  
\bibitem[Holtsmark(1919)]{1919AnP...363..577H} Holtsmark, J.\ 1919, Annalen der Physik, 363, 577
  
\bibitem[Ishiyama(2014)]{2014ApJ...788...27I} Ishiyama, T.\ 2014, \apj, 788, 27 

\bibitem[Just \& Pe{\~n}arrubia(2005)]{2005A&A...431..861J} Just, A., \& Pe{\~n}arrubia, J.\ 2005, \aap, 431, 861 


\bibitem[Kandrup(1980)]{1980PhR....63....1K} Kandrup, H.~E.\ 1980, \physrep, 63, 1 

\bibitem[Lee(1968)]{1968ApJ...151..687L} Lee, E.~P.\ 1968, \apj, 151, 687 


\bibitem[Loeb \& Zaldarriaga(2005)]{2005PhRvD..71j3520L} Loeb, A., \& Zaldarriaga, M.\ 2005, \prd, 71, 103520

\bibitem[Lovell et al.(2014)]{2014MNRAS.439..300L} Lovell, M.~R., Frenk, C.~S., Eke, V.~R., et al.\ 2014, \mnras, 439, 300 

\bibitem[Lynden-Bell(1999)]{1999PhyA..263..293L} Lynden-Bell, D.\ 1999, 
  Physica A Statistical Mechanics and its Applications, 263, 293
  
\bibitem[Navarro et al.(1997)]{1997ApJ...490..493N} Navarro, J.~F., Frenk, C.~S., \& White, S.~D.~M.\ 1997, \apj, 490, 493 



\bibitem[Oort(1950)]{1950BAN....11...91O} Oort, J.~H.\ 1950, \bain, 11, 91
\bibitem[Padmanabhan(1990)]{1990PhR...188..285P} Padmanabhan, T.\ 1990, 
\physrep, 188, 285 

\bibitem[Pe{\~n}arrubia et al.(2010)]{2010MNRAS.406.1290P} Pe{\~n}arrubia, J., Benson, A.~J., Walker, M.~G., et al.\ 2010, \mnras, 406, 1290
  
\bibitem[Pe{\~n}arrubia(2013)]{2013MNRAS.433.2576P} Pe{\~n}arrubia, J.\ 2013, \mnras, 433, 2576 


\bibitem[Pe{\~n}arrubia(2015)]{2015MNRAS.451.3537P} Pe{\~n}arrubia, J.\ 2015, \mnras, 451, 3537 


\bibitem[Pe{\~n}arrubia(2018)]{2018MNRAS.474.1482P} Pe{\~n}arrubia, J.\ 2018, \mnras, 474, 1482  (Paper I)


\bibitem[Press et al.(1992)]{1992nrfa.book.....P} Press, W.~H., et al.\ 1992, Cambridge: University Press,  2nd ed.  

\bibitem[Renaud et al.(2011)]{2011MNRAS.418..759R} Renaud, F., Gieles, M., \& Boily, C.~M.\ 2011, \mnras, 418, 759 

\bibitem[S{\'a}nchez-Conde \& Prada(2014)]{2014MNRAS.442.2271S} S{\'a}nchez-Conde, M.~A., \& Prada, F.\ 2014, \mnras, 442, 2271 

\bibitem[Schmid et al.(1999)]{1999PhRvD..59d3517S} Schmid, C., Schwarz, D.~J., \& Widerin, P.\ 1999, \prd, 59, 043517 

\bibitem[Schneider et al.(2012)]{2012MNRAS.424..684S} Schneider, A., Smith, R.~E., Macci{\`o}, A.~V., \& Moore, B.\ 2012, \mnras, 424, 684 

\bibitem[Smoluchowski(1916)]{1916ZPhy...17..557S} Smoluchowski, M.~V.\ 1916, Zeitschrift fur Physik, 17, 557 

\bibitem[Spitzer(1958)]{1958ApJ...127...17S} Spitzer, L., Jr.\ 1958, \apj, 127, 17 
\bibitem[Spitzer(1987)]{1987degc.book.....S} Spitzer, L.\ 1987, Princeton, 
NJ, Princeton University Press, 1987, 191 p.,  

\bibitem[Spitzer \& Shapiro(1972)]{1972ApJ...173..529S} Spitzer, L., Jr., \& Shapiro, S.~L.\ 1972, \apj, 173, 529

\bibitem[Springel et al.(2008)]{2008MNRAS.391.1685S} Springel, V., Wang, J., Vogelsberger, M., et al.\ 2008, \mnras, 391, 1685 

  
\bibitem[Stadel et al.(2009)]{2009MNRAS.398L..21S} Stadel, J., Potter, D., Moore, B., et al.\ 2009, \mnras, 398, L21 

\bibitem[Stref \& Lavalle(2017)]{2017PhRvD..95f3003S} Stref, M., \& Lavalle, J.\ 2017, \prd, 95, 063003
  
\bibitem[van den Bosch(2017)]{2017MNRAS.468..885V} van den Bosch, F.~C.\ 2017, \mnras, 468, 885 


\bibitem[van den Bosch \& Ogiya(2018)]{2018MNRAS.475.4066V} van den Bosch, F.~C., \& Ogiya, G.\ 2018, \mnras, 475, 4066 


\bibitem[Weinberg(1994)]{1994AJ....108.1398W} Weinberg, M.~D.\ 1994a, \aj, 108, 1398 

\bibitem[Weinberg(1994)]{1994AJ....108.1403W} Weinberg, M.~D.\ 1994b, \aj, 108, 1403 


\bibitem[Weinberg(1994)]{1994AJ....108.1414W} Weinberg, M.~D.\ 1994c, \aj, 108, 1414 


\end{thebibliography}
\end{document}